\documentclass[reqno]{article}
\usepackage[ansinew]{inputenc}
\usepackage{amsmath}
\usepackage{amssymb}
\usepackage{amscd}
\usepackage{graphicx}
\usepackage{color}
\usepackage{epsfig}
\usepackage{graphicx}
\usepackage{lscape,graphicx}

\makeatletter \@addtoreset{equation}{section} \makeatother
\renewcommand{\theequation}{\thesection.\arabic{equation}}

\makeatletter
\let\old@startsection=\@startsection
\renewcommand{\@startsection}[6]{\old@startsection{#1}{#2}{#3}{#4}{#5}{#6\mathversion{bold}}}
\makeatother

\newcommand{\beq}{\begin{equation}}
\newcommand{\eeq}{\end{equation}}
\newcommand{\beqa}{\begin{eqnarray}}
\newcommand{\eeqa}{\end{eqnarray}}
\newcommand{\ba}{\begin{array}}
\newcommand{\ea}{\end{array}}
\newcommand{\nn}{\nonumber}

\newcommand{\neqa}{\nonumber\end{eqnarray}}

\newcommand{\p}{\partial}

\newcommand{\eq}[1]{eq.(\ref{#1})}
\newcommand{\rf}[1]{(\ref{#1})}

\newcommand{\half}{\frac{1}{2}}

\newcommand{\<}{{\langle}}
\renewcommand{\>}{{\rangle}}

\newcommand{\re}{\relax{\rm I\kern-.18em R}}

\def\su2{{SU(2)}}

\def\r{{\rm r}}

\def\a{{\alpha}}

\def\({\left(}
\def\){\right)}
\def\[{\left[}
\def\]{\right]}

\def\l{\lambda}

\def\a{\alpha}
\def\b{\beta}
\def\th{\theta}

\def\({\left(}
\def\){\right)}
\def\[{\left[}
\def\]{\right]}

\def\<{\langle}
\def\>{\rangle}

\begin{document}

\renewcommand{\thefootnote}{\fnsymbol{footnote}}
\setcounter{footnote}{0}

\thispagestyle{empty}
\begin{flushright}
LPTENS-07/12\\
JINR-E2-2007-36\\
ITEP-TH-03/07
\end{flushright}
\vspace{.5cm} \setcounter{footnote}{0}
\begin{center}
{\Large{\bf Supersymmetric Bethe Ansatz and Baxter Equations from
Discrete
Hirota Dynamics}\\
   }\vspace{4mm}
{\large  Vladimir~Kazakov$^{a,\!}$\footnote{Membre de l'Institut
Universitaire de France},\hspace{1cm} Alexander~Sorin\rm
$^{b}$,\hspace{1cm}
Anton~Zabrodin\rm $^{c,d}$\\[7mm]
\large\it\small ${}^a$ Laboratoire de Physique Th\'eorique\\
de l'Ecole Normale Sup\'erieure et l'Universit\'e Paris-VI,\\
24 rue Lhomond, Paris  CEDEX 75231, France\footnote{ \tt\noindent
Email:\indent   kazakov@physique.ens.fr,
 \indent  sorin@theor.jinr.ru,
 \indent zabrodin@itep.ru}
 \vspace{3mm}\\
\large\it\small ${}^b$  Bogoliubov Laboratory of Theoretical Physics,\\
Joint Institute for Nuclear Research,\\
 141980  Dubna (Moscow Region), Russia
 \vspace{3mm}\\
\large\it\small ${}^c$  Institute of Biochemical Physics, \\
 Kosygina str. 4, 119991, Moscow, Russia
   \vspace{3mm}\\
 \large\it\small ${}^d$  Institute of Theoretical and Experimental Physics,
 \\ Bol.~Cheremushkinskaya str. 25, 117259, Moscow, Russia
}

\end{center}
\noindent\\[20mm]
\begin{center}
{\sc Abstract}\\[2mm]
\end{center}

We show that eigenvalues of the family of Baxter $Q$-operators for
supersymmetric integrable spin chains constructed with the $gl
(K|M)$-invariant $R$-matrix obey the Hirota bilinear difference
equation. The nested Bethe ansatz for super spin chains, with any
choice of simple root system, is  then treated as a discrete
dynamical system for zeros of polynomial solutions to the Hirota
equation. Our basic tool is a chain of B\"acklund transformations
for the Hirota equation connecting quantum transfer matrices. This
approach also provides a systematic way to derive the complete set
of generalized Baxter equations for super spin chains.

\newpage

\setcounter{page}{1}
\renewcommand{\thefootnote}{\arabic{footnote}}
\setcounter{footnote}{0}

\tableofcontents

\newpage

\section{Introduction}

\subsection{Motivation and background}

Supersymmetric extensions  of quantum integrable spin chains were
proposed long ago \cite{KulSk,Kulish} but the proper generalization
of the standard methods such as algebraic Bethe ansatz and Baxter
$TQ$-relations is still not so well understood, as compared to the
case of integrable models with usual symmetry algebras, and still
contains some elements of guesswork.

Bethe ansatz equations for integrable models based on superalgebras
are believed to be written according to the general ``empirical"
rules \cite{RW} applied to graded Lie algebras. Accepting this as a
departure point, one can try to reconstruct  other common
ingredients of the theory of quantum integrable systems such as
Baxter relations and fusion rules. For the super spin chains based
on the rational or trigonometric $R$-matrices, the algebraic Bethe
ansatz works rather similarly to the case of usual ``bosonic" spin
chains, but the Bethe ansatz equations for a given model do  not
have a unique form and depend on the choice of the system of simple
roots. The situation becomes even more complicated when one
considers spins in higher representations of the superalgebra, and
especially typical ones containing continuous Kac-Dynkin labels. The
systematic description of all possible Bethe ansatz equations and
$TQ$-relations becomes then a cumbersome task
\cite{Maassarani}-\cite{pfannmuller-1996-479} not completely fulfilled in
the literature. To our knowledge, a unified approach is still
missing.

In this article we propose a new approach to the supersymmetric spin
chains based on the Hirota-type relations for quantum transfer
matrices and B\"acklund transformations for them. Functional
relations between commuting quantum transfer matrices are known to
be a powerful tool for solving quantum integrable models. They are
based on the fusion rules for various irreducible representations in
the auxiliary space of the model. First examples were given in
\cite{KulResh1,Resh}. Later, these functional relations were
represented in the determinant form \cite{BR1} and in the form of
the Hirota bilinear difference equation
\cite{KlumperPearce,Kuniba-0}. In
\cite{Tsuboi-1}-\cite{Tsuboi-4} it was shown that
transfer matrices in the supersymmetric case are subject to exactly
the same functional equations as in the purely bosonic case. The
Hirota form of the functional equations has been proved to be
especially useful and meaningful \cite{KLWZ,Z1}. It is the starting
point for our construction.

The Hirota equation \cite{Hirota} is probably the most famous
equation in the theory of classical integrable systems on the
lattice. It provides a universal integrable discretization of
various soliton equations and, at the same time, serves as a
generating equation for their hierarchies. In this sense, it is a
kind of a Master equation for the theory of solitons. It covers a
great variety of integrable problems, classical and quantum.

In our approach, quantization and discretization appear to be
closely related in the sense that solutions to the quantum problems
are given in terms of the discrete classical dynamics. What
specifies  the problem are the boundary and analytic conditions for
the variables entering the Hirota equation. In applications to
quantum spin chains, these variables are parameters of the
representation of the symmetry algebra in the auxiliary space. For
representations associated with rectangular Young diagrams they are
height and length of the diagram denoted by $a$ and $s$
respectively. In the case of spin chains based on usual (bosonic)
algebras $gl(K)$, the boundary conditions are such that the
non-vanishing transfer matrices live in a strip $0\le a\le K$ in the
$(a,s)$ plane \cite{KLWZ,Z1}. In the case of superalgebras of the
type $gl(K|M)$, the strip turns into a domain of the ``fat hook"
type presented in Fig.~\ref{fig:HirotaHook}.

In the nested Bethe ansatz scheme for $gl(K)$, one successively
lowers the rank of the algebra $gl(K)\to gl(K\! -\! 1)$ (and thus
the width of the strip) until the problem gets fully ``undressed".
This purely quantum technique has a remarkable ``classical face": it
is equivalent to a chain of B\"acklund transformations for the
Hirota equation \cite{KLWZ,Z1}. They stem from the discrete zero
curvature representation and associated systems of auxiliary linear
equations. This solves the discrete Hirota dynamics in terms of
Bethe equations or the general system of Baxter's $TQ$-relations.

The aim of this article is to
extend this program to the models based on superalgebras $gl(K|M)$.
In this case, there exist two different types of B\"acklund
transformations. One of them lowers $K$ while the other one lowers
$M$. The undressing goes until the fat hook is collapsed to $K=M=0$.
The result of this procedure does not depend on the order in which
we perform these transformations but the form of the equations does.
Different orders lead to different types of Bethe ansatz equations
and Baxter's $TQ$-relations associated with Cartan matrices for
different systems of simple roots. In this way, the abundance of
various Bethe equations and $TQ$-relations is easily explained and
classified. All of them are constructed in our paper. Instead of
$K+1$ Baxter's $Q$-functions for the bosonic $gl(K)$ algebra we
recover $(K\! +\! 1)(M\! +\! 1)$ $Q$-functions (some of them are
initially fixed by the physical problem).  More than that, we
establish a new equation relating all these $Q$-functions which is
again of the Hirota type. This ``$QQ$-relation" opens the most
direct and easiest way to construct various systems of Bethe
equations. Similar relations hold for the transfer matrices at each
step of the undressing procedure.

Our construction goes through when observables in the Hilbert
space of the generalized spin chain are in arbitrary finite
dimensional representations of the symmetry (super)algebra.
We can also successfully incorporate the case of typical
representations carrying the continuous Kac-Dynkin labels,
as it is illustrated by examples of superalgebras
$gl(1|1)$ and $gl(2|1)$.

Some standard facts and notation related to superalgebras and their
representations are listed in Appendix~A. For details see
\cite{Kac}-\cite{GouldZhang}.
Throughout the paper, we use the language of the algebraic
Bethe ansatz and the quantum inverse
scattering method on the lattice developed in \cite{FT} (see also
reviews \cite{KulSk,Faddeev}
and book \cite{book}). On the other hand, we employ standard  methods of
classical theory of solitons \cite{ZMNP} and discrete integrable equations
\cite{Miwa}-\cite{Z2}.

\subsection{A sketch of the results}

\begin{figure}[t]
    \centering
        \includegraphics[angle=-90,scale=0.4]{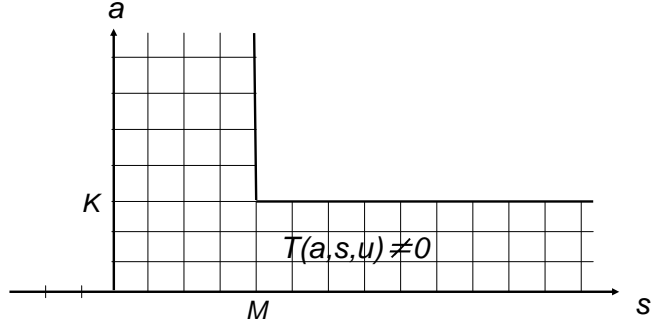}
    \caption{\it  The domain (``fat hook") of
non-vanishing transfer matrices
    $T(a,s,u)$  for the
 supersymmetric spin chain with the $gl(K|M)$ symmetry. }
    \label{fig:HirotaHook}
\end{figure}

We consider integrable generalized spin chains with
$gl(K|M)$-invariant $R$-matrix. The generating function of commuting
integrals of motion is the quantum transfer matrix
$T^{(\lambda )}(u)$, which depends on the spectral parameter
$u\in {\mathbb C}$
and the Young diagram $\lambda$.
It is obtained as the (super)trace
of the quantum monodromy matrix ${\cal T}^{(\lambda )}(u)$ in the
auxiliary space carrying the irreducible representation of the
symmetry algebra labeled by $\lambda$:
$$
T^{(\lambda )}(u)=\mbox{str}_{{\rm aux}}\, {\cal T}^{(\lambda )}(u)\,.
$$
We deal with covariant tensor irreducible representations (irreps)
of the superalgebra.

The transfer matrices for different $\lambda$ are known to be
functionally dependent. For rectangular diagrams $\lambda =(s^a)$
with $a$ rows and $s$ columns the functional relation takes the form
of the famous Hirota difference equation:
\begin{equation}
\label{HIROTA0}
  T(a,s,u \! +\! 1)T(a,s,u\! -\! 1)-
T(a,s \! +\! 1,u)T(a,s\! -\! 1,u) =T(a\! +\! 1,s,u)T(a\! -\! 1,s,u),
\end{equation}
where the transfer matrix for rectangular diagrams, after some
$a,s$-dependent shift of the spectral parameter $u$, is denoted by
$T(a,s,u)$. We see that it enters the equation as the
$\tau$-function \cite{JimboMiwa}.
Since all the $T$'s commute, the same relation
holds for their eigenvalues. We use the normalization in which all
non-vanishing $T$'s are polynomials in $u$ of degree $N$, where $N$
is the number of sites in the spin chain.

While the functional relation is the same for ordinary and
super-algebras, the boundary conditions are different. In the
$gl(K)$-case, the domain of non-vanishing $T$'s in the $(s,a)$ plane
is the half-strip $s\geq 0$, $0\leq a \leq K$. In the
$gl(K|M)$-case, the domain of non-vanishing $T$'s has the form of a
``fat hook". It is shown in Fig.~\ref{fig:HirotaHook}. To ensure
compatibility with the Hirota equation, the boundary values at $s=0$
and $a=0$ should be rather special. In our normalization,
\begin{equation}\label{normalization}
T(0,s,u)=\phi (u-s)\,, \quad \quad T(a, 0,u)=\phi (u+a)\,,
\end{equation}
where $\phi (u)=\prod_i (u-\theta_i)$ is a fixed polynomial of
degree $N$ which characterizes the spin chain.

Our goal is to solve the Hirota equation, with the boundary
conditions given above, using the classical methods of the theory of
solitons. This program for the case of ordinary Lie algebras (and
for models with elliptic $R$-matrices) was realized in \cite{KLWZ}.
In this paper, we extend it to the case of superalgebras (for models
with rational $R$-matrices).

One of the key features of soliton equations is the existence of
(auto) B\"acklund transformations (BT), i.e., transformations that
send any solution of a soliton equation to another solution of the
same equation. A systematic way to construct such transformations is
provided by considering an over-determined system of linear problems
(called auxiliary linear problems)
whose compatibility condition is the non-linear equation at hand. We
introduce two B\"acklund transformations  for the Hirota equation.
They send any solution with the boundary conditions described above
to a solution of the same Hirota equation with boundary conditions
of the same fat hook type but with different $K$ or $M$.
Specifically, one of them lowers $K$ by $1$ and the other one lowers
$M$ by $1$ leaving all other boundaries intact. We call these
transformations BT1 and BT2. Applying them successively $K+M$ times,
one comes to a collapsed domain which is a union of two lines, the
$s$-axis and the $a$-semi-axis shown in Fig.\ref{fig:HirotaHook},
meaning that the original problem gets ``undressed" to a trivial
one. This procedure appears to be equivalent to the nested Bethe
ansatz. Different orders in which we diminish $K$ and $M$ give raise
to different ``dual" systems of nested Bethe Ansatz equations. All
of them describe the same system. They correspond to different
choices of the basis of simple roots.

Let $k,m$ be indices running from $0$ to $K$ and from $0$ to $M$
respectively, and let $T_{k,m}(a,s,u)$ be the transfer matrices
obeying the Hirota equation with the boundary conditions as above
but with $K,M$ replaced by $k,m$. They are obtained from
$T_{K,M}(a,s,u)$ by a chain of BT's. Namely, $T_{k-1,m}$'s and
$T_{k,m-1}$'s are solutions to the auxiliary linear problems for the
Hirota equation for $T_{k,m}$'s:
$$
\begin{CD}
T_{k, m\! -\! 1}@<{\rm BT2}<< T_{k, m}\\
@. @VV{\rm BT1}V\\
@. \;\;\;\; T_{k\! -\! 1, m}
\end{CD}
$$
The explicit formulas of these transformations are given below
(\eq{LINPRT1} and \eq{LINPRT2}). The functions $T_{k,m}(a,s,u)$ are
polynomials in $u$ for any $a,s,k,m$ but the degree depends only on
$k,m$. Let $Q_{k,m}(u)$ be the boundary values of the
$T_{k,m}(a,s,u)$, i.e.,
$$
T_{k,m}(0,s,u)=Q_{k,m}(u-s)\,, \quad \quad
T_{k,m}(a,0,u)=Q_{k,m}(u+a)\,.
$$
However, they are not fixed for
the values of $k,m$ other than $k=K$, $m=M$ (when $Q_{K,M}(u)=\phi (u)$)
and $k=m=0$ (when $Q_{0,0}(u)$ is
put equal to $1$) but are to be determined from a solution to the
hierarchy of Hirota equations. In fact these $Q$'s are Baxter
polynomial functions whose roots obey the Bethe equations.

The result of a successive application of the transformations BT1
and BT2 does not depend on their order. This fact can be
reformulated as a discrete zero curvature condition
\begin{equation}\label{ZC0}
\hat U_{k,m+1}^{-1}\hat V_{k,m}=\hat V_{k+1,m}\hat U_{k,m}^{-1}
\end{equation}
for the shift operators in $k$ and $m$:
\begin{eqnarray} \label{NOTAT0}
\hat U_{k,m}(u)&=& \frac{Q_{k+1,m}(u)\, Q_{k,m}(u\! +\! 2)}
{Q_{k+1,m}(u\! +\! 2)\, Q_{k,m}(u)}
\, -\,\,e^{2\p_u},\nn\\
\hat V_{k,m}(u)&=& \frac{Q_{k,m}(u)\, Q_{k,m+1}(u\! +\!
2)}{Q_{k,m}(u\! +\! 2) \, Q_{k,m+1}(u)}\, -\,\, e^{2\p_u}.
\end{eqnarray}
Relation (\ref{ZC0}) is equivalent to the following  Hirota equation
for the Baxter $Q$-functions:
\begin{equation} \label{QHIROTA0}
  Q_{k,m}(u)Q_{k+1,m+1}(u+2)- Q_{k+1,m+1}(u)Q_{k,m}(u+2)=
  Q_{k,m+1}(u)Q_{k+1,m}(u+2).
\end{equation}
This equation represents our principal new result. By analogy with
Baxter's $TQ$-relations, we call eq.~(\ref{QHIROTA0}) {\it the
$QQ$-relation}. It provides the most transparent way to derive
different systems of Bethe equations for the generalized spin chain
and ``duality transformations" between them. We also show that a
number of more general Hirota equations of the similar type (i.e.,
acting in the space spanned by $k,m$ and a particular linear
combination of $a,s,u$) hold for the full set of functions
$T_{k,m}(a,s,u)$. They lead to a system of algebraic equations
for their roots which generalizes the system of Bethe equations.

The transfer matrices can be expressed through the $Q$-functions via
generalized Baxter's $TQ$-relations. A simple way to represent them
is to consider the (non-commutative) generating series of the
transfer matrices for one-row diagrams,
\begin{equation}\label{DIFFOP0}
\hat W_{k,m}(u)= Q_{k,m}^{-1}(u)\sum_{s=0}^\infty T_{k,m}(1,s,u\!
+\! s\! +\! 1)\, e^{2s\p_u},
\end{equation}
where the factor in front of the sum is put here for the proper
normalization. The operator $\hat W$ is similar to the wave (or
dressing) operator in the Toda lattice theory. We prove the
following fundamental operator relations:
\begin{equation}\label{WREL0}
\hat W _{k+1,m}=\hat U_{k,m}^{-1}\hat W _{k,m}\,, \quad \quad \hat W
_{k,m+1}=\hat V_{k,m} \hat W _{k,m}\,,
\end{equation}
which implement the B\"acklund transformations on the level of the
generating series. These relations allow one to represent $\hat
W_{K,M}$ (the quantity of prime interest) as an ordered product of
the operators $\hat U_{k,m}^{-1}$ and $\hat V_{k,m}$ along a zigzag
path from the point $(0,0)$ to the point $(K,M)$ on the $(k,m)$
lattice. This representation provides a concise form of the
generalized Baxter relations. Different zigzag paths correspond to
different choices of the basis of simple roots, i.e., to different
``dual" forms of supersymmetric Bethe equations (see
Fig.\ref{fig:UNDRESS}).

\begin{figure}[t]
    \centering
        \includegraphics[angle=-90,scale=0.4]{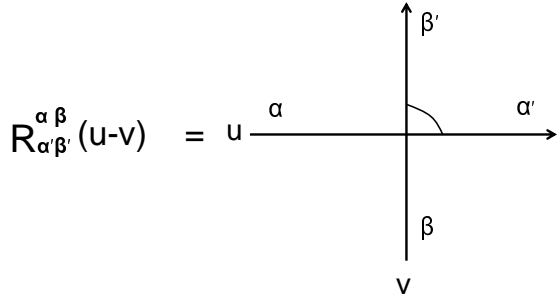}
    \caption{\it    The $R$-matrix. }
    \label{fig:Rmatr}
\end{figure}

The normalization (\ref{normalization}), with the roots $\theta_i$
of the polynomial $\phi (u)$ being in general position, implies that
the ``spins" of the inhomogeneous spin chain belong to the vector
representation of the algebra $gl(K|M)$. Higher representations can
be obtained by fusing of several copies of the vector ones.
According to the fusion procedure, the corresponding $\theta$'s must
be chosen in a specific ``string-like" way which means that the
differences between them are even integers constrained also by some
more specific requirements. This amounts to the fact that the
$Q$-functions become divisible by certain polynomials with
explicitly prescribed roots located according to a similar
string-like pattern. If one redefines the $Q$-functions dividing
them by these normalization factors, then the $QQ$-relation gets
modified and leads, in the same way as before, to the systems of
Bethe equations with non-trivial right hand sides (sometimes called
``vacuum parts").

In section 7, we specify our construction for two popular examples,
the $gl(1|1)$ and $gl(2|1)$ superalgebras. We present the
$TQ$- and $QQ$-relations for all possible types of $Q$-functions
$Q_{km}(u)$ ($0<k\le K,\quad 0<m\le M$) as well as Bethe
equations for all types of simple roots systems and
representations of the superalgebra (including typical
representations with a continuous component of the Kac-Dynkin label)
and compare them with
the results existing in the literature.
In our formalism, the construction becomes rather transparent and
algorithmic.
In this respect, our method is an interesting alternative to
the algebraic Bethe ansatz approach.

\section{Fusion relations for transfer
matrices and Hirota equation}

\subsection{Quantum transfer matrices}

\begin{figure}[t]
    \centering
        \includegraphics[angle=-90,scale=0.4]{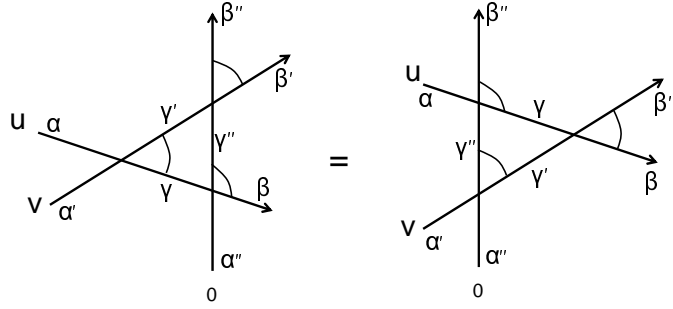}
     \caption{\it   The Yang-Baxter equation.}
    \label{fig:YBrel}
\end{figure}

Let $V= {\Bbb C}^{K}\oplus {\Bbb C}^{M}\equiv {\Bbb C}^{K|M}$ be the
graded linear space of the vector representation of the superalgebra
$gl(K|M)$. The fundamental $gl(K|M)$-invariant rational $R$-matrix
acts in $V\otimes V$ and has the form  \cite{KulSk,Kulish}
\begin{equation}\label{F1}
R(u)=u I + 2 {\mit\Pi}=u I+2\sum_{\gamma ,\gamma' =1}^{K+M}
(-1)^{p(\gamma')} E_{\gamma \gamma'}\otimes E_{\gamma' \gamma} \,.
\end{equation}
Here $I$ is the identity operator and $\mit\Pi$ is the
super-permutation, i.e., the operator such that ${\mit\Pi}(x\otimes
y)=(-1)^{p(x)p(y)}y\otimes x$ for any homogeneous vectors $x,y\in
V$, and $E_{\alpha \beta}$ are the generators of the (super)algebra.
In components, they read
$$
I^{\alpha \beta}_{\alpha' \beta'}= \delta_{\alpha
\alpha'}\delta_{\beta \beta'}\,, \quad \quad {\mit\Pi}^{\alpha
\beta}_{\alpha' \beta'}= (-1)^{p(\alpha ) p(\beta )} \delta_{\alpha
\beta'} \delta_{\alpha' \beta}\,, \quad \quad (E_{\gamma
\gamma'})^{\alpha}_{\alpha'}= \delta_{\alpha \gamma}\delta_{\alpha'
\gamma'}\,.
$$
Note that $\mit\Pi^2 =I$, as in the ordinary case. The notation
$p(x)$ is used to denote parity of the object $x$ (see Appendix A).
The $R(u)$ has only even matrix elements. The complex variable $u$
is called the spectral parameter. The $R$-matrix obeys the graded
Yang-Baxter equation (Fig.~\ref{fig:YBrel}):
\begin{equation}\label{F2}
\begin{array}{ll}
&\displaystyle{ \sum_{\gamma \gamma' \gamma''} (-1)^{p(\gamma'
)(p(\alpha'')+p(\gamma''))} R^{\alpha \alpha'}_{\gamma \gamma'}(u-v)
R^{\gamma \alpha''}_{\beta \gamma''}(u) R^{\gamma' \gamma''}_{\beta'
\beta''}(v)}
\\&\\
=& \displaystyle{ \sum_{\gamma \gamma' \gamma''} (-1)^{p(\gamma'
)(p(\beta'')+p(\gamma''))} R^{\alpha' \alpha''}_{\gamma'
\gamma''}(v) R^{\alpha \gamma''}_{\gamma \beta''}(u) R^{\gamma
\gamma'}_{\beta \beta'}(u-v)}
\end{array}
\end{equation}
 This is the key relation to construct a
family of commuting operators, as is outlined below.

\begin{figure}[t]
    \centering
        \includegraphics[angle=-90,scale=0.4]{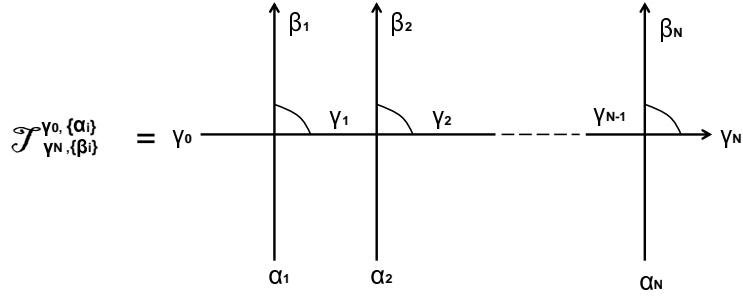}
    \caption{\it   The monodromy  matrix. }
    \label{fig:Tmatr}
\end{figure}

The $R$-matrix is an operator in the tensor product of two linear
spaces (not necessarily isomorphic). It is customary to call one of
them {\it quantum} and the other one {\it auxiliary} space. In
Fig.~\ref{fig:Rmatr} they are associated with the vertical and the
horizontal lines respectively. Let us fix a set of $N$ complex
numbers $\theta_i$ and consider a chain of $N$ fundamental
$R$-matrices $R(u-\theta_i)$ with the common auxiliary space $V$.
Multiplying  them as linear operators in $V$ along the chain, we get
an operator in $V^{\otimes N}$ which is called quantum monodromy
matrix. In components, it can be written as
\begin{equation}
{\cal T}^{\gamma_0, \{\alpha_i\}}_{\gamma_{N}, \{\beta_i\}}(u) =
\!\!\! \sum_{\gamma_1, \ldots , \gamma_{N-1}} R^{\gamma_{N-1}
\alpha_N}_{\gamma_{N} \beta_N}(u\! -\! \theta_N) \ldots
 R^{\gamma_1 \alpha_2}_{\gamma_2 \beta_2}(u\! -\! \theta_2)
R^{\gamma_0 \alpha_1}_{\gamma_1 \beta_1}(u\! -\! \theta_1)
\,(-1)^{\sum_{i=2}^{N}[p(\a_i)+p(\b_i)]\sum_{j=1}^{i-1} p(\a_j)}
\end{equation}
It is usually regarded as an operator-valued matrix in the auxiliary
space (with indices $\gamma_0$, $\gamma_N$), with the matrix
elements being operators in the quantum space $V^{\otimes N}$ (see
Fig.~\ref{fig:Tmatr}). Setting $\gamma_0 = \gamma_N =\gamma$ and
summing with the sign factor $(-1)^{p(\gamma)}$, we get the
supertrace of the quantum monodromy matrix in the auxiliary space,
\begin{equation}\label{F3}
T(u)=\, \mbox{str}\, {\cal T}(u)= \sum_{\gamma}
(-1)^{p(\gamma)}{\cal T}^{\gamma}_{\gamma}(u)
\end{equation}
which is an operator in the quantum space (indices of the quantum
space are omitted). It is called the quantum transfer matrix. Using
the graded Yang-Baxter equation, it can be shown that the $T(u)$
commute at different $u$: $[T(u), \, T(u')]=0$. Their
diagonalization is the subject of one or another version of Bethe
ansatz.

The (inhomogeneous) integrable ``spin chain" or a vertex model on
the square lattice is characterized by the symmetry algebra
$gl(K|M)$ and the parameters $\theta_i$ (inhomogeneities at the
sites or ``rapidities"). Given such a model, the family of transfer
matrices $T(u)$ (\ref{F3}) can be included into a larger family of
commuting operators. It is constructed by means of the fusion
procedure.

\subsection{Fusion procedure}

Using the $R$-matrix (\ref{F1}) as a building block, it is possible
to generate other $gl(K|M)$-invariant solutions to the (graded)
Yang-Baxter equation. They are operators in tensor products of two
arbitrary finite dimensional irreps of the symmetry algebra.
We consider covariant tensor irreps which are labeled (although in
a non-unique way, see Appendix A) by Young diagrams $\lambda$.
Let $V_{\lambda}$ be the corresponding representation space, then
the $R$-matrix acts in $V_{\lambda}\otimes V_{\lambda'}$.
The construction of this $R$-matrix from the elementary one is
referred to as {\it fusion procedure}
\cite{KulResh1,KRS}, \cite{Cherednik0}-\cite{NazTar}.

Here we consider fusion in the auxiliary space, which allows one to
construct, by taking (super)trace in this space, a large family of
transfer matrices commuting with $T(u)$. In words, the fusion
procedure consists in tensor multiplying several copies of the
fundamental $R$-matrix (\ref{F1}) in the auxiliary space and
subsequent projection onto an irrep of the $gl(K|M)$ algebra. As a
result, one obtains an $R$-matrix $R^{(\lambda )}(u)$ whose quantum
space is still $V$ (the space of the vector representation of
$gl(K|M)$) and whose auxiliary space is the space $V_{\lambda}$ of a
higher irrep $\lambda$ of this algebra. The fact that the $R$-matrix
obtained in this way obeys the (graded) Yang-Baxter equation follows
from the observation \cite{KulResh1,Cherednik}
that the projectors onto higher irreps can be
realized as products of the fundamental $R$-matrices (\ref{F1})
taken at special values of the spectral parameter. These values
correspond to the degeneration points of the $R$-matrices.

In particular, the $R$-matrix (\ref{F1}) has two degeneration points
$u=\pm 2$. Indeed, one can easily see that
\begin{equation}
\det R(u)= (u+2)^{d_{+}} (u-2)^{d_{-}}\,, \quad
d_{\pm}=\frac{1}{2}((K+M)^2 \pm (K-M))\,,
\end{equation}
\begin{equation}\label{PROJECTOR}
R(\pm 2)=\pm 2(I\pm  {\mit\Pi} )  = \pm 4 P_{\pm}\,,
\end{equation}
where $P_{\pm}$ are projectors onto the symmetric and antisymmetric
subspaces in $V\otimes V$. The dimensions of these subspaces are
equal to $d_{\pm}$, and the projectors are complimentary, i.e.,
$P_{+}P_{-}=P_{-}P_{+}=0$.  Therefore, to get the $R$-matrix with
the auxiliary space carrying the $\lambda = (2^1)$ irrep
(respectively, the $\lambda =(1^2)$ irrep), one should take the
tensor product $ R(u+2) \otimes R(u)$ (respectively, $ R(u-2)
\otimes R(u)$) in the auxiliary space and apply the projector
$P_{+}$ (respectively, $P_{-}$):
$$
R^{(2^1)}(u)=P_{+}\Bigl [ R(u+2) \otimes R(u)\Bigr ] P_{+} \,, \quad
R^{(1^2)}(u)=P_{-}\Bigl [ R(u-2) \otimes R(u)\Bigr ] P_{-}\,.
$$
Here, the tensor product notation still implies the usual matrix
product in the quantum space. Note that $R^{(2^1)}(-2)$ vanishes
identically since the projector $P_{+}$ gets multiplied by the
complementary projector $P_{-}$ coming from one of the $R$-matrices
in the tensor product. Similarly, $R^{(1^2)}(2)=0$.

\begin{figure}[t]
    \centering
        \includegraphics[angle=-90,scale=0.4]{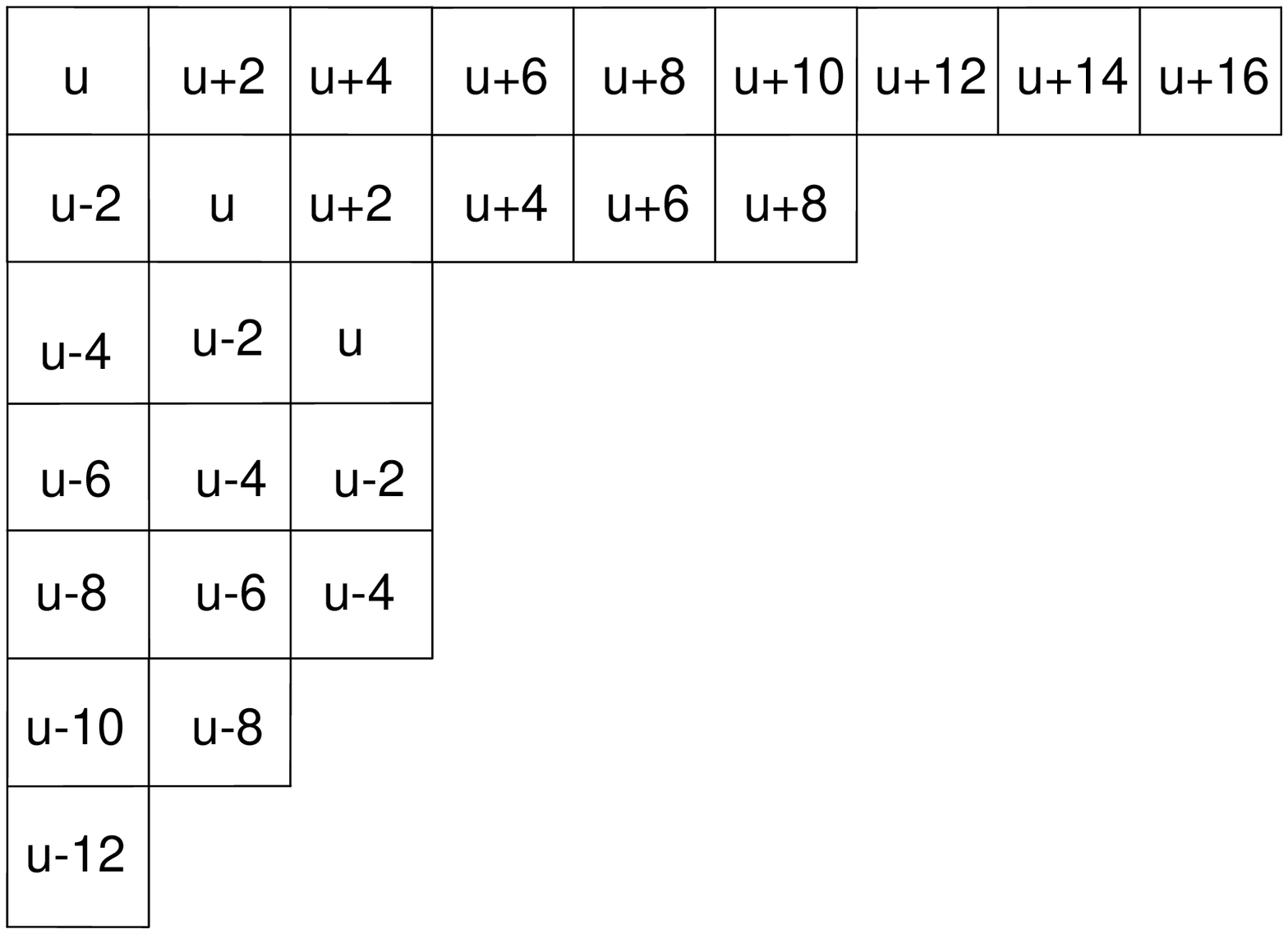}
    \caption{\it The Young diagram decorated by spectral parameters.}
    \label{fig:DecoYoung}
\end{figure}

In a more general case, the procedure is as follows. Let $\lambda$
be the Young diagram associated with a given irrep of the
algebra\footnote{For superalgebras this may be a delicate point
since this correspondence is in general not one-to-one.}. Let
$n=|\lambda |$ be the number of boxes in the diagram and let
$$
P_{\lambda}: \left ({\Bbb C}^{K|M}\right )^{\otimes n}
\longrightarrow  V_{\lambda}
$$
be the projector onto the space of the irreducible representation.
Write in the box with coordinates $(i,j)$ ($i$-th line counting from
top to bottom and $j$-th column counting from left to right) the
number $u_{ij}=u -2(i-j)$, as is shown in Fig.~\ref{fig:DecoYoung}.
Then $R^{(\lambda )}(u)$ is constructed as
\begin{equation}\label{fusedR}
R^{(\lambda )}(u)=P_{\lambda} \Bigl [ \bigotimes_{(ij)\in
\lambda}^{\longleftarrow} R(u_{ij})\Bigr ]\,\, P_{\lambda}
\end{equation}
Here, the tensor factors are placed {\it from right to left} in the
lexicographical order, i.e., elements of the first row from left to
right, then elements of the second row from left to right, etc. For
example, for the diagram $\lambda = (3,2)$ the product under
the projectors reads $R(u)\otimes R(u-2)\otimes R(u+4 ) \otimes
R(u+2)\otimes R(u)$.

Omitting further details of the fusion procedure (see, e.g.,
\cite{Nazarov1,NazTar}), we now describe the structure of zeros of
the fused $R$-matrix $R^{(\lambda )}(u)$ which is essential for what
follows. As we have seen, the $R$-matrix $R^{(\lambda )}(u)$ is
obtained by fusing $n=|\lambda |$ fundamental $R$-matrices in the
auxiliary space. Then, by the construction, any matrix element of
the $R^{(\lambda )}(u)$ is a polynomial in $u$ of degree $\leq n$.
However, one can see that at some special values of the spectral
parameter the complimentary projectors get multiplied and the matrix
$R^{(\lambda )}(u)$ vanishes as a whole thing. In other words, the
matrix elements have a number of common zeros. We call them
``trivial zeros". A more detailed analysis shows that the number of
trivial zeros is $n-1$ (counting with multiplicities) and thus a
scalar polynomial of degree $n-1$ can be factored out. Namely, this
polynomial is equal to the product of the factors $u_{ij}=u-2(i-j)$
over the boxes of the diagram $\lambda$ with $(i,j)\in \lambda$
except $i=j=1$.

We consider here only the representations of the type
$\l=(s^a)$ corresponding to rectangular Young diagrams with $s$ rows
and $a$ columns (rectangular irreps). In this case
$$
R^{(s^a)}(u)=r_{a,s}(u)\, {\sf R}^{(s^a)}(u)\,,
$$
where matrix elements of ${\sf R}^{(s^a)}(u)$ are polynomials of
degree $\leq 1$ and
\begin{equation}\label{F4}
r_{a,s}(u)=u^{-1}\prod_{j=1}^{s}\prod_{l=1}^{a} (u+2j-2l)\,, \quad
\quad a,s\geq 1,
\end{equation}
is the polynomial of degree $as-1$. For a future reference, we give
its representation through the Barnes function $G(z)$ (a unique
entire function such that $G(z+1)=\Gamma (z)G(z)$ and $G(1)=1$):
\begin{equation}\label{F5}
r_{a,s}(u)=2^{as}\, u^{-1}\,\frac{G(\frac{1}{2}u+s+1)\,
G(\frac{1}{2}u-a+1)}{G(\frac{1}{2}u+1) \,\,G(\frac{1}{2}u+s-a+1)}
\,.
\end{equation}

The quantum monodromy matrix ${\cal T}^{(\lambda )}$ with the
auxiliary space $V_{\lambda}$ is defined by the same formula as
before but with the $R$-matrix $R^{(\lambda )}$. The transfer matrix
is
\begin{equation}\label{F6}
 T^{(\lambda )}(u)=\, \mbox{str}_{V_{\lambda}}
 {\cal T}^{(\lambda )} (u)
\end{equation}
The transfer matrices commute for any $\lambda$ and $u$:
$[T^{(\lambda )}(u),\, T^{(\lambda ' )}(u')]=0$. This commuting
family of operators extends the family (\ref{F3}) which corresponds
to the one-box diagram $\lambda$. For the empty diagram $\lambda
=\emptyset$ we formally put $T^{(\emptyset )}(u)=1$. For rectangular
diagrams $\lambda = s^a$ we use the special notation $T^{(\lambda
)}(u):= T^{a}_{s}(u)$.

\subsection{Functional relations for transfer matrices}

The transfer matrices are functionally dependent. It appears that
all $T^{(\lambda )}(u)$ can be expressed through $T^{1}_{s}(u)$ or
$T^{a}_{1}(u)$ by means of the nice determinant formulas due to
Bazhanov and Reshetikhin \cite{BR1}. They are the same for all
(super)algebras of the type $gl(K|M)$ with $K,M\geq 0$
\cite{Tsuboi-1}:
\begin{equation}\label{F7a}
\begin{array}{lll}
T^{(\lambda )}(u)&=&\displaystyle{\det_{1\leq i,j \leq \lambda_{1}'}
 T^{1}_{\lambda_j +i-j}(u-2i+2)}
 \\&&\\
 &=&  \displaystyle{\det_{1\leq i,j \leq \lambda_1}
 T^{\lambda_{j}' +i-j}_{1}(u+2i-2)}\,.
 \end{array}
 \end{equation}
For rectangular diagrams, they read:
\begin{equation}\label{F7}
\begin{array}{lll}
T^{a}_{s}(u)&=&\displaystyle{\det_{1\leq i,j \leq a}
 T^{1}_{s+i-j}(u-2i+2)}
 \\&&\\
 &=&  \displaystyle{\det_{1\leq i,j \leq s}
 T^{a+i-j}_{1}(u+2i-2)} \,.
 \end{array}
 \end{equation}
These formulas should be supplemented by the ``boundary conditions"
$T^{0}_{s}(u)=T^{a}_{0}(u)=1$, $T^{-n}_{s}(u)=T^{a}_{-n}(u)=0$ for
all $a,s \geq 0$ and $n\geq 1$. Since all the transfer matrices
commute, they can be diagonalized simultaneously, and the same
relations are valid for any of their eigenvalues. Keeping this in
mind, we will often refer to the transfer matrices as scalar
functions and call them ``$T$-functions".

Applying the Jacobi identity for determinants to formulas (\ref{F7})
(see, e.g., \cite{Kuniba-1,KLWZ}), one obtains a closed  functional
relation between transfer matrices for rectangular diagrams,
equivalent to Hirota difference equation:
\begin{equation}\label{F8}
T^{a}_{s}(u-2) T^{a}_{s}(u)- T^{a}_{s+1}(u-2) T^{a}_{s-1}(u) =
T^{a-1}_{s}(u-2)T^{a+1}_{s}(u)
\end{equation}
The bilinear form of the functional relations was discussed in
\cite{KlumperPearce}, \cite{Kuniba-0}, \cite{KLWZ},
\cite{Kuniba-1}-\cite{Zhou}. Below we illustrate this
relation by simple examples.

\subsubsection{Simple examples of the functional equation}

\begin{figure}[t]
    \centering
        \includegraphics[angle=-90,scale=0.4]{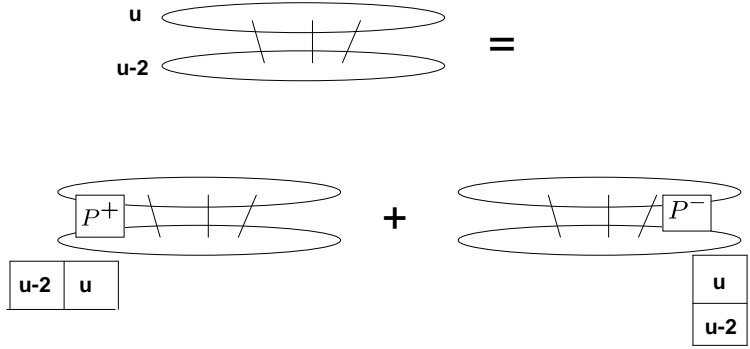}
    \caption{\it Fusion of two fundamental transfer matrices and the
corresponding Hirota equation. As usual, each line carries the
vector representation, each crossing corresponds to the insertion of
the $R$-matrix, and the closed lines correspond to taking traces in
the auxiliary space. The insertions of projectors $P^{\pm}$ are
interpreted as  insertions of the $R$-matrices with given values of
spectral parameters.}
    \label{fig:FUS11}
\end{figure}

\begin{figure}[t]
    \centering
        \includegraphics[angle=-90,scale=0.4]{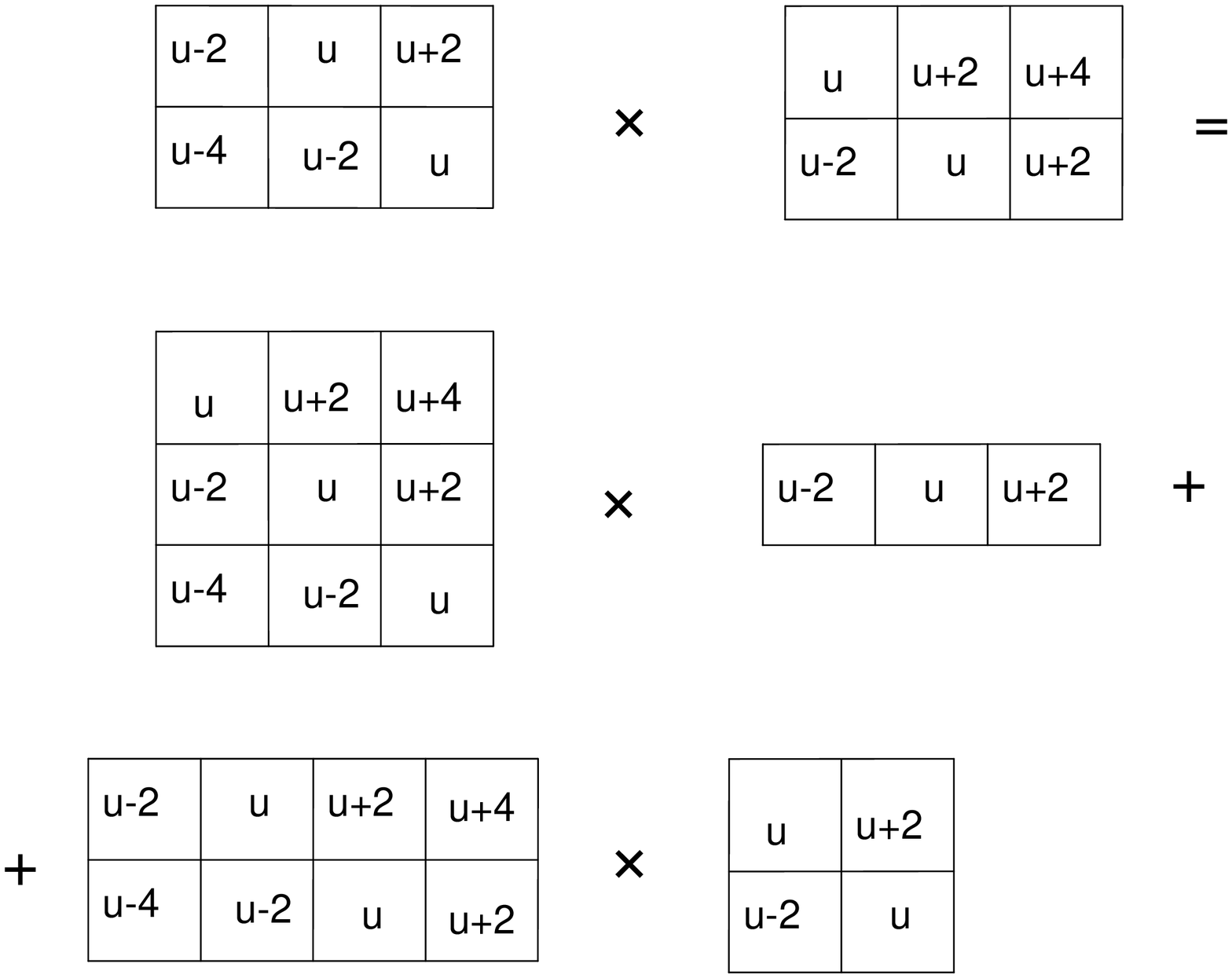}
    \caption{\it The Hirota-type relation
illustrated by rectangular Young diagrams
decorated by spectral parameters. To obtain the first term in the r.h.s.,
one takes the first row of the second diagram in the l.h.s.
and puts it on the top of the first one.
To obtain the second term in the r.h.s.,
one takes the first column of the first diagram in the l.h.s.
and attaches it to the second one from the left.}
    \label{fig:FUS32}
\end{figure}

To figure out the general pattern, it is useful to consider the
simplest case of the fusion of two transfer matrices with
fundamental representations in the auxiliary space. The proof is
summarized in Fig.~\ref{fig:FUS11}. Let us represent the first term
in \eq{F8} as
\begin{eqnarray}
T_{1}^{1}(u-2)T_{1}^{1}(u)= {\rm tr}_{{\rm aux}\,, \, V\otimes
V}\[{\cal T}_{20}(u-2){\cal T}_{10}(u)\]
\end{eqnarray}
The index $0$ denotes the common quantum space represented in the
figure by several vertical lines, the indices $1,2$ denote the two
copies of the auxiliary space $V$. Inserting $I=P_+ + P_-$ inside
the trace, we get two terms. Using the projector property
$P_{\pm}^2=P_{\pm}$, we immediately see that the term with $P_{+}$
is equal to $T_{2}^{1}(u-2)$, by the definition of the latter. The
term with $P_{-}$ does not literally coincide with the definition of
$T_{1}^{2}(u)$ since the order of the horizontal lines is reversed.
However, plugging $P_{-}=-\frac{1}{4}R(-2)$ and using the
Yang-Baxter equation to move the vertical lines to the other side of
the $R$-matrix $R(-2)$, we come to the equivalent graph with the
required order of the horizontal lines. Finally, we get
\begin{eqnarray}\label{HIR111}
T_{1}^{1}(u-2)T_{1}^{1}(u) =T_{2}^{1}(u-2) + T_{1}^{2}(u)\,.
\end{eqnarray}
Hence we have reproduced the simplest case of \eq{F8}.

Let us comment on the general case.
An illustrative
example is given in fig.\ref{fig:FUS32}.
The rectangular
Young diagrams are decorated by the values of the spectral parameter
\begin{equation}\label{DECOR}
u_{ij}=u-2(i-j),
\end{equation}
as is explained above.
Each box of such decorated diagrams corresponds to a line
characterized by a given value of the spectral parameter. All these
lines cross each other and the $R$-matrices are associated to
the crossing points. The order of the crossings is irrelevant due to the
Yang-Baxter equation.
The reshuffling of spectral parameters in the second
and third terms of equation (\ref{F8})
(corresponding to the second and the third lines
in the figure) is done by the exchange
of a row or a column between the two diagrams of the first line. The
decoration of the resulting Young diagrams follows the  rule
(\ref{DECOR}). Of course this is not a proof, just an illustration.

\subsubsection{Analogy with formulas for characters
and identities for symmetric functions}

Equations (\ref{F7a}) are spectral parameter dependent versions
of the second Weyl formula for characters
of (super)groups (see
\cite{BahaBalantekin:1980qy}):
\begin{equation} \label{SUPERJ}
\chi_\l(w_1,\dots,w_K;v_1,\dots,v_M)=\det_{i,j=1,\dots,K+M}
S_{\lambda_i+i-j}\,.
\end{equation}
In the theory of symmetric functions such formulas are known as the
Jacobi-Trudi determinant identities. Here $\lambda_i$ are the
lengths of the rows of the diagram $\l$, $S_{n}$ are super
Schur polynomials defined as
\begin{equation}
\label{SCHUR}  \frac{\prod_{m=1}^M \(1-z v_m\)}{\prod_{k=1}^K \(1-z
w_k\)} =\sum_{n=1}^\infty z^n S_{n},
\end{equation}
and $w_k$, $v_m$ are eigenvalues of the $A$- and $D$-parts
of the diagonalized element of the supergroup in the matrix
realization of the type (\ref{ABCD}).

For the rectangular irrep with $\l=s^a$ eq.~(\ref{SUPERJ}) reads
\begin{equation}
\label{SSQ} \chi_{s^a}\equiv \chi (a,s) =\det_{i,j=1,\dots,a}
S_{s+i-j}.
\end{equation}
The characters $\chi_{s^a}$ satisfy the bilinear relation
\begin{equation}
\label{JACOBY}
\chi^2(a,s)=\chi(a+1,s)\chi(a-1,s)+\chi(a,s+1)\chi(a,s-1)
\end{equation}
which follows from the Jacobi identity for determinants.
It is the spectral parameter independent version of the Hirota
equation.

\subsubsection{Normalization and boundary conditions}

A simple redefinition of the $T$-functions by a shift of the
spectral parameter,
\begin{equation}\label{F9}
T^{(a,s)}(u)\equiv T^{a}_{s}(u-s+a)
\end{equation}
brings equation (\ref{F8}) to the form
\begin{equation}\label{F10}
T^{(a,s)}(u+1) T^{(a,s)}(u-1)- T^{(a, s+1)}(u) T^{(a, s-1)}(u) =
T^{(a+1, s)}(u) T^{(a-1, s)}(u)
\end{equation}
which appears to be {\it completely symmetric} with respect to
interchanging of $a$ and $s$. This is the famous Hirota bilinear
difference equation \cite{Hirota} which is the starting point of our
approach to the Bethe ansatz and generalized Baxter equations in
this paper. For convenience, we also give the determinant
representation (\ref{F7}) in terms of $T^{(a,s)}(u)$:
\begin{equation}\label{F12}
\begin{array}{lll}
T^{(a,s)}(u)&=&\displaystyle{\det_{1\leq i,j \leq a}
 T^{(1, \, s+i-j)}(u+a+1-i-j)}
 \\&&\\
 &=&  \displaystyle{\det_{1\leq i,j \leq s}
 T^{(a+i-j, \, 1)}(u+s+1-i-j)}\,.
 \end{array}
 \end{equation}

Let us comment on the meaning of equation (\ref{F10}). On the first
glance, there is no much content in this equation. Its general
solution (with the boundary conditions fixed above) is just given by
formulas (\ref{F12}) with arbitrary functions $T^{(1, s)}(u)$ or
$T^{(a, 1)}(u)$. However, in the problem of interest these functions
are by no means arbitrary. They are to be found from certain
analytic conditions. For the finite spin chains with finite
dimensional representations at each site these conditions simply
mean that $T^{(a,s)}(u)$ must be a polynomial of degree $asN$, where
$N$ is the length of the chain, with $(as-1)N$ fixed zeros. These
zeros are just the trivial zeros coming from the fusion procedure.
Their location is determined by the scalar factor $r_{a,s}(u)$
(\ref{F4}) of the $R$-matrix. We thus see that the polynomial
$T^{(a,s)}(u)$ for all $a,s\geq 1$ must be divisible by the
polynomial
$$
\prod_{i=1}^{N}r_{a,s}(u-s+a-\theta_i):= \Phi (a,s,u)\,.
$$
This constraint makes the problem non-trivial.

Let us introduce the function
\begin{equation}\label{F13}
\phi (u)=\prod_{i=1}^{N}(u-\theta_i)\,,
\end{equation}
in terms of which the scalar polynomial factor is written as
$$
\Phi (a,s,u)=\prod_{i=1}^{N}r_{a,s}(u-s+a-\theta_i)=
\phi^{-1}(u-s+a)\prod_{j=1}^{s}\prod_{l=1}^{a} \phi (u-s+a+2j-2l)\,,
\quad a,s\geq 1.
$$
The representation through the Barnes function (\ref{F5}) allows us
to extend this formula to all values of $a$ and $s$:
\begin{equation}\label{F14}
 \Phi (a,s,u)=2^{asN}\phi^{-1}(u\!-\! s\! +\! a)\,
 \prod_{i=1}^{N}
\frac{G\left (\frac{1}{2}(u+s+a)+1-\theta_i \right )\,
 G\left (\frac{1}{2}(u-s-a)+1-\theta_i \right )}{G\left
 (\frac{1}{2}(u-s+a)+1-\theta_i \right )\,
 G\left (\frac{1}{2}(u+s-a)+1-\theta_i \right )}
 \end{equation}
Extracting it from the $T^{(a,s)}(u)$, we introduce the $T$-function
\begin{equation}\label{F15}
 T(a,s,u)=\Phi^{-1}(a,s,u)\, T^{(a,s)}(u)
 \end{equation}
 which is a polynomial in $u$ of degree $N$ for all
 $a,s\geq 0$. Note that at $a=0$ or $s=0$ equation
 (\ref{F14}) yields $\Phi (0,s,u)=1/\phi(u-s)$,
 $\Phi (a,0,u)=1/\phi (u+a)$, so
 \begin{equation}\label{F16}
 T(0,s,u)=\phi (u-s)\,,
 \quad \quad T(a,0,u)=\phi (u+a)\,.
 \end{equation}

It is important to note that the renormalized $T$-function
$T(a,s,u)$ obeys the same Hirota equation (\ref{F10}) as the
$T$-function $T^{(a,s)}(u)$. Indeed, it easy to check that the
transformation
\begin{equation}\label{gaugef}
T^{(a,s)}(u)\longrightarrow f_0 (u \! +\! s \! +\!a)f_1 (u \! +\! s
\! -\!a) f_2 (u \! -\! s \! +\!a) f_3 (u \! -\! s \! -\!a)
T^{(a,s)}(u),
\end{equation}
where $f_i$ are arbitrary functions, leaves the form of the
equation unchanged. Equation (\rf{F14}) shows that the function
$\Phi (a,s,u)$ is precisely of this form (the factor $2^{asN}$ is
easily seen to be of this form, too).

The main difference between the ``bosonic" $gl(K)$ and
supersymmetric $gl(K|M)$ cases is in the boundary conditions for the
transfer matrices in the $(s,a)$-plane. For the algebra $gl(K)$ the
rectangular Young diagrams live in the half-band $s\ge 0$, $0\le
a\le K$, while for the superalgebra the rectangular diagrams live in
the domain shown below in Fig.~\ref{fig:Hirota}. The Hirota equation
with boundary conditions of this type will be our starting point for
the analysis of the inhomogeneous quantum integrable super spin
chains and it will allow us to obtain the full hierarchy of Baxter
relations, the new Hirota equation for the Baxter functions  and the
nested Bethe ansatz equations for all possible choices of simple
root systems for $gl(K|M)$. This naturally generalizes the known
relations for the spin chains based on the $gl(K)$ algebra.

\section{Hierarchy of   Hirota equations}

Throughout the rest of the paper we deal with rectangular irreps
only and use the normalization (\ref{F15}), where all the ``trivial"
zeros of the transfer matrix are excluded. This normalization was
used in \cite{KLWZ,Z1}, for the supersymmetric case see
\cite{Tsuboi-4}.

\subsection{Hirota equation and boundary conditions for
superalgebra $gl(K|M)$}

\begin{figure}[t]
    \centering
        \includegraphics[angle=-90,scale=0.6]{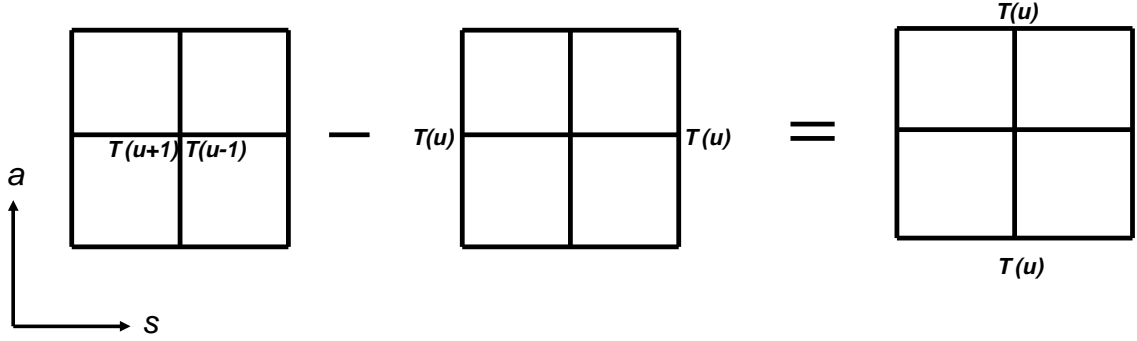}
    \caption{\it  Hirota equation in the $(s,a)$-plane.}
    \label{fig:HirotaEQ}
\end{figure}

As we have seen in the previous section, the functional relations
for transfer matrices of integrable quantum spin chains with
``spins" belonging to representations of the superalgebra $gl(K|M)$
can be written \cite{Tsuboi-1} in the form of the same Hirota
equation as in the case of the ordinary Lie algebra $gl (K)$:
\begin{equation}
\label{HIROTA}
  T(a,s,u \! +\! 1)T(a,s,u\! -\! 1)-
T(a,s \! +\! 1,u)T(a,s\! -\! 1,u) =T(a\! +\! 1,s,u)T(a\! -\! 1,s,u).
\end{equation}
It can be schematically drawn in the $(s,a)$-plane as is shown in
Fig.~\ref{fig:HirotaEQ}. All the non-vanishing $T(a,s,u)$'s are
polynomials in $u$ of one and the same degree $N$ equal to the
number of sites in the spin chain.

\begin{figure}[t]
    \centering
        \includegraphics[angle=-90,scale=0.4]{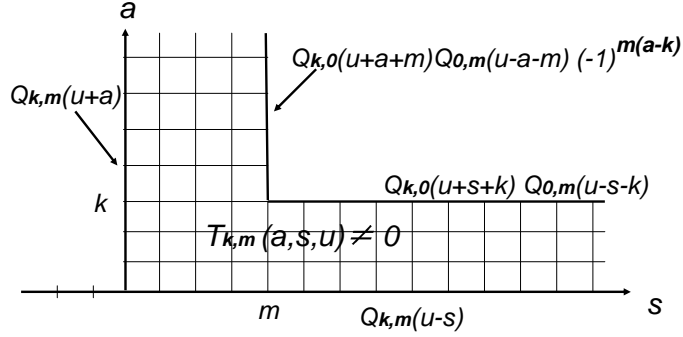}
    \caption{\it   Boundary conditions for the Hirota equation in
the case of the superalgebra $gl(k|m)$. Note that $T(0,s,u)
=Q_{k,m}(u-s)\neq 0$ on the whole axis $-\infty < s < +\infty$.}
    \label{fig:Hirota}
\end{figure}

To distinguish solutions relevant to Bethe ansatz for a given
(super)algebra, one should specify the boundary conditions in the
discrete variables $a$ and $s$. For the superalgebra $gl(K|M)$ one
can see \cite{Martin,Tsuboi-1} that
\begin{eqnarray}
\label{BCOND1}
  T(a,s,u)=0\quad  {\rm if:}\quad  \mbox{(i)} \;\; a<0
  \quad {\rm or} \quad \mbox{(ii)} \;\; s<0 \;\; {\rm and} \;\;
  a\ne 0, \quad {\rm or} \quad \mbox{(iii)} \;\; a>K \;\; {\rm and}\;\;  s>M.
\end{eqnarray}
(see Fig.~\ref{fig:Hirota}, where we use the letters $k,m$ rather
than $K,M$ for later references).  The latter requirement, that
$T(a,s,u)$ vanishes if simultaneously $a>K$ and $s>M$, comes from
the fact that the Young superdiagrams for $gl(K|M)$ containing a
rectangular subdiagram with $K+1$ rows and $M+1$ columns are
illegal, i.e., the corresponding representations vanish \cite{BMR}.
Note that we want the Hirota equation to be valid in the whole
$(s,a)$ plane, not just in the quadrant $a,s\geq 1$. This is why we
have to require that $T(0, s, u)$ does not vanish identically on the
negative $s$-axis, otherwise the Hirota equation would break down at
the origin $a=s=0$.

The boundary  values of $T(a,s,u)$ are rather special. For example,
at $a=0$ eq. (\ref{HIROTA}) converts into
$$
T(0,s,u \! +\! 1)T(0,s,u\! -\! 1) = T(0,s \! +\! 1,u)T(0,s\! -\!
1,u)
$$
which is a discrete version of the d'Alembert equation with the
general solution $T(0,s,u)=f_{+}(u+s) f_{-}(u-s)$ where $f_{\pm}$
are arbitrary functions. In the normalization (\ref{F15}) we have
$f_{+}(u)=1$, $f_{-}(u)=\phi (u)$ (see (\ref{F16})). Similarly, the
boundary function at the half-axis $s=0, a\geq 0$ is normalized to
depend on $u+a$ only in which case it has to be equal to
$f_{-}(u+a)=\phi (u+a)$. As soon as this is fixed, there is no more
freedom left, and the boundary functions at the interior
boundaries\footnote{We call the boundaries at $a=0$ and $s=0$,
$a\geq 0$ {\it exterior} and the boundaries inside the right upper
quadrant in Fig.~\ref{fig:Hirota} {\it interior} ones.} are in
general products of a function of $u+s$ and a function of $u-s$ on
the horizontal line (respectively, of $u+a$ and $u-a$ on the
vertical line). One more thing to be taken into account is the
identification (up to a sign) of the $T$-functions on the two
interior boundaries:
\begin{equation} \label{IDENT}
  T(K,M+n,u)=(-1)^{nM} T(K+n,M,u),\quad  n\ge 0.
\end{equation}
This equality reflects the fact that the two rectangular Young
diagrams of the shapes $\left ( (M\! +\! n)^{K}\right )$ and
$(M^{K+n})$ correspond to the same representation of the algebra
$gl(K|M)$ with the Kac-Dynkin label $b_1 = \ldots =b_{K-1}=0$, $b_K
= M+n$, $b_{K+1}=\ldots = b_{K+M-1}=0$ \cite{BMR}. Note also that
every point with integer coordinates inside the domain in
Fig.~\ref{fig:Hirota} corresponds to an atypical representation
while the points on the interior boundaries correspond to typical
ones, and in this respect the coordinate along the boundary can be
treated as a continuous number.

Summarizing, we can write:
\begin{eqnarray}
\label{BCOND2}
  T(0,s,u)&=&Q_{K,M}(u-s),\quad  -\infty<s<\infty,\nn\\
  T(a,0,u)&=&Q_{K,M}(u+a),\quad  0\le a <\infty,\nn\\
  T(K,s,u)&=&Q_{K,0}(u+s+K)Q_{0,M}(u-s-K),\quad  M\le s<\infty,\nn\\
  T(a,M,u)&=&(-1)^{M(a-K)} Q_{K,0}(u+a+M)
Q_{0,M}(u-a-M),\quad  K \le a <\infty,
\end{eqnarray}
where the polynomial boundary function $Q_{K,M}(u)=\phi (u)$ is
regarded as a fixed input characterizing the quantum space of the
spin chain while the polynomials $Q_{K,0}(u)$ and $Q_{0,M}(u)$ are
to be determined from the solution to the Hirota equation. At
$K=M=0$ the domain of non-vanishing $T$'s shrinks to the axis $a=0$
and the half-axis $s=0$, $a\geq 0$, and the ``gauge" freedom allows
one to put $Q_{0,0}(u)=1$. It should be noted that the
identification (\ref{IDENT}) does not yet imply the coincidence of
the $Q$-functions in the third and the fourth lines of eq.
(\ref{BCOND2}) (i.e., on the horizontal and vertical parts of the
interior boundary). In fact the specific form of the boundary
conditions given in eq. (\ref{BCOND2}) (as well as the sign factor
$(-1)^{M(a-K)}$) is determined by a consistency with a more general
hierarchy of Hirota equations connecting the $T$-functions for
different values of $K$ and $M$. The uniqueness of the boundary
conditions (\ref{BCOND2}) will be justified in the next subsection.

In the case of the usual Lie algebra $gl(K)$ ($M=0$) the boundary
conditions (\ref{BCOND2}) become the same as the ones imposed in
\cite{Z1} (in the original paper \cite{KLWZ} a gauge equivalent
version was used). The domain of non-vanishing $T$'s in
Fig.~\ref{fig:Hirota} degenerates so that the vertical strip
collapses to a line. Therefore, the $T$-functions on the interior
boundaries cease to be dynamical variables and become equal, up to a
shift of the argument, to the fixed function $Q_{K,0}=\phi (u)$
characterizing the spin chain (see \cite{KLWZ,Z1}).

One of the possible setups of the problem is the following: we fix
the polynomial $Q_{K,M}(u)$ and then solve the Hirota equation with
the aforementioned boundary and analytic conditions. The result is a
finite set of solutions for $T(a,s,u)$ which yield the spectrum of
eigenvalues of the quantum transfer matrix. We proceed by
constructing a hierarchy of Hirota equations connecting neighboring
``levels" of the array in Fig.~\ref{fig:Hirota}, i.e., equations
connecting $T$-functions for which $K$ or $M$ differ by $1$. The
existence of such a hierarchy follows from classical integrability
of the Hirota equation. Decreasing $K$ and $M$ by $1$, one can
``undress" step by step the original $gl(K|M)$ problem to an empty
problem formally corresponding to $gl(0|0)$. This procedure appears
to be equivalent to the hierarchial (nested) Bethe ansatz.

\subsection{Auxiliary linear problems and
B\"acklund transformations}

Like almost all known nonlinear integrable equations, the Hirota
equation (\ref{HIROTA}) serves as a compatibility condition for
over-determined linear problems \cite{SS,KLWZ}. To introduce them,
it is convenient to pass to the new variables
\begin{equation}\label{LP1}
\begin{array}{l}
p=\frac{1}{2}(u-s-a)\\ \\
q=\frac{1}{2}(u+s+a)\\ \\
r=\frac{1}{2}(-u-s+a)
\end{array}
\end{equation}
We call them ``chiral" or ``light-cone" variables while the original
ones will be refered to as ``laboratory" variables. Here are the
formulas for the inverse transformation,
\begin{equation}\label{LP2}
a=q+r\,, \quad s=-p-r\,, \quad u=p+q
\end{equation}
and for the transformation of the vector fields:
\begin{equation}\label{LP3a}
\p_p =\p_u - \p_s\,, \quad \p_q =\p_u + \p_a\,, \quad \p_r =\p_a -
\p_s
\end{equation}

We set $\tau (p,q,r)= T(q+r,\,  -p-r, \, p+q)$ and introduce the
following linear problems for an auxiliary function $\psi = \psi
(p,q,r)$:
\begin{equation}\label{LP3}
\begin{array}{l}
\displaystyle{ \left ( e^{\p_r}+ \frac{\tau (p\! +\! 1, r\! +\!
1)\,\, \tau}{\tau (p+1)\tau (r+1)}
\right ) \psi = \psi (p+1)} \\ \\
\displaystyle{ \left ( e^{\p_r}- \frac{\tau (q\! +\! 1, r\! +\!
1)\,\, \tau}{\tau (q+1)\tau (r+1)} \right ) \psi = \psi (q+1)}
\end{array}
\end{equation}
where we indicate explicitly only those variables that are subject
to shifts. The compatibility means that the difference operators
$$
e^{-\p_p}\left ( e^{\p_r}+ \frac{\tau (p\! +\! 1, r\! +\! 1)\,\,
\tau}{\tau (p+1)\tau (r+1)} \right ) \quad \mbox{and} \quad \;
e^{-\p_q}\left ( e^{\p_r}- \frac{\tau (q\! +\! 1, r\! +\! 1)\,\,
\tau}{\tau (q+1)\tau (r+1)} \right )
$$
commute. This leads to the relation
$$
\tau (p+1)\tau (q+1, r+1)+ \tau (q+1)\tau (p+1, r+1)= h(2p, 2q)\tau
(r+1)\tau (p+1, q+1)
$$
where $h$ can be an arbitrary function of $p$ and $q$. In the
original variables this equation reads
$$
T(a+1)T(a-1)+T(s+1)T(s-1)=h( u\! -\! s\! -\! a, u\! +\! s\! +\! a)\,
T(u+1)T(u-1)
$$From the boundary conditions (\ref{BCOND2}) at $a=0$ or $s=0$
it follows that $h=1$ and we obtain the Hirota equation
(\ref{HIROTA}).

An advantage of the ``light-cone" variables is their separation in
the linear problems: the first problem does not involve $q$ while
the second one does not involve $p$. However, in contrast to the
``laboratory" variables $a,s,u$, they have no immediate physical
meaning. Coming back to the ``laboratory" variables, we set $\psi
(p,q,r)=\Psi (q+r, \, -p-r, \, p+q)$ and rewrite the linear problems
(\ref{LP3}) in the form
\begin{equation}\label{LP4}
\begin{array}{l}
\displaystyle{\left [e^{\p_a - \p_s} + \frac{T(a\! -\! 1, s\! +\! 1,
u) T(a, s\! -\! 1, u\! +\! 1)}{T(a, s, u)\, T (a\! -\! 1, s, u\! +\!
1)}\right ] \Psi (a\! -\! 1, s\! +\! 1, u)=
\Psi (a\! -\! 1,s, u\! +\! 1)}\\ \\
\displaystyle{ \left [e^{\p_a - \p_s} - \frac{T(a\! -\! 1, s\! +\!
1, u)T(a\! +\! 1, s, u\! +\! 1)}{T(a, s, u)\, T (a,s\! +\! 1, u\!
+\! 1) }\right ] \Psi (a\! -\! 1, s\! +\! 1, u)=\Psi (a, s\! +\! 1,
u\! +\! 1)}
\end{array}
\end{equation}
Because the $T$-functions can vanish identically at some $a,s$, we
eliminate the denominators by passing to the new auxiliary function
$F=T\Psi$, in terms of which we have
\begin{equation}\label{LP5}
\begin{array}{l}
\displaystyle{ T(a\! -\! 1, s, u \! +\! 1)F(a, s, u) \, +\, T(a, s\!
-\! 1, u\! +\! 1)F(a\! -\! 1, s\! +\! 1, u)
=T(a, s, u)F(a\! -\! 1,s,u \! +\! 1)}\\ \\
\displaystyle{ T(a,s\! +\! 1,u \! +\! 1)F(a, s, u) \, -\, T(a\! +\!
1, s, u\! +\! 1)F(a\! - \! 1, s\! +\! 1, u) = T(a, s, u)F(a,s\! +\!
1,u \! +\! 1)}
\end{array}
\end{equation}
Note that the second equation can be obtained from the first one by
the transformation $T(a,s,u)\longrightarrow (-1)^{as}T(-s, -a, u)$
(and the same for $F$) which leaves the Hirota equation invariant.
Note also that the pair of equations can be written in a matrix form
as follows:
\begin{equation}\label{LP51}
\left ( \begin{array}{cc}
T(a\! -\! 1, s, u)& T(a, s\! -\! 1, u)\\ &\\
T(a, s\! +\! 1, u) & -T(a\! +\! 1, s, u)
\end{array} \right )
\left ( \begin{array}{c} F(a,\, s,\, u\! -\! 1)\\ \\ F(a\! -\! 1,s\!
+\! 1,u\! -\! 1)
\end{array} \right ) = T(a,s,u\! -\! 1)
\left ( \begin{array}{c}
F(a \! -\! 1, s, u)\\ \\
F(a, s\! +\! 1, u)
\end{array} \right ).
\end{equation}

\begin{figure}[t]
    \centering
        \includegraphics[angle=-90,scale=0.4]{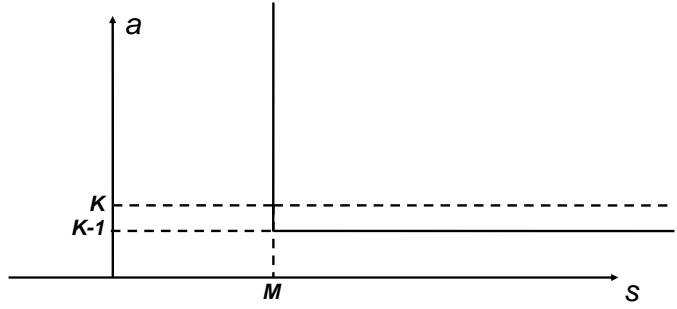}
    \caption{\it  Boundary conditions for the first
    B\"acklund transformation in the $(s,a)$ plane.}
    \label{fig:SHIFT1}
\end{figure}

A remark on the symmetry properties of the linear problems is in
order. One can see that while the Hirota equation (\ref{HIROTA})
written for the function $(-1)^{\frac{1}{2}(a^2 +s^2)}T(a,s,u)$ is
form-invariant with respect to any permutation of the variables
$a,s,u$ and changing sign of any variable, the system of two linear
problems (\ref{LP5}) is not. To make the symmetry explicit, we
multiply both sides of eq. (\ref{LP51}) by the matrix inverse to the
one in the left hand side and use the Hirota equation for $T$'s. In
this way we get another pair of linear problems,
\begin{equation}\label{LP52}
\begin{array}{l}
\displaystyle{ T(a\! +\! 1, s\! +\! 1, u)F(a, s, u) \, -\, T(a\! +\!
1, s, u\! +\! 1)F(a, s\! +\! 1, u\! -\! 1)
=T(a, s, u)F(a\! +\! 1,s\! +\! 1, u)}\\ \\
\displaystyle{ T(a, s, u\! +\! 1) F(a, s, u\! -\! 1)\, -\, T(a,\,
s\! -\! 1,\, u)\, F(a,\, s\! +\! 1,\, u) \, =\, T(a\! +\! 1,\, s, u)
\, F(a\! -\! 1,\, s, u)}
\end{array}
\end{equation}
which are equivalent to (and thus compatible with) the pair
(\ref{LP5}) by construction. The set of four linear problems
(\ref{LP51}), (\ref{LP52}) possesses the required symmetry (note
that the second equation in (\ref{LP52}) is symmetric by itself, and
its structure resembles the Hirota equation). Furthermore, the
Hirota equation can be derived as a compatibility condition for any
two linear problems of these four, and the other two hold
automatically. The four linear problems (\ref{LP51}), (\ref{LP52})
can be combined into a single matrix equation:
\begin{equation}\label{LP53}
\left (
\begin{array}{cccc}
0 & T(a,s,u\! -\! 1) & - T(a,s\! +\! 1,u) & T(a\! +\! 1,s,u)
\\ &&&\\
-T(a,s,u\! -\! 1) & 0 & T(a\! -\! 1, s,u) & T(a,s\! -\! 1,u)
\\ &&& \\
T(a,s\! +\! 1,u) & -T(a\! -\! 1,s,u) & 0 & -T(a,s,u\! +\! 1)
\\ &&& \\
-T(a\! +\! 1,s,u) & -T(a,s\! -\! 1, u) & T(a,s,u\! +\! 1) & 0
\end{array} \right )
\left (
\begin{array}{c}
F(a\! -\! 1,\, s,\, u) \\ \\ F(a,\, s\! +\! 1,\, u) \\ \\
F(a,\, s,\, u\! -\! 1) \\ \\ F(a\! -\! 1, s\! +\! 1, u\! -\! 1)
\end{array}
\right ) =0.
\end{equation}
The Hirota equation implies that the determinant of the
antisymmetric matrix in the left hand side vanishes. If this holds,
the rank of this matrix equals 2, so there are two linearly
independent solutions to the linear problem (\ref{LP53}). Their
meaning will be clarified below. The symmetric form of the linear
problems was suggested in \cite{Shinzawa-1}.
 For more
details on the linear problems for the Hirota equation and their
symmetries see \cite{SS,Z2,Shinzawa-1,Shinzawa-2} and Appendix B.

\begin{figure}[t]
    \centering
        \includegraphics[angle=-90,scale=0.5]{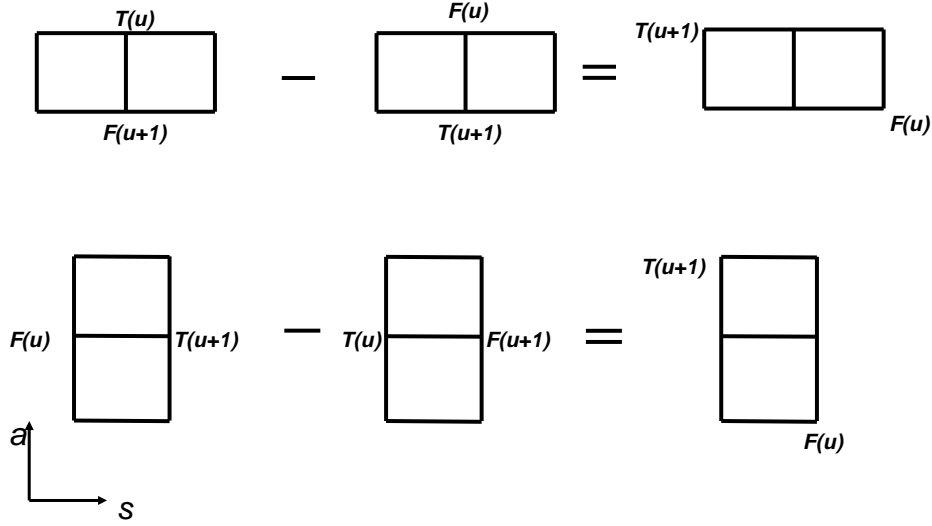}
    \caption{\it Two equations for the first B\"acklund
transformation (BT1)  in the $(s,a)$ plane: the
    variables change within  a ``brick" consisting of two
neighboring squares. }
    \label{fig:HfigLP1}
\end{figure}

There is a remarkable duality  between $T(a,s,u)$ and $F(a, s, u)$
\cite{SS,KLWZ}: one can exchange the roles of the functions $T$, $F$
and treat eqs. (\ref{LP5}) as an over-determined system of linear
problems for the function $T$ with coefficients $F$. Their
compatibility condition is the same Hirota equation for $F$:
\begin{eqnarray}
F(a,s,u+1)F(a, s,u-1)- F(a, s+1,u)F(a, s-1,u)=
F(a+1,s,u)F(a-1,s,u)\,. \label{LP6}
\end{eqnarray}
This can be seen by rewriting (\ref{LP5}) in yet another equivalent
form. Namely, shifting $a \to a+1$ in the first equation and $s \to
s-1$ in the second, we represent the two equations in the matrix
form
\begin{equation}\label{LP7}
\left ( \begin{array}{cc}
F(a\! +\! 1, s, u)& F(a, s\! +\! 1, u)\\ &\\
F(a, s\! -\! 1, u) & -F(a\! -\! 1, s, u)
\end{array} \right )
\left ( \begin{array}{c} T(a,\, s,\, u\! +\! 1)\\ \\
T(a\! +\! 1,s\! -\! 1,u\! +\! 1)
\end{array} \right ) = F(a,s,u\! +\! 1)
\left ( \begin{array}{c}
T(a \! +\! 1, s, u)\\ \\
F(a, s\! -\! 1, u)
\end{array} \right )
\end{equation}
which is to be compared with (\ref{LP51}). It is obvious that they
differ by the substitution $T\leftrightarrow F$ and changing signs
of all variables. Therefore, we get the same Hirota equation for
$F$. We thus conclude that any solution to the linear problems
(\ref{LP5}), where the $T$-function obeys the Hirota equation,
provides an auto-B\"acklund transformation, i.e., a transformation
that sends a solution of the nonlinear integrable equation to
another solution of the same equation. In what follows we call them
simply B\"acklund transformations (BT) and distinguish two types of
them.

Let us rewrite the linear problems (\ref{LP5}) changing the order of
the terms and shifting the variables:
\begin{eqnarray}
\label{LINPR1}
  T(a\! +\! 1,s,u)F(a,s,u\! +\! 1)-
  T(a,s,u\! +\! 1)F(a\! +\! 1,s,u) &=&
  T(a\! +\! 1,s\! -\! 1,u\! +\! 1)F(a,s\! +\! 1,u),\nn\\
   T(a,s\! +\! 1,u\! +\! 1)F(a,s,u)-
   T(a,s,u)F(a,s\! +\! 1,u\! +\! 1) &=& T(a\! +\! 1,s,u\! +\! 1)
   F(a\! -\! 1,s\! +\! 1,u).
\end{eqnarray}
These equations are graphically represented in
Fig.~\ref{fig:HfigLP1} in the $(s,a)$ plane. Given polynomials
$T(a,s,u)$ obeying the Hirota equation, we are going to seek for
polynomial solutions for $F$.

It is easy to see  that equations (\ref{LINPR1}) are not compatible
if one imposes the boundary conditions for $F(a,s,u)$ and $T(a,s,u)$
of the fat hook type with the same $K$ and $M$. Indeed, applying
them in the corner point of the interior boundary, one sees that the
boundary values must vanish identically. However, it is
straightforward to verify that equations (\ref{LINPR1}) are
compatible with the boundary conditions of the following two types.
The boundary conditions for $F(a,s,u)$ can be either
\begin{eqnarray}
\label{BCONDF1}
  F(a,s,u)=0\quad  {\rm if:}\quad  \mbox{(i)} \;\; a<0
  \quad {\rm or} \quad \mbox{(ii)} \;\; s<0 \;\; {\rm and} \;\;
  a\ne 0, \quad {\rm or} \quad \mbox{(iii)} \;\; a>K\! -\! 1
  \;\; {\rm and}\;\;  s>M,
\end{eqnarray}
or
\begin{eqnarray}
\label{BCONDF1a}
  F(a,s,u)=0\quad  {\rm if:}\quad  \mbox{(i)} \;\; a<0
  \quad {\rm or} \quad \mbox{(ii)} \;\; s<0 \;\; {\rm and} \;\;
  a\ne 0, \quad {\rm or} \quad \mbox{(iii)} \;\; a>K
  \;\; {\rm and}\;\;  s>M+1\,,
\end{eqnarray}
which are again of the fat hook type but with the shifts $K\to K-1$
or $M\to M+1$.

\subsubsection{First B\"acklund transformation}

\begin{figure}[t]
    \centering
        \includegraphics[angle=-90,scale=0.4]{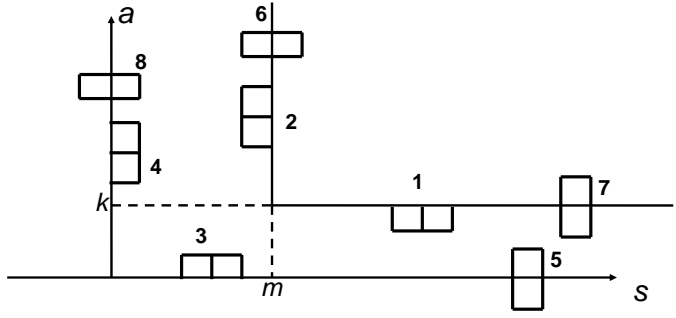}
    \caption{\it  Particular cases of application of the first
    B\"acklund transformation (BT1)
    in the form depicted in Fig.~\ref{fig:HfigLP1}.}
    \label{fig:LINPAP1}
    \end{figure}

The {\it first B\"acklund transformation} (BT1) $T\longrightarrow F$
is given by the linear problems (\ref{LINPR1}) with boundary
conditions (\ref{BCONDF1}). Moreover, the linear problems are also
compatible with the more specific form of the boundary conditions
(\ref{BCOND2}), where $K$ is replaced by $K-1$ and $M$ remains the
same. To see this, one should check all cases when one of the three
terms in the linear equations vanishes, which imposes certain
constraints on the boundary functions. The first equation at $s=0$
(the ``brick" in position 8 in Fig.~\ref{fig:LINPAP1}), with $T(a,
0,u)$ as in (\ref{BCOND1}), states that $F(a,0,u)$ depends only on
$u+a$. Therefore, we can set $F(a,0,u)=Q_{K-1, M}(u+a)$, which at
this step is just the notation. Similarly, the second equation at
$a=0$ (position 5) implies that $F(0,s,u)$ depends only on $u-s$. At
$s=0$ it must equal $Q_{K-1, M}(u)$, therefore, $F(0,s,u)=Q_{K-1,
M}(u-s)$.

The consistency of the last two boundary conditions in
(\ref{BCOND1}) follows from a similar analysis at the interior
boundaries. Indeed, at $a=K-1,\,\, s>M$ (position 1) the first
equation becomes
\begin{equation}\label{LINPRcPrim1}
\begin{array}{ll}
  &Q_{K,0}(u+s+K)Q_{0,M}(u-s-K) F(K-1,s,u+1)\\ &\\
 =& Q_{K,0}(u+s+K)Q_{0,M}(u-s-K+2)F(K-1,s+1,u)\,.
\end{array}
\end{equation}
This equation fixes the $(u-s)$-dependent factor in $F(K-1,s,u)$ to
be $Q_{0,M}(u-s-(K-1))$ while no restrictions on the
$(u+s)$-dependent factor emerge. One is free to call it
$Q_{K-1,0}(u+s +K-1)$, so that
\begin{equation}\label{HARMSOL}
  F(K-1,s,u)=Q_{K-1,0}(u+s+K-1)Q_{0,M}(u-s-K+1),\quad  M\le s<\infty,
\end{equation}
i.e., the boundary condition  for $F(a,s,u)$ on the half-line
$a=K-1,\,\, M\le s<\infty$ takes the same form as the 3-rd one from
\eq{BCOND2} for $T(a,s,u)$ on the half-line $a=K,\,\, M\le s<\infty$
(see  Fig.~\ref{fig:SHIFT1}). Finally, one can check that the first
equation in (\ref{LINPR1}) at $s=M$, $a\geq K$ (position 6) is
consistent with
\begin{equation}\label{HARMSOL1}
  F(a,M,u)=(-1)^{M(a-K+1)}Q_{K-1,0}(u+a+M)
Q_{0,M}(u-a-M),\quad  K-1\le a<\infty,
\end{equation}
and does not bring any new constraints. The sign factor is not fixed
by this argument. To fix it, one should require consistency with the
second B\"acklund transformation which we consider in the next
subsection.

\subsubsection{Second B\"acklund transformation}

\begin{figure}[t]
    \centering
        \includegraphics[angle=-90,scale=0.5]{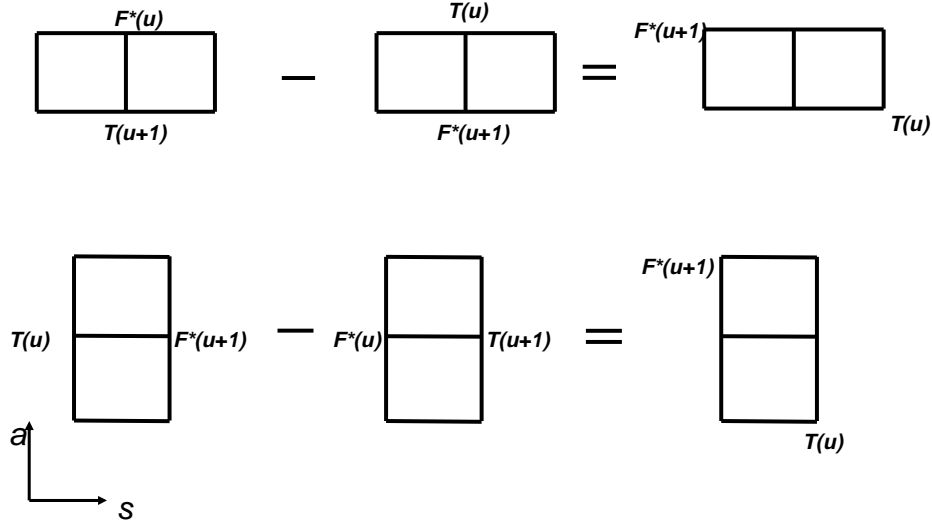}
    \caption{\it Two equations of the second pair of linear problems
in the $(s,a)$ plane. }
    \label{fig:HfigLP2}
\end{figure}

As is seen from eq. (\ref{BCONDF1a}), the transformations generated
by the linear problems (\ref{LINPR1}) are not able to move the
vertical part of the interior boundary from the right to the left.
For our purpose, we need a transformation which would be able to
decrease $M$. Using the duality between $T$ and $F$ explained in the
end of Sect. 3.2, one can introduce B\"acklund transformations of
the required type. The {\it second B\"acklund transformation} (BT2)
$T\longrightarrow  F^*$ is obtained from \eq{LINPR1} by exchanging
$T\to F^*$ and $F\to T$:
\begin{eqnarray}
\label{LINPR2}
  F^*(a\! +\! 1,s,u)T(a,s,u\! +\! 1)-
 F^* (a,s,u\! +\! 1)T(a\! +\! 1,s,u) &=&
F^*(a\! +\! 1,s\! -\! 1,u\! +\! 1)T(a,s\! +\! 1,u),\nn\\
   F^*(a,s\! +\! 1,u\! +\! 1)T(a,s,u)-
F^*(a,s,u)T(a,s\! +\! 1,u\! +\! 1) &=& F^*(a\! +\! 1,s,u\! +\!
1)T(a\! -\! 1,s\! +\! 1,u).
\end{eqnarray}
These equations are represented graphically in
Fig.~\ref{fig:HfigLP2} in the $(s,a)$ plane. As is argued in Sect.
3.2, their compatibility condition is the Hirota equation
(\ref{HIROTA}) for $T$. If it holds, then any solution $F^*$ obeys
the same Hirota equation and thus provides an auto-B\"acklund
transformation.

Given a solution $T(a,s,u)$ with the boundary conditions
(\ref{BCOND2}), equations (\ref{LINPR2})  are compatible with the
following boundary condition for $F^*(a,s,u)$:
\begin{eqnarray}
\label{BCONDF2} F^*(a,s,u)=0 \quad  {\rm if:}\quad  \mbox{(i)} \;\;
a<0
  \quad {\rm or} \quad \mbox{(ii)} \;\; s<0 \;\; {\rm and} \;\;
  a\ne 0, \quad {\rm or} \quad \mbox{(iii)} \;\; a>K
  \;\; {\rm and}\;\;  s>M-1,
\end{eqnarray}
which directly follow from (\ref{BCONDF1a}) and differ from those
for $T$ (\ref{BCOND1}) by the shift $M\to M-1$. Moreover, they are
also compatible with the more specific form of the boundary
conditions (\ref{BCOND2}), where $M$ is replaced by $M-1$ and $K$
remains the same. In complete analogy with BT1, one verifies this by
applying BT2 on the exterior boundaries (positions 8 and 5 in
Fig.~\ref{fig:LINPAP1}) and on the interior ones (positions 7 and
2). In particular, on the half-line $s=M-1,\, K\le a<\infty$ we have
\begin{equation}\label{HARMSOLS2}
F^*(a,M-1,u)=(-1)^{(M-1)(a-K)} Q_{K,0}(u+a+M-1) Q_{0,M-1}(u-a-M+1),
\quad  K \le a<\infty,
\end{equation}
the sign being uniquely fixed from the second equation of
(\ref{LINPR2}) applied in position 2. This means that  the boundary
condition for $F^*(a,s,u)$ on the half-line $s=M-1, \,\, K\le
a<\infty$ takes the same form as the one for $T(a,s,u)$ on the
half-line $s=M, \,\, K\le a < \infty$ (see the 4-th equation in
(\ref{BCOND2}) and  Fig.~\ref{fig:SHIFT2}).

Our final goal is to ``undress", using the transformations BT1 and
BT2, the original problem by collapsing the region where the Hirota
equation acts: $K\to 0,\,\,M\to 0$.

\subsection{Hierarchy of Hirota equations and linear problems}

\begin{figure}[t]
    \centering
        \includegraphics[angle=-90,scale=0.4]{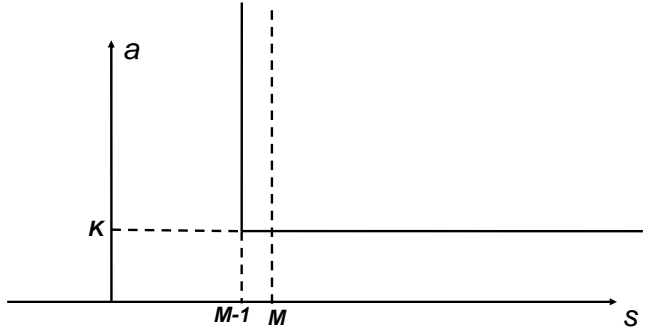}
    \caption{\it  Boundary conditions for the second
    B\"acklund transformation in the $(s,a)$ plane.}
    \label{fig:SHIFT2}
\end{figure}

We see now that by applying BT1 or BT2 to a solution to the Hirota
equation with the boundary conditions (\ref{BCOND2}) we can shift
the interior boundaries as $K\to K-1$ or $M\to M-1$
(Fig.~\ref{fig:SHIFT1} and Fig.~\ref{fig:SHIFT2}, respectively). In
this way we come to the problem for $F(a,s,u)$ or $F^*(a,s,u)$ of
the same kind as the original problem for $T(a,s,u)$. Repeating
these steps several times, we arrive at the hierarchy of the
functions $T_{k,m}(a,s,u)$ such that
\begin{equation}
\label{HIERT}
\begin{array}{l}
  T_{K,M}(a,s,u)\equiv T(a,s,u), \\ \\
   T_{K-1,M}(a,s,u)\equiv F(a,s,u), \\ \\
 T_{K,M-1}(a,s,u)\equiv  F^*(a,s,u),
   \quad {\rm etc.},
\end{array}
\end{equation}
all of them satisfying the Hirota equation
\begin{equation}
\label{HIROTATT}
  T_{k,m}(a,s,u+1)T_{k,m}(a,s,u-1)-T_{k,m}(a,s+1,u)T_{k,m}(a,s-1,u)
  =T_{k,m}(a+1,s,u)T_{k,m}(a-1,s,u)\,,
\end{equation}
where $k=0, \dots K$ and $m=0, \dots, M$. The  boundary conditions
are as  follows
\begin{eqnarray}
\label{BCOND1T} T_{k,m}(a,s,u)=0 \quad  {\rm if:}\quad  \mbox{(i)}
\;\; a<0
  \quad {\rm or} \quad \mbox{(ii)} \;\; s<0 \;\; {\rm and} \;\;
  a\ne 0, \quad {\rm or} \quad \mbox{(iii)} \;\; a>k
  \;\; {\rm and}\;\;  s>m,
\end{eqnarray}
and
\begin{eqnarray}
\label{BCOND2T}
  T_{k,m}(0,s,u)&=&Q_{k,m}(u-s),\quad  -\infty<s<\infty,\nn\\
  T_{k,m}(a,0,u)&=&Q_{k,m}(u+a),\quad  0\le a <\infty,\nn\\
  T_{k,m}(k,s,u)&=&Q_{k,0}(u\! +\! s\! +\! k)\,
Q_{0,m}(u\! -\! s\! -\! k),\quad  m\le s<\infty,\nn\\
  T_{k,m}(a,m,u)&=&
(-1)^{m(a-k)}\, Q_{k,0}(u\! +\! a\! +\! m)\, Q_{0,m}(u\! -\! a\! -\!
m),\quad  k \le a <\infty,
\end{eqnarray}
(see Fig.~\ref{fig:Hirota}). The function $Q_{K,M}(u)$ is a fixed
polynomial of degree $N$ which depends on the choice of the quantum
(super) spin chain with the $gl(K|M)$ symmetry. We set
$Q_{0,0}(u)=1$ (it corresponds to an empty chain). The other
polynomial functions $Q_{k,m}(u)$ will be determined from polynomial
solutions to the hierarchy of Hirota equations.

The linear problem (\ref{LINPR1}) generates a chain of B\"acklund
transformations BT1:
\begin{equation}
\label{LINPRT1}
\begin{array}{c}
  T_{k,m}(a+1,s,u)T_{k-1,m}(a,s,u+1)-T_{k,m}(a,s,u+1)T_{k-1,m}(a+1,s,u)
\\ \\
=\,\, T_{k,m}(a+1,s-1,u+1)T_{k-1,m}(a,s+1,u), \\ \\
  T_{k,m}(a,s+1,u+1)T_{k-1,m}(a,s,u)-T_{k,m}(a,s,u)T_{k-1,m}(a,s+1,u+1)
\\ \\
=\,\,  T_{k,m}(a+1,s,u+1)T_{k-1,m}(a-1,s+1,u),\\ \\
\end{array}
\end{equation}
($k=1,\dots,K$). The linear problem (\ref{LINPR2}) generates a chain
of B\"acklund transformations BT2:
\begin{equation}
\label{LINPRT2}
\begin{array}{c}
  T_{k,m-1}(a+1,s,u)T_{k,m}(a,s,u+1)-T_{k,m-1}(a,s,u+1)T_{k,m}(a+1,s,u)
\\ \\
=\,\,  T_{k,m-1}(a+1,s-1,u+1)T_{k,m}(a,s+1,u), \\ \\
   T_{k,m-1}(a,s+1,u+1)T_{k,m}(a,s,u)-T_{k,m-1}(a,s,u)T_{k,m}(a,s+1,u+1)
\\ \\
=\,\,    T_{k,m-1}(a+1,s,u+1)T_{k,m}(a-1,s+1,u)
\end{array}
\end{equation}
($m=1,\dots , M$). Note that (\ref{LINPRT2}) differs from
(\ref{LINPRT1}) only by the ``direction" of the B\"acklund flow: it
is $k\to k-1$ in (\ref{LINPRT1}) and $m\to m+1$ in (\ref{LINPRT2}).
In the 5D linear space with coordinates $a,s,u,k,m$,
the four equations (\ref{LINPRT1}), (\ref{LINPRT2}) act
in the hyper-planes
$$
\begin{array}{c}
\{m=\mbox{const}\}\, \cap \, \{u+s+a=\mbox{const}\}, \\ \\
\{m=\mbox{const}\}\, \cap \, \{u-s-a=\mbox{const}\}, \\  \\
\{k=\mbox{const}\}\, \cap \, \{u+s+a=\mbox{const}\}, \\  \\
\{k=\mbox{const}\}\, \cap \, \{u-s-a=\mbox{const}\},
\end{array}
$$
respectively. Thus each of them is actually a dynamical
equation in three variables rather than five. It is easy to
see that all of them can be transformed to the standard form of
the Hirota equation (\ref{AH}) by linear changes of variables.

The two B\"acklund transformations can be unified in a matrix
equation of the type (\ref{LP53}). Let ${\Bbb T}_{k,m}(a,s,u)$ be
the antisymmetric matrix from the left hand side in (\ref{LP53}) at
the level $k,m$:
\begin{equation}\label{LP532}
{\Bbb T}_{k,m}(a,s,u)= \left (
\begin{array}{cccc}
0 & T_{k,m}(a,s,u\! -\! 1) & - T_{k,m}(a,s\! +\! 1,u) & T_{k,m}(a\!
+\! 1,s,u)
\\ &&&\\
-T_{k,m}(a,s,u\! -\! 1) & 0 & T_{k,m}(a\! -\! 1, s,u) &
T_{k,m}(a,s\! -\! 1,u)
\\ &&& \\
T_{k,m}(a,s\! +\! 1,u) & -T_{k,m}(a\! -\! 1,s,u) & 0 &
-T_{k,m}(a,s,u\! +\! 1)
\\ &&& \\
-T_{k,m}(a\! +\! 1,s,u) & -T_{k,m}(a,s\! -\! 1, u) & T_{k,m}(a,s,u\!
+\! 1) & 0
\end{array} \right ),
\end{equation}
then the B\"acklund transformations BT1 and $(\mbox{BT2})^{-1}$ (the
transformation inverse to BT2) in the symmetric form are obtained as
the first and the second columns of the matrix equation
\begin{equation}\label{LP531}
{\Bbb T}_{k,m}(a,s,u) \left (
\begin{array}{cc}
T_{k\! -\! 1,m}(a\! -\! 1,\, s,\, u)& T_{k, m\! +\! 1}(a\! -\! 1,\,
s,\, u)
\\ &\\
T_{k\! -\! 1,m}(a,\, s\! +\! 1,\, u)& T_{k, m\! +\! 1}(a,\, s\! +\!
1,\, u)
\\& \\
T_{k\! -\! 1,m}(a,\, s,\, u\! -\! 1)& T_{k, m\! +\! 1}(a,\, s,\, u\!
-\! 1)
\\ &\\
T_{k\! -\! 1,m}(a\! -\! 1, s\! +\! 1, u\! -\! 1)& T_{k, m\! +\!
1}(a\! -\! 1, s\! +\! 1, u\! -\! 1)
\end{array}
\right ) =0.
\end{equation}
The first and the second equations in (\ref{LINPRT1}),
(\ref{LINPRT2}) are obtained in the second and the first lines of
the matrix equation (\ref{LP531}) respectively. As we have seen
above, the rank of the $4\times 4$ matrix ${\Bbb T}_{k,m}$ is $2$,
so it has two linearly independent zero eigenvectors. Now we see
that they correspond to the two independent transformations shifting
either $k$ or $m$.

\subsection{Bilinear equations for the $T$-functions
with different $k$ and $m$}

In this subsection we derive additional bilinear equations
(\ref{BT1,2-1})-(\ref{BT1,2-3n}) for
the functions $T_{k,m}(a,s,u)$ in which {\it both} indices
$k,m$ undergo shifts by $\pm 1$. They are also of the Hirota type.
A special case of them
(equation (\ref{bilEqQ})) is particularly important.
It provides a bilinear relation for the $Q$-functions
(the ``$QQ$-relation") which will be also
derived in section 4 by other means.

All these additional bilinear relations follow from
compatibility of equations (\ref{LINPRT1}), (\ref{LINPRT2})
with shifts in $k$ and in $m$. To see this more explicitly,
we note that these equations admit a remarkable
alternative representation. Namely, it is straightforward to verify
that the first equations in (\ref{LINPRT1}), (\ref{LINPRT2}) can be
identically rewritten as

\begin{equation}\label{alt1}
\frac{T_{k-1,m}(a\! +\! 1,s,u)}{T_{k-1,m}(a,s\! +\! 1,u)}= \left [
\frac{T_{k-1,m}(a,s,u\! +\! 1)T_{k,m}(a,s\! +\! 1,u)}
{T_{k-1,m}(a,s\! +\! 1,u)T_{k,m}(a,s,u\! +\! 1)} - e^{\p_u -
\p_s}\right ] \frac{T_{k,m}(a\! +\! 1,s,u)}{T_{k,m}(a,s\! +\!
1,u)}\,,
\end{equation}
\begin{equation}\label{alt2}
\frac{T_{k,m+1}(a\! +\! 1,s,u)}{T_{k,m+1}(a,s\! +\! 1,u)}= \left [
\frac{T_{k,m+1}(a,s,u\! +\! 1)T_{k,m}(a,s\! +\! 1,u)}
{T_{k,m+1}(a,s\! +\! 1,u)T_{k,m}(a,s,u\! +\! 1)} - e^{\p_u -
\p_s}\right ] \frac{T_{k,m}(a\! +\! 1,s,u)}{T_{k,m}(a,s\! +\!
1,u)}\,,
\end{equation}

\noindent
while the second equations as

\begin{equation}\label{alt3}
\frac{T_{k+1,m}(a,s\! +\! 1,u)}{T_{k+1,m}(a\! +\! 1,s,u)}= \left [
\frac{T_{k+1,m}(a,s,u\! -\! 1)T_{k,m}(a\! +\! 1,s,u)}
{T_{k+1,m}(a\! +\! 1,s,u)T_{k,m}(a,s,u\! -\! 1)} + e^{-\p_u -
\p_a}\right ] \frac{T_{k,m}(a,s\! +\! 1,u)}{T_{k,m}(a\! +\!
1,s,u)}\,,
\end{equation}
\begin{equation}\label{alt4}
\frac{T_{k,m-1}(a,s\! +\! 1,u)}{T_{k,m-1}(a\! +\! 1,s,u)}= \left [
\frac{T_{k,m-1}(a,s,u\! -\! 1)
T_{k,m}(a\! +\! 1,s,u)}{T_{k,m-1}(a\! +\! 1,s,u)T_{k,m}(a,s,u\! -\! 1)}
+ e^{-\p_u - \p_a}\right ]
\frac{T_{k,m}(a,s\! +\! 1,u)}{T_{k,m}(a\! +\! 1,s,u)}\,.
\end{equation}

\noindent
In this form they appear as linear problems for the difference
operators in the brackets together with particular solutions. One
can notice that they again look like auxiliary linear problems for
the Hirota equation but for a different choice of the variables. For
example, equations (\ref{alt1}), (\ref{alt2}) should be compared
with eq. (\ref{A1}) from Appendix B, where we identify $\tau$ with
$T$ and choose the variables as $p_1 = -k$, $p_2 =m$,
$p_3=\frac{1}{2}(u-s)$ (note that the variables $a$ and
$\frac{1}{2}(u+s)$ are kept constant in the $T$-functions entering
the difference operator in (\ref{alt1}), (\ref{alt2})). We see that
the two equations coincide with the two corresponding linear problems
from (\ref{A1}) (with $\lambda_1 =-\lambda_2 =1$), with
$$
\psi_{k,m}(a,s,u) =
\frac{T_{k,m}(a\! +\! 1,s,u)}{T_{k,m}(a,s\! +\! 1,u)}
$$
being their common solution. Because they
hold for any $k,m$, the function $\psi_{k-1,m+1}(a,s,u)$
can be represented in two different ways as follows:

$$
\left [ \frac{T_{k-1,m+1}(a,s,u\! +\! 1)T_{k-1,m}(a,s\! +\! 1,u)}
{T_{k-1,m+1}(a,s\! +\! 1,u)T_{k-1,m}(a,s,u\! +\! 1)} - e^{\p_u -
\p_s}\right ] \left [ \frac{T_{k-1,m}(a,s,u\! +\! 1)T_{k,m}(a,s\!
+\! 1,u)} {T_{k-1,m}(a,s\! +\! 1,u)T_{k,m}(a,s,u\! +\! 1)} - e^{\p_u
- \p_s}\right ]  \psi_{k,m}
$$

$$
=\,\,\,
\left [ \frac{T_{k-1,m+1}(a,s,u\!
+\! 1)T_{k,m+1}(a,s\! +\! 1,u)} {T_{k-1,m+1}(a,s\! +\!
1,u)T_{k,m+1}(a,s,u\! +\! 1)} - e^{\p_u - \p_s}\right ]
\left [ \frac{T_{k,m+1}(a,s,u\! +\! 1)T_{k,m}(a,s\! +\! 1,u)}
{T_{k,m+1}(a,s\! +\! 1,u)T_{k,m}(a,s,u\! +\! 1)} - e^{\p_u -
\p_s}\right ] \psi_{k,m}\,.
$$

\noindent
Opening the brackets, one finds that the terms multiplied by
$\psi_{k,m}(a,s,u)$ and $\psi_{k,m}(a,s\! -\! 2,u\! +\! 2)$
cancel automatically while the terms proportional to
$\psi_{k,m}(a,s\! -\! 1,u\! +\! 1)$ give a non-trivial relation
connecting the $T$-functions with different $k$ and $m$.
It has the form
\begin{equation}\label{bileqT}
\begin{array}{c}
T_{k,m}(a, s\! +\! 1, u)T_{k+1, m+1}(a,s, u\! +\! 1) -
T_{k,m}(a,s, u\! +\! 1) T_{k+1, m+1}(a, s\! +\! 1, u)
\\ \\
=\,\,\, f_{k,m}(a, u\! +\! s)\, T_{k+1, m}(a,s, u\! +\! 1) T_{k,
m+1}(a, s\! +\! 1, u)\,,
\end{array}
\end{equation}
where $f_{k,m}(a, u\! +\! s)$ is an arbitrary function of $k,m$ and
$a, \, u+s$. Comparing it with a similar equation obtained
in the same way from the other pair of linear problems
(\ref{alt3}), (\ref{alt4}), one can see that it actually depends
on the combination $u+s-a$ as well as on $k,m$ (see Appendix C
for details).
To fix it, we take $s=m$, so the first term of
the equation vanishes (at $a\geq k+1$),
and use boundary conditions (\ref{BCOND2T}).
This fixes the function $f_{k,m}$ to be $1$.

Clearly,
eq.~(\ref{bileqT}) (with $f_{k,m}(a, u\! +\! s)=1$)
is the Hirota equation of the form (\ref{AH}) in
the variables $-k$, $m$ and $\frac{1}{2}(u-s)$. The Hirota equation
implies the compatibility of the linear problems, i.e., the discrete
zero curvature condition for the difference operators in
(\ref{alt1}), (\ref{alt2}) holds true. In our situation, it appears
to be equivalent to the ``weak form" of this condition, i.e., with
the operators being applied to a particular solution of the linear
problems. In other words, the compatibility of the linear problems
(which in general means existence of a continuous family of common
solutions) follows, in our case, from the existence of just one
common solution (cf. \cite{Krichever06}).

There are other equations of the same type. It is
convenient to write them all as the following chain of
equalities:

\begin{eqnarray}
\label{BT1,2-1}
1&=&\frac{T_{k,m}(a,s+1,u)
T_{k+1,m+1}(a,s,u+1)-T_{k,m}(a,s,u+1)T_{k+1,m+1}(a,s+1,u)}
{T_{k,m+1}(a,s+1,u)T_{k+1,m}(a,s,u+1)}\\
\label{BT1,2-4}
&=&\frac{T_{k,m}(a-1,s,u)T_{k+1,m+1}(a,s,u+1)-
T_{k,m}(a,s,u+1)T_{k+1,m+1}(a-1,s,u)}
{T_{k,m+1}(a-1,s,u)T_{k+1,m}(a,s,u+1)}\\
\label{BT1,2-3}
&=&\frac{T_{k,m}(a-1,s,u)T_{k+1,m+1}(a,s-1,u)-
T_{k,m}(a,s-1,u)T_{k+1,m+1}(a-1,s,u)}
{T_{k,m+1}(a-1,s,u)T_{k+1,m}(a,s-1,u)}\,.
\end{eqnarray}

\noindent
These equations have the same structure. In each equation, one of
the variables $a,s,u$ enters as a parameter. More precisely,
they act in the hyper-planes
$$
\begin{array}{c}
\{a=\mbox{const}\}\, \cap \, \{u+s=\mbox{const}\}, \\ \\
\{s=\mbox{const}\}\, \cap \, \{u-a=\mbox{const}\}, \\ \\
\{u=\mbox{const}\}\, \cap \, \{a+s=\mbox{const}\}, \\ \\
\end{array}
$$
respectively.
The first equality in this chain is already proved. The
proof of the other two is straightforward: one should pass to common
denominator, to group together similar terms and to use equations
(\ref{LP531}). In fact this chain can be continued by three more
equations of a similar but different structure:

\begin{eqnarray}
\label{BT1,2-5}
1&=&\frac{T_{k,m}(a,s-1,u)T_{k+1,m+1}(a,s,u+1)-
T_{k,m}(a,s,u+1)T_{k+1,m+1}(a,s-1,u)}
{T_{k,m+1}(a-1,s,u)T_{k+1,m}(a+1,s-1,u+1)}\\
\label{BT1,2-2}
&=&\frac{T_{k,m}(a,s,u+1)T_{k+1,m+1}(a+1,s,u)-
T_{k,m}(a+1,s,u)T_{k+1,m+1}(a,s,u+1)}
{T_{k,m+1}(a,s+1,u)T_{k+1,m}(a+1,s-1,u+1)}\\
\label{BT1,2-3n}
&=&\frac{T_{k,m}(a,s-1,u)T_{k+1,m+1}(a+1,s,u)-
T_{k,m}(a+1,s,u)T_{k+1,m+1}(a,s-1,u)}
{T_{k,m+1}(a,s,u-1)T_{k+1,m}(a+1,s-1,u+1)}\,.
\end{eqnarray}

\noindent
They are proved in the same manner. There is no need to prove the
first equality separately since one can just continue the chain of
equations (\ref{BT1,2-1})-(\ref{BT1,2-3}) proving that
(\ref{BT1,2-3}) is equal to (\ref{BT1,2-5}).
Other forms of these equations and more details can be found
in Appendix C.
Equations
(\ref{BT1,2-5})-(\ref{BT1,2-3n}) act in the hyper-planes
$$
\begin{array}{c}
\{a-k+m =\mbox{const}\} \, \cap \, \{u-s-a=\mbox{const}\}, \\ \\
\{s+k-m=\mbox{const}\} \, \cap \, \{u+s+a=\mbox{const}\}, \\ \\
\{u-k+m=\mbox{const}\} \, \cap \, \{u+s-a=\mbox{const}\}, \\  \\
\end{array}
$$
respectively.
Therefore, equations (\ref{BT1,2-1})-(\ref{BT1,2-3}) and
(\ref{BT1,2-5})-(\ref{BT1,2-3n}) are actually equations in
three variables rather than five. They can be transformed to the
standard form of the Hirota equation (\ref{AH}) by linear changes
of variables.

We note that equations (\ref{BT1,2-1})-(\ref{BT1,2-3}) can be
written in the following concise form:
\begin{equation}\label{C143}
\left ( e^{\p_p}-e^{\p_q}\right ) \frac{T_{k+1, m+1}}{T_{km}}\, =\,
\left (e^{\p_p}\frac{T_{k+1, m}}{T_{km}}\right ) \left
(e^{\p_q}\frac{T_{k, m+1}}{T_{km}}\right )\,.
\end{equation}
Here $T_{k,m}\equiv T_{k,m}(a,u,s)$ and $(p,q)$ stands for any one
of the pairs $(u,s)$, $(u,-a)$ and $(a,s)$.

Restricting the bilinear equations to the boundaries
of the fat hook domain, one obtains new equations which
include the $Q$-functions. For example,
setting $a=0$ in eq.~(\ref{bileqT})
(or eq.~(\ref{BT1,2-1})) and using (\ref{BCOND2T}), we get
the $QQ$-relation mentioned in the Introduction:
\begin{eqnarray}
\label{bilEqQ}
Q_{k,m}(u)Q_{k+1,m+1}(u+2)-
Q_{k,m}(u+2)Q_{k+1,m+1}(u)= Q_{k,m+1}(u)Q_{k+1,m}(u+2)\,.
\end{eqnarray}
In the next section, it will be derived by other means.
Another new relation, which is of a mixed
($TQ$ and $QQ$) type,
is the particular case of eq.~(\ref{BT1,2-3n}) at $a=0, s=1$:
\begin{eqnarray}
\label{bilEqTQ}
T_{k+1,m+1}(1,1,u)Q_{k,m}(u)-
T_{k,m}(1,1,u)Q_{k+1,m+1}(u)= Q_{k,m+1}(u-2)Q_{k+1,m}(u+2)\,.
\end{eqnarray}

\section{$TQ$- and $QQ$-relations}

In this section we partially solve the ``undressing" problem for the
hierarchy of the $T$-functions $T_{k,m}(a,s,u)$ and derive the
generalized Baxter equations ($TQ$-relations) which express
$T(1,s,u)$, $T(a,1,u)$ through the Baxter functions $Q_{k,m}(u)$.
This is done by constructing an operator generating series for the
$T$-functions and factorizing it into an ordered product of first
order difference operators, with coefficients being ratios of the
$Q$-functions\footnote{For the bosonic case this was done in
\cite{BR1,Kuniba-1,Kuniba-2,Kuniba-3,KLWZ}. For the
supersymmetric case
such equations were conjectured in \cite{Tsuboi-1} (see also
\cite{Beisert:2005di}).}.
These operators
obey a discrete zero curvature condition which leads to
a bilinear relation for the functions $Q_{k,m}$ with
different values of $k,m$ (the $QQ$-relation).

\subsection{Operator generating series and generalized Baxter
relations}

We start by introducing the following difference operators of
infinite order:
\begin{eqnarray}\label{DIFFOP}
\hat W_{k,m}(u)&=& \sum_{s=0}^\infty
\frac{T_{k,m}(1,s,u\! +\! s\! +\! 1)}{Q_{k,m}(u)}\, e^{2s\p_u},\nn\\
\check W_{k,m}(u) &=&\sum_{a=0}^\infty (-1)^a e^{2a\p_u}\,
\frac{T_{k,m}(a,1,u\! -\! a\! +\! 1)}{Q_{k,m}(u+2)}
\end{eqnarray}
which represent operator generating series for the
transfer matrices corresponding to one-row or one-column
Young diagrams. The denominators are introduced for the
proper normalization.
Let us show that the difference operators
\begin{eqnarray} \label{NOTAT}
\hat U_{k,m}(u)&=& \frac{Q_{k+1,m}(u)\, Q_{k,m}(u\! +\! 2)}
{Q_{k+1,m}(u\! +\! 2)\, Q_{k,m}(u)}
\, -\,\,e^{2\p_u},\nn\\
\hat V_{k,m}(u)&=& \frac{Q_{k,m}(u)\, Q_{k,m+1}(u\! +\!
2)}{Q_{k,m}(u\! +\! 2) \, Q_{k,m+1}(u)}\, -\,\, e^{2\p_u}.
\end{eqnarray}
shift the level indices $k,m$ of the $\hat W_{k,m}(u)$ and $\check
W_{k,m}(u)$. Namely, we are going to prove the following operator
relations:
\begin{eqnarray}\label{WREL}
\hat W_{k,m}(u)&=&  \hat U_{k,m}(u)     \hat W_{k+1,m}(u),\nn\\
\hat W_{k,m+1}(u) &=&  \hat V_{k,m}(u)    \, \hat    W_{k,m}(u),
\end{eqnarray}
\begin{eqnarray}\label{CHWREL}
\check W_{k\! +\! 1,m}(u)&=& \check W_{k,m}(u)   \,\hat U_{k,m}(u),\nn\\
\check W_{k,m}(u) &=& \check  W_{k,m\! +\! 1}(u) \, \hat V_{k,m}(u).
\end{eqnarray}

\begin{figure}[t]
    \centering
        \includegraphics[angle=-90,scale=0.4]{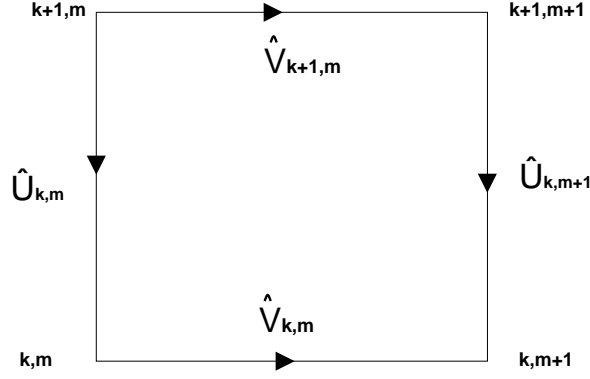}
    \caption{\it   The discrete zero curvature condition
    $\hat U_{k,m+1}  \hat V_{k+1,m} =  \hat V_{k,m} \hat U_{k,m}$.}
    \label{fig:ZeroCurv}
\end{figure}

To prove the first equation in (\ref{WREL}), we write
\begin{eqnarray}\label{REWR}
&&\hat W_{k-1}(u)+e^{2\p_u}\hat W_{k}(u)= \hat W_{k-1}(u)+
\hat W_{k}(u+2)e^{2\p_u}\nn\\
&&=\,\,\,\, \frac{T_{k-1}(1,0,u+1)}{Q_{k-1}(u)}+\sum_{s=1}^\infty
\[\frac{T_{k-1}(1,s,u\! +\! s\! +\! 1)}{Q_{k-1}(u)}+
\frac{T_{k}(1,s\! -\! 1,u\! +\! s\! +\! 2)}{Q_{k}(u+2)}\, \]
e^{2s\p_u}\,.
\end{eqnarray}
(Here and below in the proof we omit the second index $m$ since it
is the same everywhere.) To transform the expression in the square
brackets, we use the first equation of the BT1 at $a=0$ (position 3
in Fig.~\ref{fig:LINPAP1}):
\begin{equation} \label{REARR2}
  T_k(1,s,u)Q_{k-1}(u\! -\! s\! +\! 1)-T_{k-1}(1,s,u)
Q_{k}(u\! -\! s\! +\! 1) =
  T_k(1,s\! -\! 1,u\! +\! 1)Q_{k-1}(u\! -\! s\! -\! 1).
\end{equation}
Shifting it $u\to u+s+1$ and dividing both sides by
$Q_{k}(u+2)Q_{k-1}(u)$, we obtain
$$
  \frac{T_k(1,s,u+s+1)Q_{k-1}(u+2)}{Q_{k}(u+2)Q_{k-1}(u)}\, -\,
\frac{T_{k-1}(1,s,u+s+1)}{Q_{k-1}(u)}\, =\,
  \frac{T_k(1,s-1,u+s+2)}{Q_{k}(u+2)}\,.
$$
Using this, we rewrite the term in the square brackets under the sum
in \eq{REWR} as
$$
\frac{Q_{k}(u)\, Q_{k-1}(u\! +\! 2)}{Q_{k}(u\! +\! 2)\,
Q_{k-1}(u)}\,\,
  \frac{T_k(1,s,u+s+1)}{Q_{k}(u)}
$$
and continue the equality (\ref{REWR}):
\begin{eqnarray}\label{REWR2}
\hat W_{k-1}(u)+e^{2\p_u}\hat
W_{k}(u)&=&\frac{T_{k-1}(1,0,u+1)}{Q_{k-1}(u)}+ \frac{Q_{k}(u)\,
Q_{k-1}(u\! +\! 2)}{Q_{k}(u\! +\! 2)\, Q_{k-1}(u)} \sum_{s=1}^\infty
\frac{T_{k}(1,s,u\! +\! s\! +\! 1)}{Q_{k}(u)} \, e^{2s\p_u}\nn\\
&=& \frac{Q_{k}(u)\, Q_{k-1}(u\! +\! 2)}{Q_{k}(u\! +\! 2)\,
Q_{k-1}(u)} \sum_{s=0}^\infty \frac{T_{k}(1,s,u+s+1)}{Q_{k}(u)} \,
e^{2s\p_u}.
\end{eqnarray}
In the last step we have noticed that the $s=0$ term of the sum
multiplied by the ratio of $Q$'s is just equal to
$T_{k-1}(1,0,u+1)/Q_{k-1}(u)$. The sum in the r.h.s. is $\hat
W_{k}(u)$, so the first equality in (\ref{WREL}) is proved. The
proof of the three other equations in (\ref{WREL}-\ref{CHWREL}) is completely
similar.

Combining equations (\ref{WREL}),(\ref{CHWREL}), we see that
$$
\check W_{k-1,m}(u)  \hat W_{k-1,m}(u)= \check W_{k,m}(u)\hat
W_{k,m}(u)= \check W_{k,m+1}(u)  \hat W_{k,m+1}(u)\,,
$$
i.e., the operator $\check W_{k,m}(u)\hat W_{k,m}(u)$ does not
depend on $k,m$. Note that $\hat W_{0,0}(u)=\check W_{0,0}(u)=1$ as
operators, since all the terms  in (\ref{DIFFOP}) are zero except
the first one, which is $1$ thanks to the ``boundary conditions"
$T_{0,0}(1,0,u+1)=Q_{0,0}(u+2)=1,T_{0,0}(0,1,u+1)=Q_{0,0}(u)=1$.
Therefore, we conclude that the operators $\hat W_{k,m}$ and $\check
W_{k,m}$ are mutually inverse\footnote{ Using the determinant
representation (\ref{F7}), it is not
 difficult to derive this fact directly
 from their definitions (\ref{DIFFOP}).}:
\begin{equation}\label{WWONE}
\check W_{k,m}(u) \hat W_{k,m}(u)=  1.
\end{equation}
In addition, applying equations (\ref{WREL}), (\ref{CHWREL}) many
times, we arrive at the following operator relations:
\begin{equation}\label{WZEROM}
\begin{array}{lll}
\hat W_{k,m}&=&  \hat U^{-1}_{k-1,m}\dots \hat U^{-1}_{0,m}\, \hat
V_{0,m-1}\dots \hat V_{0,0} ,
\\&&\\
\check W_{k,m}&=&  \hat V^{-1}_{0,0}\dots \hat V^{-1}_{0,m-1} \,\hat
U_{0,m}\dots \hat U_{k-1,m},
\end{array}
\end{equation}
where we have skipped $u$ since it is the same everywhere. Taking
equations (\ref{DIFFOP}), (\ref{WZEROM}) at $k=K$, $m=M$, we obtain
the ``non-commutative generating functions" for the transfer
matrices in the basic representations $(s^1)$ or $(1^a)$:
\begin{eqnarray}\label{BAXTER}
\sum_{s=0}^\infty\frac{T_{K,M}(1,s,u+s+1)}{Q_{K,M}(u)}\,
e^{2s\p_u}&=&
 \hat U^{-1}_{K-1,M}(u)\dots  \hat U^{-1}_{0,M}(u) \,\hat V_{0,M-1}(u)\dots
\hat V_{0,0}(u)  ,\nn\\
\sum_{a=0}^\infty (-1)^a\frac{T_{K,M}(a,1,u+a+1)}{Q_{K,M}(u+2a+2)}\,
e^{2a\p_u}&=& \hat V^{-1}_{0,0}(u)\dots  \hat V^{-1}_{0,M-1}(u)
\,\hat U_{0,M}(u)\dots \hat U_{K-1,M}(u).
\end{eqnarray}
Expanding the right hand sides in powers of $e^{2\p_u}$ and
comparing the coefficients, one obtains a set of generalized Baxter
relations between $T$'s and $Q$'s. In principle, these formulas
solve our original problem: they give solutions to the Hirota
equation in terms of the $Q$-functions representing the boundary
conditions at each level $k,m$.

\subsection{Zero curvature condition and $QQ$-relation}

The $Q$-functions are polynomials whose roots obey the Bethe
equations. Contrary to the case of bosonic $gl(K)$ algebras, the
Bethe equations for superalgebras admit many different forms. They
correspond to all possible ``undressing paths" in the $(k,m)$ plane.
Their equivalence can be established by means of certain ``duality
transformations" \cite{Tsuboi-3}.

Here we suggest an easy transparent argument to derive all these
systems of Bethe equations and the corresponding duality
transformations. Namely, we are going to show that the functions
$Q_{k,m}(u)$ obey their own Hirota equation. Given an undressing
path, it immediately produces the chain of Bethe equations. The
duality transformation is nothing else than the discrete zero
curvature condition for the operators (\ref{NOTAT}) on the $k,m$
lattice.

\begin{figure}[t]
    \centering
        \includegraphics[angle=-90,scale=0.5]{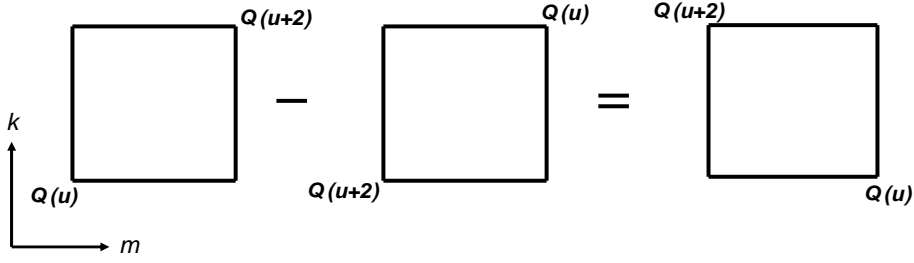}
    \caption{\it  The Hirota equation for the Baxter functions
    $Q_{k,m}(u)$ in the $(k,m)$ plane. }
    \label{fig:QHIR}
\end{figure}

Equations (\ref{WREL}) imply $ \hat W_{k-1,m+1}=  \hat U_{k-1,m+1}
\hat V_{k,m} \hat W_{k,m}= \hat V_{k-1,m} \hat U_{k-1,m} \hat
W_{k,m} $ which gives the discrete zero curvature condition
\begin{equation}\label{ZEROC}
  \hat U_{k,m+1}  \hat V_{k+1,m} =  \hat V_{k,m} \hat U_{k,m}.
\end{equation}
We remark that it looks a bit non-symmetric because $\hat V$ shifts
$m$ by $+1$ while $\hat U$ shifts $k$ by $-1$. Being written through
$\hat U^{-1}$ and $\hat V$, the zero curvature condition acquires
the standard symmetric form
\begin{equation}\label{ZEROC1}
  \hat U_{k,m+1}^{-1}  \hat V_{k,m} =  \hat V_{k+1,m} \hat
U_{k,m}^{-1}.
\end{equation}
As a consequence of it, the following bilinear relation for the
$Q's$ is valid:
\begin{equation} \label{QHIROTA}
  Q_{k,m}(u)Q_{k+1,m+1}(u+2)- Q_{k+1,m+1}(u)Q_{k,m}(u+2)=
  Q_{k,m+1}(u)Q_{k+1,m}(u+2).
\end{equation}
It was already derived in section 3.4 as a particular case
of more general ``$TT$-relations"
(\ref{BT1,2-1}), (\ref{BT1,2-4})
(see (\ref{bilEqQ})).
This is the Hirota equation in ``chiral" variables (see Appendix B).
Strictly speaking, the zero curvature condition (\ref{ZEROC})
implies \eq{QHIROTA} up to an additional factor in the r.h.s.
depending on $k,m$ which remains unfixed by this argument. However,
this factor can always be eliminated by an appropriate normalization
of $Q_{k,m}(u)$'s:
\begin{equation} \label{POLYNORMQ}
  Q_{k,m}(u)=  A _{k,m}\prod_{j=1}^{J_{k,m}} \(u-u_j^{(k,m)}\)\,,
\end{equation}
i.e., by choosing the coefficients $A_{k,m}$. Moreover, the
result of section 3.4 shows that the boundary conditions
(\ref{BCOND2T}) already imply the normalization in which
the $QQ$-relation has the form (\ref{QHIROTA})
(see the argument right after eq.~(\ref{bileqT})).

The zero curvature condition allows us to represent equations
(\ref{BAXTER}) in a more general form. Given an arbitrary zigzag
path $\gamma_{K,M}$ from $(K,M)$ to $(0,0)$, the r.h.s. of these
equations becomes the ordered  product of the shift operators along
it (see Fig.~\ref{fig:UNDRESS}):
\begin{equation}\label{BAXTERARB}
\begin{array}{rcl}
\displaystyle{ \sum_{s=0}^\infty \,
\frac{T_{K,M}(1,s,u+s+1)}{Q_{K,M}(u)}\, e^{2s\p_u}}&=&
\displaystyle{ \prod\limits_{({\bf x,n})\in
\gamma_{K,M}}^{\longleftarrow} \hat V_{({\bf x,n})}(u)}\,,
 \\&&\\
 \displaystyle{
\sum_{a=0}^\infty (-1)^a\frac{T_{K,M}(a,1,u+a+1)}{Q_{K,M}(u+2a+2)}\,
e^{2a\p_u}}&=& \displaystyle{ \prod\limits_{({\bf x,n})\in
\gamma_{K,M}}^{\longrightarrow} \hat V^{-1}_{({\bf x,n})}(u)}\,.
\end{array}
\end{equation}
Here we use the natural notation: ${\bf x}$ is the vector on the
lattice with coordinates $(k,m)$ ($k$ is the vertical coordinate and
$m$ is the horizontal coordinate!), ${\bf n}=(-1,0)$ or $(0, -1)$ is
the unit vector looking along the next step of the path. In other
words, $({\bf x}, {\bf n})$ is the (oriented) edge of the path from
$(K,M)$ to $(0,0)$ starting at the point ${\bf x}$ and looking in
the direction ${\bf n}$. In the first equation, the shift operators
are ordered from the last edge of the path (ending at the origin) to
the first one while in the second equation the order is opposite. In
fact the zero curvature condition implies that equations
(\ref{BAXTERARB}) remain true for any path leading from $(K,M)$ to
the origin provided the shift operators are chosen as follows:
\begin{equation}\label{VUREL}
\begin{array}{llll}
 \hat V_{({\bf x,n})}(u)&=&  \hat U_{k-1,m}^{-1}(u), & {\bf n}= ( -1,0),
 \\&&&\\
  \hat V_{({\bf x,n})}(u)&=&  \hat U_{k,m}(u), &  {\bf n}= (1,0),
  \\&&&\\
  \hat V_{({\bf x,n})}(u) &=&  \hat V_{k,m}^{-1}(u), & {\bf n}=(0,1),
  \\&&&\\
  \hat V_{({\bf x,n})}(u) &=&  \hat V_{k,m-1}(u), &  {\bf n}= (0,-1).
  \end{array}
\end{equation}
Some simple examples of equations (\ref{BAXTERARB}) are given in
Section 8.

\section{Bethe equations}

The bilinear $QQ$-relation (\ref{QHIROTA}) obtained
in the previous section (see also section 3.4 for an
alternative derivation) gives the easiest and the
most transparent way to derive different systems of Bethe
equations and to prove their equaivalence.
In a similar way, the generalized
$TT$-relations (\ref{BT1,2-1})-(\ref{BT1,2-3}) can be used
to derive a new system of Bethe-like equations for roots of the
polynomials $T_{k,m}(a,s,u)$.

\subsection{Bethe equations for roots of $Q$'s}

To derive the system of equations for zeros of the polynomials
$Q_{k,m}$ (\ref{POLYNORMQ}), we put $u$ in the Hirota equation
(\ref{QHIROTA}) successively equal to $u_{j}^{(k,m)}$,
$u_{j}^{(k+1,m+1)}\! -\! 2$, $u_{j}^{(k,m)}\! -\! 2$,
$u_{j}^{(k+1,m+1)}$, $u_{j}^{(k,m+1)}$ and $u_{j}^{(k+1,m)}\! -\!
2$,  each corresponding to a zero of one of the six $Q$-functions
in the equation. After proper shifts of $k$ and $m$ such that the
arguments of the $Q$-functions become $u_{j}^{(k,m)}$ and
$u_{j}^{(k,m)}\pm 2$, we get the relations
\begin{equation}\label{BE1}
\begin{array}{rcll}
-\, Q_{k,m}\left (u_{j}^{(k,m)} +2\right ) Q_{k+1, m+1}\left
(u_{j}^{(k,m)}\right )&=& Q_{k, m+1}\left (u_{j}^{(k,m)}\right )
Q_{k+1,m}\left (u_{j}^{(k,m)} +2\right )  &\quad \quad ({\rm a})
\\&&&\\
-\, Q_{k,m}\left (u_{j}^{(k,m)} -2\right ) Q_{k-1, m-1}\left
(u_{j}^{(k,m)}\right )&=& Q_{k, m-1}\left (u_{j}^{(k,m)}\right )
Q_{k-1,m}\left (u_{j}^{(k,m)} -2\right ) &\quad \quad ({\rm b})
\\&&&\\
Q_{k,m}\left (u_{j}^{(k,m)} -2\right ) Q_{k+1, m+1}\left
(u_{j}^{(k,m)}\right )&=& Q_{k+1,m}\left (u_{j}^{(k,m)}\right )
Q_{k, m+1}\left (u_{j}^{(k,m)}-2\right ) &\quad \quad ({\rm c})
\\&&&\\
Q_{k,m}\left (u_{j}^{(k,m)} +2\right ) Q_{k-1, m-1}\left
(u_{j}^{(k,m)}\right )&=& Q_{k-1, m}\left (u_{j}^{(k,m)}\right )
Q_{k,m-1}\left (u_{j}^{(k,m)} +2\right ) &\quad \quad ({\rm d})
\\&&&\\
 Q_{k,m-1}\left (u_{j}^{(k,m)} \right )
 Q_{k+1,m}\left (u_{j}^{(k,m)}+2\right )&=&
 Q_{k,m-1}\left (u_{j}^{(k,m)} +2\right )
 Q_{k+1,m}\left (u_{j}^{(k,m)}\right )   &\quad \quad ({\rm e})
\\&&&\\
 Q_{k,m+1}\left (u_{j}^{(k,m)} \right )
 Q_{k-1,m}\left (u_{j}^{(k,m)}-2\right )&=&
 Q_{k,m+1}\left (u_{j}^{(k,m)} -2\right )
 Q_{k-1,m}\left (u_{j}^{(k,m)}\right )  &\quad \quad ({\rm f})\,.
\end{array}
\end{equation}
Here $1\leq k \leq K-1$, $1\leq m \leq M-1$ and $j$ runs from $1$ to
$J_{k,m}$. This is the (over)complete set of Bethe equations for our
problem. Their consistency is guaranteed by the Hirota equation
(\ref{QHIROTA}). To convert them into a more familiar form, let us
divide eq.~(a) by eq.~(c) and eq.~(b) by eq.~(d). Using also eqs.
(e) and (f), it is easy to rewrite the system in the form where each
group of equations contains the $Q$-functions at three neighboring
sites. In this way we obtain the following sets of equations:

\begin{equation}\label{Bethe1}
\begin{array}{lll}
\begin{CD}
\bullet  \\ @VVV \\ \bullet \\@VVV \\ \bullet
\end{CD}
&\phantom{aaaaa}&\displaystyle{ \frac{Q_{k-1,m}\left
(u_{j}^{(k,m)}\right ) Q_{k,m}\left (u_{j}^{(k,m)}-2\right )
Q_{k+1,m}\left (u_{j}^{(k,m)}+2\right )}{ Q_{k-1,m}\left
(u_{j}^{(k,m)}-2\right ) Q_{k,m}\left (u_{j}^{(k,m)}+2\right )
Q_{k+1,m}\left (u_{j}^{(k,m)}\right )}=\, -\, 1}\,,
\end{array}
\end{equation}

\vspace{3mm}

\begin{equation}\label{Bethe2}
\begin{array}{lll}
\begin{CD}
 \bullet @<<< \bullet @<<< \bullet
\end{CD}
&\phantom{aa}&\displaystyle{ \frac{Q_{k,m+1}\left
(u_{j}^{(k,m)}\right ) Q_{k,m}\left (u_{j}^{(k,m)}-2\right )
Q_{k,m-1}\left (u_{j}^{(k,m)}+2\right )}{ Q_{k,m+1}\left
(u_{j}^{(k,m)}-2\right ) Q_{k,m}\left (u_{j}^{(k,m)}+2\right )
Q_{k,m-1}\left (u_{j}^{(k,m)}\right )}=\, -\, 1}\,,
\end{array}
\end{equation}

\vspace{3mm}

\begin{equation}\label{Bethe5}
\begin{array}{lll}
\begin{CD}
@. \bullet \\ @. @VVV\\ \bullet @<<<  \bullet
\end{CD}
&\phantom{aaaaaaaaaaaaaaaaaaaaa}&\displaystyle{ \frac{Q_{k+1,m}\left
(u_{j}^{(k,m)}\right ) Q_{k,m-1}\left (u_{j}^{(k,m)}+2\right )}{
Q_{k+1,m}\left (u_{j}^{(k,m)}+2\right ) Q_{k,m-1}\left
(u_{j}^{(k,m)}\right )}=\,  1}\,,
\end{array}
\end{equation}

\vspace{3mm}

\begin{equation}\label{Bethe6}
\begin{array}{lll}
\begin{CD}
\bullet @<<< \bullet \\  @VVV @. \\ \bullet @.
\end{CD}
&\phantom{aaaaaaaaaaaaaaaaaaaaa}&\displaystyle{ \frac{Q_{k,m+1}\left
(u_{j}^{(k,m)}\right ) Q_{k-1,m}\left (u_{j}^{(k,m)}-2\right )}{
Q_{k,m+1}\left (u_{j}^{(k,m)}-2\right ) Q_{k-1,m}\left
(u_{j}^{(k,m)}\right )}=\,  1}\,.
\end{array}
\end{equation}

\noindent These equations are valid at any point of the $(k,m)$
lattice and do not depend on the choice of the undressing zigzag
path. The figures show the sites of the $(k,m)$ lattice ($k$ and $m$
are the vertical and horizontal coordinates respectively) which are
connected by the corresponding Bethe equation. The point $(k,m)$ is
the one between the other two. The edges of the $(k,m)$ lattice are
represented by the arrows which show the directions of the
transformations BT1 and $\mbox{BT2}$. For completeness, we also
present here two other equations derived from (\ref{BE1}):

\begin{equation}\label{Bethe3}
\begin{array}{lll}
\begin{CD}
 \bullet @. \\  @VVV  @. \\ \bullet @<<<  \bullet
\end{CD}
&\phantom{aaa}&\displaystyle{ \frac{Q_{k,m+1}\left
(u_{j}^{(k,m)}\right ) Q_{k,m}\left (u_{j}^{(k,m)}-2\right )
Q_{k+1,m}\left (u_{j}^{(k,m)}+2\right )}{ Q_{k,m+1}\left
(u_{j}^{(k,m)}-2\right ) Q_{k,m}\left (u_{j}^{(k,m)}+2\right )
Q_{k+1,m}\left (u_{j}^{(k,m)}\right )}=\, -\, 1}\,,
\end{array}
\end{equation}

\vspace{3mm}

\begin{equation}\label{Bethe4}
\begin{array}{lll}
\begin{CD}
\bullet @<<< \bullet \\ @. @VVV\\ @.  \bullet
\end{CD}
&\phantom{aaa}&\displaystyle{ \frac{Q_{k-1,m}\left
(u_{j}^{(k,m)}\right ) Q_{k,m}\left (u_{j}^{(k,m)}-2\right )
Q_{k,m-1}\left (u_{j}^{(k,m)}+2\right )}{ Q_{k-1,m}\left
(u_{j}^{(k,m)}-2\right ) Q_{k,m}\left (u_{j}^{(k,m)}+2\right )
Q_{k,m-1}\left (u_{j}^{(k,m)}\right )}=\, -\, 1}\,,
\end{array}
\end{equation}

\vspace{3mm}

\noindent It is clear from the figures that these patterns are
forbidden for a zigzag path.

\begin{figure}[t]
    \centering
        \includegraphics[angle=-90,scale=0.4]{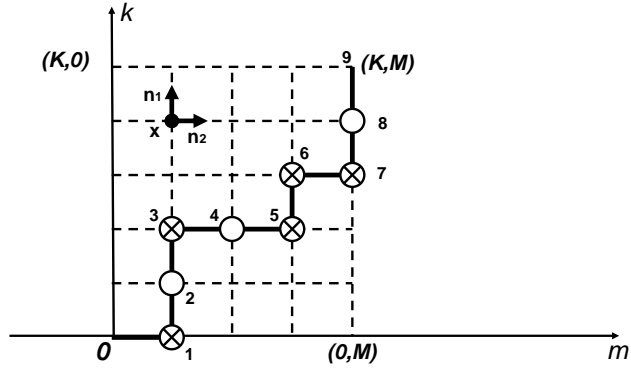}
    \caption{\it  The ``undressing" procedure by means of BT1 and BT2
in the $(s,a)$ plane
    along an arbitrary zigzag path $\gamma_{K,M}$. This path defines
    the
    Cartan matrix which characterizes the basis of  generators of the
    Cartan subalgebra.}
    \label{fig:UNDRESS}
\end{figure}

Let us show how to reduce this 2D array of Bethe equations to a
chain. Suppose one fixes a particular zigzag path $\gamma_{K,M}$
from $(K,M)$ to $(0,0)$. Then, for each vertex of the path (except
for the first and the last ones), one writes the Bethe equations
according to the configuration of the path around the vertex. Let us
enumerate vertices of the path by numbers from $0$ to $K+M$ so that
the point $(k,m)\in \gamma_{K,M}$ acquires the number $k+m$. Set
$$
Q_{k,m}(u)\equiv Q_{k+m}(u)=
A_{k+m}\prod_{j=1}^{J_{k+m}}(u-u_{j}^{(k+m)})\,, \quad \quad
(k,m)\in \gamma_{K,M} \,.
$$
Let us also enumerate edges of the path by numbers from $1$ to $K+M$
so that the edge joining the $(n-1)$-th and the $n$-th vertex
acquires the number $n$. (Note that in the course of undressing one
passes these edges in the inverse order.) To the $n$-th edge of the
path (oriented according to the direction of the undressing, i.e.,
either from the north to the south or from the east to the west) we
assign the sign factor according to the rule:
$$
p_n = \left \{
\begin{array}{l}
+1 \quad \mbox{if the $n$-th edge looks to the south}
\\ \\
-1 \quad \mbox{if the $n$-th edge looks to the west}
\end{array}\right.
$$
Then the system of the Bethe equations along the path can be written
as follows:
\begin{equation}\label{Bethe7}
\frac{Q_{n+1}\left ( u_{j}^{(n)}+2p_{n+1}\right ) Q_{n}\left (
u_{j}^{(n)}-2p_{n+1}\right ) Q_{n-1}\left ( u_{j}^{(n)}\right )}{
Q_{n+1}\left ( u_{j}^{(n)}\right ) Q_{n}\left (
u_{j}^{(n)}+2p_{n}\right ) Q_{n-1}\left ( u_{j}^{(n)}-2p_n\right
)}\, =\, (-1)^{\frac{1+p_n p_{n+1}}{2}},
\end{equation}
where $n$ runs from $1$ to $K+M-1$. The boundary conditions are
$Q_{0}(u)=1$, $Q_{K+M}(u)=\phi (u)$. Any chain of Bethe equations
includes $K+M-1$ equations for the roots of $K+M-1$ polynomials
$Q_{k,m}(u)$ picked along a path as in Fig.~\ref{fig:UNDRESS}. All
the other $[(K+1) (M+1)-2]-(K+M-1)=K M$ $Q$-functions inside the
$K\times M$ rectangle can be expressed through them by iterations of
the Hirota equation (\ref{QHIROTA}).

This form of the Bethe equations agrees with the general one
suggested in \cite{RW}. To see this, let us redefine the
$Q$-functions by the shift of the spectral parameter:
\begin{equation}\label{Bethe88}
Q_{k,m}(u)=\check Q_{k+m} \left ( u-k+m\right )\,,
\end{equation}
which is equivalent to
\begin{equation}\label{Bethe8}
Q_n(u)=\check Q_n \left ( u-\sum_{\alpha =1}^{n}p_{\alpha}\right
)\,.
\end{equation}
In terms of the roots
\begin{equation}\label{Bethe9}
\check u_j^{(n)}=u_j^{(n)}-\sum_{\alpha =1}^{n}p_{\alpha}\,.
\end{equation}
of the polynomial $\check Q_n (u)$ the system of Bethe equations
(\ref{Bethe7}) acquires a concise form
\begin{equation}\label{Bethe10}
\prod_{b=1}^{K+M} \frac{\check Q_b \left ( \check
u_{j}^{(a)}-K_{ab}\right )}{ \check Q_b \left ( \check
u_{j}^{(a)}+K_{ab}\right )}= (-1)^{\frac{K_{aa}}{2}}\,, \quad \quad
a=1, \ldots , K+M-1\,,
\end{equation}
where
\begin{equation}\label{Bethe11}
K_{ab}=(p_a +p_{a+1})\delta_{a,b}- p_{a+1}\delta_{a+1,b}-p_a
\delta_{a, b+1}
\end{equation}
is the Cartan matrix for the simple root system corresponding to the
chosen undressing path (see, for example, \cite{Tsuboi-3}). So it is
natural to think of the Kac-Dynkin diagram for the superalgebras as
a zigzag path on the $(k,m)$ plane, as shown in
Fig.\ref{fig:UNDRESS}.

Let us give a remark on the duality transformations. There are
$\frac{(K+M)!}{K!M!}$ different ways to choose the undressing path
$\gamma_{K,M}$ (Fig.~\ref{fig:UNDRESS}), and hence there are as many
chains of Bethe equations for $gl(K|M)$ algebra, all of them
describing the same system but for different choices of the simple
roots basis.  The transformation from one basis to another can be
(and sometimes is) called the duality transformation meaning that
the two descriptions of the model are equivalent and in a sense dual
to each other. For particular low rank superalgebras such
transformations were discussed in
\cite{Bares,Korepin,Essler:1992uc,GS,Woynarovich}, and for general
superalgebras in \cite{Tsuboi-3} (see also \cite{Beisert:2005di}).
In various solid state applications of supersymmetric integrable
models (for example, the t-J model) this transformation corresponds
to the so-called ``particle-hole" duality. It is clear that any
duality transformation can be decomposed into a chain of elementary
ones. The elementary duality transformation consists in swithing two
neighboring orthogonal edges of the path (joining at a ``fermionic"
node of the Kac-Dynkin diagram) to another pair of such edges
surrounding the same face of the lattice. It corresponds to
replacing \eq{Bethe5} at the roots $u=u^{(k,m)}_j$ by \eq{Bethe5} at
the roots $u=u^{(k+1,m-1)}_j$ or vice versa, which induces also the
subsequent change of the Bethe equations at two neighboring nodes.
On the operator level, the elementary duality transformation
consists in replacing $\hat V^{-1}_{k,m}\hat U_{k,m+1}$ by $\hat
U_{k,m}\hat V^{-1}_{k+1,m}$ in the products (\ref{BAXTER}),
according to the zero curvature condition (\ref{ZEROC}).

\subsection{Bethe-like equations for roots of $T$'s}

Actually, roots of all the polynomial $T$-functions $T_{k,m}(a,s,u)$
obey a system of algebraic equations which generalize the
Bethe equations (\ref{Bethe7}).
They can be derived along the same lines using, instead
of the $QQ$-relation (\ref{bilEqQ}) or (\ref{QHIROTA}),
the bilinear $TT$-relations
(\ref{BT1,2-1})-(\ref{BT1,2-3n}). Fixing an undressing
zigzag path $\gamma_{K,M}$, we set
$$
T_{k,m}(a,s,u)\equiv T_{k+m}(a,s,u)
=A_{k+m}(a,s)\prod_{j=1}^{J_{k+m}}(u-u_{j}^{(k+m)}(a,s))\,,
\quad (k,m)\in \gamma_{K,M}\,.
$$
At $a=s=0$ the root $u_{j}^{(n)}(0,0)$ coincides with
$u_{j}^{(n)}$ from the previous subsection.
Repeating all
the steps leading to the Bethe
equations (\ref{Bethe7}), we arrive at the following
Bethe-like equations:

\begin{equation}
\label{TBethe2}
\frac{T_{n+1}(a,s \! -\! p_{n+1},u_* \! +\! p_{n+1})
~T_{n}(a,s\! +\! p_{n+1},u_* \! -\! p_{n+1})
~T_{n-1}(a,s,u_* )}
{T_{n+1}(a,s,u_* )
~T_{n}(a,s\! -\! p_{n},u_* \! +\! p_{n}) ~T_{n-1}(a,s\! +\!
p_{n},u_* \! -\! p_{n})}\,
=\, (-1)^{\frac{1+p_n p_{n+1}}{2}}\,,
\end{equation}

\begin{equation}
\label{TBethe3}
\frac{T_{n+1}(a\! +\! p_{n+1},s,u_* \! +\! p_{n+1})
~T_{n}(a\! -\! p_{n+1},s,u_* \! -\! p_{n+1})
~T_{n-1}(a,s,u_* )}
{T_{n+1}(a,s,u_* )
~T_{n}(a\! +\! p_{n},s,u_* \! +\! p_{n})
~T_{n-1}(a\! -\! p_{n},s,u_* \! -\! p_{n})}\,
=\, (-1)^{\frac{1+p_n p_{n+1}}{2}}\,,
\end{equation}

\begin{equation}
\label{TBethe4}
\frac{T_{n+1}(a\! -\! p_{n+1},s\! +\! p_{n+1},u_* )
~T_{n}(a\! +\! p_{n+1},s\! -\! p_{n+1},u_* )
~T_{n-1}(a,s,u_* )}
{T_{n+1}(a,s,u_* )
~T_{n}(a\! -\! p_{n},s\! +\! p_{n},u_*)
~T_{n-1}(a\! +\! p_{n},s\! -\! p_{n},u_* )}\,
=\, (-1)^{\frac{1+p_n p_{n+1}}{2}}\,,
\end{equation}

\begin{equation}
\label{TBethe5}
\frac{T_{n+1}(a-p_{n+1},s+p_{n+1},u_*)
~T_{n}(a,s-p_{n+1},u_*-p_{n+1})
~T_{n-1}(a+p_{n},s,u_*+p_{n})}
{T_{n+1}(a-p_{n+1},s,u_*-p_{n+1})
~T_{n}(a,s+p_{n},u_*+p_{n})
~T_{n-1}(a+p_{n},s-p_{n},u_*)}\,
=\, (-1)^{\frac{1+p_n p_{n+1}}{2}}\,,
\end{equation}

\begin{equation}
\label{TBethe6}
\frac{T_{n+1}(a-p_{n+1},s+p_{n+1},u_*)
~T_{n}(a+p_{n+1},s,u_*-p_{n+1})
~T_{n-1}(a,s-p_{n},u_*+p_{n})}
{T_{n+1}(a,s+p_{n+1},u_*-p_{n+1})
~T_{n}(a-p_{n},s,u_*+p_{n})
~T_{n-1}(a+p_{n},s-p_{n},u_*)}\,
=\, (-1)^{\frac{1+p_n p_{n+1}}{2}}\,,
\end{equation}

\begin{equation}
\label{TBethe7}
\frac{T_{n+1}(a-p_{n+1},s,u_*-p_{n+1})
~T_{n}(a+p_{n+1},s+p_{n+1},u_*)
~T_{n-1}(a,s-p_{n},u_*+p_{n})}
{T_{n+1}(a,s+p_{n+1},u_*-p_{n+1})
~T_{n}(a-p_{n},s-p_{n},u_*)
~T_{n-1}(a+p_{n},s,u_*+p_{n})}\,
=\, (-1)^{\frac{1+p_n p_{n+1}}{2}}\,,
\end{equation}

\noindent where $u_* \equiv u_{j}^{(n)}(a,s)$. The
values of $a,~s$ and $n$ are
assumed to be such that none of the $T$-functions is
identically zero. At the boundaries $a=0$ or $s=0$ equations
(\ref{TBethe2}), (\ref{TBethe3})
coincide with the Bethe equations (\ref{Bethe7}).


\section{Algorithm for integration
of the Hirota equation}

In this section we develop a general algorithm
to solve the Hirota equation (\ref{HIROTATT})
expressing the functions $T_{k,m}(a,s,u)$ through the boundary
functions $Q_{k,m}(u)$ (\ref{BCOND2T}). We note that
it gives an operator realization of the combinatorial rules
given in \cite{Tsuboi-1}.

\subsection{Shift operators}

Our starting point is the alternative representation of the
first and second B\"acklund transformations given by
equations (\ref{alt1})-(\ref{alt4}) which we rewrite here
in a slightly different form.
Equations  (\ref{alt1}), (\ref{alt2}) read
\begin{eqnarray}
\label{LPT1eq1}
T_{k-1,m}(a,s,u)&=&\widehat H_{k^-,m}(a-1,s,u)~T_{k,m}(a,s,u),\\
T_{k,m+1}(a,s,u)&=&\widehat H_{k,m^+}(a-1,s,u)~T_{k,m}(a,s,u)\,,
\label{LPT2eq1}
\end{eqnarray}
and equations (\ref{alt3}), (\ref{alt4}) read
\begin{eqnarray}
\label{LPT1eq2}
T_{k+1,m}(a,s,u)&=&\widehat H_{k^+,m}(a,s-1,u)~T_{k,m}(a,s,u),\\
T_{k,m-1}(a,s,u)&=&\widehat H_{k,m^-}(a,s-1,u)~T_{k,m}(a,s,u)\,.
\label{LPT2eq2}
\end{eqnarray}
Here the difference operators $\widehat H_{k^{\pm},m}(a,s,u)$ and
$\widehat H_{k,m^{\pm}}(a,s,u)$ are given by
\begin{eqnarray}
\label{H1-} &&\widehat H_{k^-,m}(a,s,u):=
\frac{T_{k-1,m}(a,s,u+1)}{T_{k,m}(a,s,u+1)}
-\frac{T_{k-1,m}(a,s+1,u)}{T_{k,m}(a,s,u+1)}~e^{\partial_u-\partial_s},\\
\label{H1+} &&\widehat H_{k,m^+}(a,s,u):=
\frac{T_{k,m+1}(a,s,u+1)}{T_{k,m}(a,s,u+1)}
-\frac{T_{k,m+1}(a,s+1,u)}{T_{k,m}(a,s,u+1)}~e^{\partial_u-\partial_s}, \\
\label{H2+} &&\widehat H_{k^+,m}(a,s,u):=
\frac{T_{k+1,m}(a,s,u-1)}{T_{k,m}(a,s,u-1)}
+\frac{T_{k+1,m}(a+1,s,u)}{T_{k,m}(a,s,u-1)}~e^{-(\partial_u+\partial_a)},\\
\label{H2-} &&\widehat H_{k,m^-}(a,s,u):=
\frac{T_{k,m-1}(a,s,u-1)}{T_{k,m}(a,s,u-1)}
+\frac{T_{k,m-1}(a+1,s,u)}{T_{k,m}(a,s,u-1)}~e^{-(\partial_u+\partial_a)}.
\end{eqnarray}
As we have seen in section 3.4,
these equations are equivalent to equations (\ref{LINPRT1}) and
(\ref{LINPRT2}) which have been used to define the B\"acklund
transformations BT1  and BT2.

The operators introduced above
obey, by construction, the ``weak" zero curvature conditions
\begin{eqnarray}
\label{z-c1a} &&T_{k\mp 1,m \pm 1}=  \widehat H_{k\mp 1,m^{\pm}}
\widehat H_{k^{\mp},m}T_{k,m}
=\widehat H_{k^{\mp},m \pm 1}\widehat H_{k,m^{\pm}}T_{k,m},
\end{eqnarray}
where $\widehat H_{k^{-},m}\equiv \widehat H_{k^{-},m}(a-1,s,u)$,
$\widehat H_{k^{+},m}\equiv \widehat H_{k^{+},m}(a,s-1,u)$,
$\widehat H_{k,m^{+}}\equiv \widehat H_{k,m^{+}}(a-1,s,u)$,
$\widehat H_{k,m^{-}}\equiv \widehat H_{k,m^{-}}(a,s-1,u)$ and
$T_{k,m}\equiv T_{k,m}(a,s,u)$.
As is pointed out in section 3.4, they imply the ``strong",
operator form of these conditions:
\begin{eqnarray}
\label{z-c1}
\widehat H_{k\mp 1,m^{\pm}} ~\widehat H_{k^{\mp},m}
=\widehat H_{k^{\mp},m \pm 1} ~\widehat H_{k,m^{\pm}}\,.
\end{eqnarray}

The operators $\widehat H_{k^{\pm},m}$,
$\widehat H_{k,m^{\pm}}$
generalize the shift operators $\hat U_{k,m}$, $\hat V_{k,m}$
introduced in section 4. We hold the same name for them.
Comparing to $\hat U_{k,m}$, $\hat V_{k,m}$, they act to
functions of three variables, not just to functions of the
spectral parameter $u$, and involve non-trivial shifts
in two independent directions.
However, the shift operators at $a=0$ or $s=0$ are effectively
one-dimensional since they
do not depend on $u+s$ (or $u-a$):
\begin{eqnarray}
\label{H;a,s=0} &&\widehat H_{k^-,m}(0,s,u)=
\frac{Q_{k-1,m}(u-s+1)}{Q_{k,m}(u-s+1)}
-\frac{Q_{k-1,m}(u-s-1)}{Q_{k,m}(u-s+1)}~e^{\partial_u-\partial_s},\nn\\
&&\widehat H_{k,m^+}(0,s,u)= \frac{Q_{k,m+1}(u-s+1)}{Q_{k,m}(u-s+1)}
-\frac{Q_{k,m+1}(u-s-1)}{Q_{k,m}(u-s+1)}~e^{\partial_u-\partial_s},\nn\\
&&\widehat H_{k^+,m}(a,0,u) =
\frac{Q_{k+1,m}(u+a-1)}{Q_{k,m}(u+a-1)}
+\frac{Q_{k+1,m}(u+a+1)}{Q_{k,m}(u+a-1)}~e^{-(\partial_u+\partial_a)},\nn\\
&&\widehat H_{k,m^-}(a,0,u) =
\frac{Q_{k,m-1}(u+a-1)}{Q_{k,m}(u+a-1)}
+\frac{Q_{k,m-1}(u+a+1)}{Q_{k,m}(u+a-1)}~e^{-(\partial_u+\partial_a)}\,.
\end{eqnarray}
They are functionals of $Q_{k,m}(u)$ only. The first (last) two of them,
when restricted to the functions of $u-s$ ($u+a$), are equivalent to
(adjoint) operators $\hat U_{k,m}$ and $\hat V_{k,m}$
($\hat U^{*}_{k,m}$ and $\hat V^{*}_{k,m}$) respectively. More precisely,
\begin{eqnarray}
\label{sim-trans} \frac{1}{Q_{k,m}(u\! -\! s\! -\! 1)}~ \widehat
H_{(k+1)^-,m} (0,s,u) ~Q_{k+1,m}(u\! -\! s\! -\! 1)&
\rightarrow & \hat U_{k,m}(u\! -\! s\! -\! 1),\nn\\
 \frac{1}{Q_{k,m+1}(u\! -\! s\! -\! 1)}~
 \widehat H_{k,m^+}(0,s,u) ~Q_{k,m}(u\! -\! s\! -\! 1)&
\rightarrow & \hat V_{k,m}(u\! -\! s\! -\! 1),\nn\\
\frac{(-1)^{-a}}{Q_{k+1,m}(u\! +\! a\! +\! 1)}~ \widehat
H_{k^+,m} (a,0,u)~(-1)^a~Q_{k,m}(u\! +\! a\! +\! 1)&
\rightarrow & \hat U^{*}_{k,m}(u\! +\! a\! -\! 1),\nn\\
\frac{(-1)^{-a}}{Q_{k,m}(u\! +\! a\! +\! 1)}~
 \widehat H_{k,(m+1)^-}(a,0,u)~(-1)^a ~Q_{k,m+1}(u\! +\! a\! +\! 1)&
\rightarrow & \hat V^{*}_{k,m}(u\! +\! a\! -\! 1),
\end{eqnarray}
where it is implied that the operators in the l.h.s. act
on functions of $u-s$ ($u+a$).

A simple inspection shows that the shift operators can be
written as
\begin{eqnarray}
\label{H1-n} &&\widehat H_{k^-,m}(a,s,u)=
\frac{T_{k-1,m}(a,s,u+1)}{T_{k,m}(a,s,u+1)}~\widehat h^{(1)}_{k-1,m}(a,s+1,u),\\
\label{H1+n} &&\widehat H_{k,m^+}(a,s,u)=
\frac{T_{k,m+1}(a,s,u+1)}{T_{k,m}(a,s,u+1)}~\widehat h^{(1)}_{k,m+1}(a,s+1,u), \\
\label{H2+n} &&\widehat H_{k^+,m}(a,s,u)=
\frac{T_{k+1,m}(a,s,u-1)}{T_{k,m}(a,s,u-1)}~\widehat h^{(2)}_{k+1,m}(a+1,s,u),\\
\label{H2-n} &&\widehat H_{k,m^-}(a,s,u)=
\frac{T_{k,m-1}(a,s,u-1)}{T_{k,m}(a,s,u-1)}~\widehat
h^{(2)}_{k,m-1}(a+1,s,u)
\end{eqnarray}
where
\begin{eqnarray}
\label{h-def1} &&\widehat h^{(1)}_{k,m}(a,s,u):= ~T_{k,m}(a,s,u)
~\left (1-e^{\partial_u-\partial_s}\right )~\frac{1}{T_{k,m}(a,s,u)},\\
&&\widehat h^{(2)}_{k,m}(a,s,u):= ~T_{k,m}(a,s,u)
~\left (1+e^{-(\partial_u+\partial_a)}\right )~\frac{1}{T_{k,m}(a,s,u)}.
\label{h-def2}
\end{eqnarray} From this representation it is obvious
that they have nontrivial kernels,
$T_{k,m}(a,s,u)f^{(1)}_{k,m}(a,u-a+s)$ and $T_{k,m}(a,s,u)(-1)^a
f^{(2)}_{k,m}(s,u-a+s)$ respectively, so their common kernel
is $T_{k,m}(a,s,u)(-1)^a f^{(3)}_{k,m}(u-a+s)$, where
$f^{(i)}_{k,m}$ are arbitrary functions of their arguments.
Modulo these kernels the shift operators
$\widehat H_{k^{\pm},m}(a,s,u)$
and $\widehat H_{k,m^{\pm}}(a,s,u)$ can be inverted.
We have:
\begin{eqnarray}
\label{H1-n-inv} &&\widehat H^{-1}_{(k+1)^-,m}(a,s,u)=
T_{k,m}(a,s+1,u) ~(1 -
e^{\partial_u-\partial_s})^{-1}~\frac{1}{T_{k,m}(a,s+1,u)}
~\frac{T_{k+1,m}(a,s,u+1)}{T_{k,m}(a,s,u+1)}\nn\\
&&= T_{k,m}(a,s+1,u)
~\sum_{j=0}^{\infty}~\frac{1}{T_{k,m}(a,s-j+1,u+j)}
~\frac{T_{k+1,m}(a,s-j,u+j+1)}{T_{k,m}(a,s-j,u+j+1)}~e^{j(\partial_u-\partial_s)}
\end{eqnarray}
and
\begin{eqnarray}
\label{H2-n-inv} &&\widehat H^{-1}_{k,(m+1)^-}(a,s,u)=
T_{k,m}(a+1,s,u) ~(1 +
e^{-(\partial_u+\partial_a)})^{-1}~\frac{1}{T_{k,m}(a+1,s,u)}
\frac{T_{k,m+1}(a,s,u-1)}{T_{k,m}(a,s,u-1)}\nn\\
&&= T_{k,m}(a+1,s,u)~\sum_{j=0}^{\infty}
~\frac{(-1)^j}{T_{k,m}(a-j+1,s,u-j)}
\frac{T_{k,m+1}
(a-j,s,u-j-1)}{T_{k,m}(a-j,s,u-j-1)}~e^{-j(\partial_u+\partial_a)}.
\end{eqnarray}
Equations (\ref{LPT1eq1}) and
(\ref{LPT2eq2}) rewritten in the following equivalent form
\begin{eqnarray}
\label{LPT1eq1-inv} T_{k+1,m}(a,s,u)=\widehat
H^{-1}_{(k+1)^-,m}(a-1,s,u)~T_{k,m}(a,s,u)\,,
\end{eqnarray}
\begin{eqnarray}
T_{k,m+1}(a,s,u)=\widehat
H^{-1}_{k,(m+1)^-}(a,s-1,u)~T_{k,m}(a,s,u)
\label{LPT2eq2-inv}
\end{eqnarray}
will be useful for integration of the Hirota equation.

As it is clear from the explicit expressions (\ref{H1-n-inv}),
(\ref{H2-n-inv}), the inverse shift operators acting on the function
$T_{k,m}(a,s,u)$ in eqs. (\ref{LPT1eq1-inv}), (\ref{LPT2eq2-inv})
are represented by sums of fractions whose numerators and
denominators are products of $T$'s containing both positive and
negative values of the arguments $a$ and/or $s$. The same is also
true for products of the shift operators $\widehat
H_{k^{\pm},m}(a,s,u)$ and $\widehat H_{k,m^{\pm}}(a,s,u)$ since the
operators $e^{-\partial_s}$ and $e^{-\partial_a}$  lower values of
$a$ and $s$. The functions $T_{k,m}(a,s,u)$ are equal to zero at
negative integer values of $s$ or $a$ according to the boundary
conditions (\ref{BCOND1T}). Therefore, the numerators and
denominators of some ratios could simultaneously become zero at some
values of $s$ or $a$. We have to define their values in a way
consistent with the hierarchy of Hirota equations. One way to do
that is to analytically continue the $T$-functions to negative
values $a= - n + \epsilon$ and/or $s = -l + \delta$, where $n, l \in
{\Bbb N}$ and $\epsilon, \delta \in {\Bbb R}$ tend to $0$. One can
straightforwardly verify that the behavior
\begin{eqnarray}
\label{LimitT}
&& T_{k,m}(- n + \epsilon, s,u)= O(\epsilon^n),\nn\\
&& T_{k,m}(a,- l + \delta, u) = O(\delta^l), \nn\\
&& T_{k,m}(- n + \epsilon, - l + \delta,u) = O(\epsilon^n \delta^l)
\end{eqnarray}
is consistent with the Hirota equation and equations
(\ref{LINPRT1}), (\ref{LINPRT2}). We then notice that both
operators $\widehat H^{-1}_{k^{-},m}(a,s,u)$ and $\widehat
H^{-1}_{k,m^{-}}(a,s,u)$, as well as their products, are nonsingular
when acting on the functions $T_{k,m}(a,s,u)$. Actually, the
behavior (\ref{LimitT}) is equivalent to the following prescription
to define the series of the form (\ref{H1-n-inv}) or
(\ref{H2-n-inv}) acting on $T_{k,m}(a,s,u)$: fractions containing
$T$'s at negative values of $s$ and/or $a$ do not give any
contribution to the sum. Thus, for finite positive values of $s$ and
$a$, the inverse shift operators $\widehat H^{-1}_{k^{-},m}(a,s,u)$
and $\widehat H^{-1}_{k,m^{-}}(a,s,u)$ acting on $T_{k,m}(a,s,u)$
contain a {\it finite} sum of nonzero terms.  We use this
prescription in what follows.

Now, we are ready to present a general algorithm of
reconstructing the functions $T_{k,m}(a,s,u)$ in terms of $Q_{k,m}(u)$
based on equations (\ref{LPT2eq1}),
(\ref{LPT1eq2}) and (\ref{LPT1eq1-inv}), (\ref{LPT2eq2-inv}).


\subsection{Integration of the Hirota equation}

Equations (\ref{LPT1eq2}) and (\ref{LPT1eq1-inv}) ((\ref{LPT2eq1})
and (\ref{LPT2eq2-inv})) are recurrence relations allowing one to
express the functions $T_{k,m}(a,s,u)$ in terms of the same
functions but with smaller values of $k,m$ and/or $a,s$:
\begin{eqnarray}
\label{p-,m} T_{k,m}(a,s,u)
&=&\Big(\prod_{p=n+1}^{k}\widehat H^{-1}_{p^-,m}(a-1,s,u)\Big)~T_{n,m}(a,s,u),\\
\label{k,q+} T_{k,m}(a,s,u)
&=& \Big(\prod_{q=l}^{m-1} \widehat H_{k,q^+}(a-1,s,u)\Big)~T_{k,l}(a,s,u),\\
\label{p+,m} T_{k,m}(a,s,u)
&=&\Big(\prod_{p=n}^{k-1}\widehat H_{p^+,m}(a,s-1,u)\Big)~T_{n,m}(a,s,u),\\
\label{k,q-} T_{k,m}(a,s,u)&=& \Big(\prod_{q=l+1}^{m} \widehat
H^{-1}_{k,q^-}(a,s-1,u)\Big)~T_{k,l}(a,s,u).
\end{eqnarray}
Here $k,l,m,n$ are positive integer numbers such that
$$0\leq l \leq m-1, \quad 0\leq n \leq k-1,
\quad 0\leq k \leq K, \quad 0\leq m \leq M,$$ and $\prod_{n=l}^{m}
\widehat O_n \equiv \widehat O_{m}\ldots  \widehat O_l$ by
definition. Substituting $n=a$, $l = 0$ into eqs. (\ref{p-,m}),
(\ref{k,q+}), and $n=0$, $l = s$ into eqs. (\ref{p+,m}),
(\ref{k,q-}), we obtain:
\begin{eqnarray}
\label{K0-p-,0} T_{K,0}(a,s,u)
&=&\Big(\prod_{p=a+1}^{K}\widehat H^{-1}_{p^-,0}(a-1,s,u)\Big)~T_{a,0}(a,s,u), \\
\label{KM-K,q+} T_{K,M}(a,s,u) &=& \Big(\prod_{q=0}^{M-1} \widehat
H_{K,q^+}(a-1,s,u)\Big)~ T_{K,0}(a,s,u).
\end{eqnarray}
(where $0 < a < K$, $0 \leq s < \infty$) and
\begin{eqnarray}
\label{KM-p+,M} T_{K,M}(a,s,u)
&=&\Big(\prod_{p=0}^{K-1}\widehat H_{p^+,M}(a,s-1,u)\Big)~T_{0,M}(a,s,u),\\
\label{0M-0,q-} T_{0,M}(a,s,u)&=& \Big(\prod_{q=s+1}^{M} \widehat
H^{-1}_{0,q^-}(a,s-1,u)\Big)~T_{0,s}(a,s,u)
\end{eqnarray}
(where $0 \leq  a < \infty$, $0 < s < M$). This representation
allows us to write down the formulas for the whole set of the
nonzero $T$-functions which do not belong to the boundaries:
\begin{eqnarray}
\label{T-KM-a} T_{K,M}(a,s,u) &=& \Big(\prod_{q=0}^{M-1} \widehat
H_{K,q^+}(a-1,s,u)\Big)~\Big(\prod_{p=a+1}^{K}\widehat
H^{-1}_{p^-,0}(a-1,s,u)\Big)~T_{a,0}(a,s,u)
\end{eqnarray}
(where $0 < a < K$, $0 \leq s < \infty$) and
\begin{eqnarray}
\label{T-KM-s} T_{K,M}(a,s,u) &=&\Big(\prod_{p=0}^{K-1}\widehat
H_{p^+,M}(a,s-1,u)\Big) ~\Big(\prod_{q=s+1}^{M} \widehat
H^{-1}_{0,q^-}(a,s-1,u)\Big)~T_{0,s}(a,s,u)
\end{eqnarray}
(where $0 \leq  a < \infty$, $0 < s < M$). Note that the functions
$T_{a,0}(a,s,u)$  and $T_{0,s}(a,s,u)$ entering these equations are
boundary functions:
\begin{eqnarray}
\label{Ta0} T_{a,0}(a,s,u)= Q_{a,0}(u+a+s) \quad (0 \leq s <
\infty), \quad T_{a,0}(a,s,u)=0 \quad (-\infty \leq s < 0)
\end{eqnarray}
and
\begin{eqnarray}
\label{T0s} T_{0,s}(a,s,u)= (-1)^{as}Q_{0,s}(u-a-s) \quad (0 \leq a <
\infty), \quad T_{0,s}(a,s,u)=0 \quad (-\infty \leq a < 0),
\end{eqnarray}
according to the boundary conditions (\ref{BCOND2T}).

Let us put $a=1$ in eq.~(\ref{T-KM-a}),
\begin{eqnarray}
\label{T1s} T_{K,M}(1,s,u) = \Big(\prod_{q=0}^{M-1} \widehat
H_{K,q^+}(0,s,u)\Big)~ \Big(\prod_{p=2}^{K}\widehat
H^{-1}_{p^-,0}(0,s,u)\Big)~T_{1,0}(1,s,u) \quad (0 \leq s < \infty,
\quad 1 < K),
\end{eqnarray}
and $s=1$ in eq.~(\ref{T-KM-s}),
\begin{eqnarray}
\label{Ta1} T_{K,M}(a,1,u)=\Big(\prod_{p=0}^{K-1}\widehat
H_{p^+,M}(a,0,u)\Big)~ \Big(\prod_{q=2}^{M} \widehat
H^{-1}_{0,q^-}(a,0,u)\Big)~T_{0,1}(a,1,u) \quad (0 \leq a < \infty,
\quad 1 < M).
\end{eqnarray}
Since the shift operators entering these equations are functionals
of the boundary functions $Q_{k,m}(u)$ only, we immediately obtain
explicit expressions for $T_{k,m}(1,s,u)$ and $T_{k,m}(a,1,u)$ in
terms of the boundary functions $Q_{k,m}(u)$.

Let us
rewrite the solutions (\ref{T1s}) and (\ref{Ta1}) in a slightly different
form:
\begin{eqnarray}
\label{T1s-equiv-1} &&T_{K,M}(1,s,u) = \Big(\prod_{q=0}^{M-1}
\widehat H_{K,q^+}(0,s,u)\Big)~ \Big(\prod_{p=1}^{K}\widehat
H^{-1}_{p^-,0}(0,s,u)\Big)~T_{0,0}(1,s,u),\\
\label{Ta1-equiv-1} &&T_{K,M}(a,1,u)=\Big(\prod_{p=0}^{K-1}\widehat
H_{p^+,M}(a,0,u)\Big)~ \Big(\prod_{q=1}^{M} \widehat
H^{-1}_{0,q^-}(a,0,u)\Big)~T_{0,0}(a,1,u).
\end{eqnarray}
Taking into account the explicit form of $T_{0,0}$,
\begin{eqnarray}
\label{T00} &&T_{0,0}(a,s,u)=1 \quad \mbox{at} \quad (i) \quad a=0
\quad \mbox{or} \quad (ii) \quad s=0 \quad \mbox{and} \quad a > 0
\nn\\
&&T_{0,0}(a,s,u)=0 \quad \mbox{at} \quad (i) \quad a\neq 0 \quad
\mbox{and} \quad s\neq 0 \quad \mbox{or} \quad (ii) \quad s=0 \quad
\mbox{and} \quad a < 0,
\end{eqnarray}
it is easy to see that these solutions can be represented
as the generating series
\begin{eqnarray}
\label{T1s-equiv-2} \sum_{n=0}^{\infty}
T_{K,M}(1, n ,u-s+n) ~e^{n
(\partial_u-\partial_s)} &=&\Big(\prod_{q=0}^{M-1} \widehat
H_{K,q^+}(0,s,u)\Big)~
\Big(\prod_{p=1}^{K}\widehat H^{-1}_{p^-,0}(0,s,u)\Big),\\
\label{Ta1-equiv-2} \sum_{n=0}^{\infty}T_{K,M}
(n,1,u+a-n) ~e^{-n (\partial_u+\partial_a)}
&=&\Big(\prod_{p=0}^{K-1}\widehat H_{p^+,M}(a,0,u)\Big)~
\Big(\prod_{q=1}^{M} \widehat H^{-1}_{0,q^-}(a,0,u)\Big).
\end{eqnarray}
Indeed, it is clear that the operator in the r.h.s. of
(\ref{T1s-equiv-2}) is expended in the powers of $e^{\p_u - \p_s}$
as is written in the l.h.s., with some coefficients. To fix them,
one applies the both sides to the function $T_{0,0}(1,s,u)$ and
takes into account that $e^{n(\p_u -\p_s)}T_{0,0}(1,s,u)=
T_{0,0}(1,s-n,u+n)=0$ unless $n=s$. The same
argument works for the second equality.
Using (\ref{sim-trans}), one can see that the operator relations
(\ref{T1s-equiv-2}), (\ref{Ta1-equiv-2}) are identical to
(\ref{BAXTER}).

Once the functions $T_{K,M}(1,s,u)$ and $T_{K,M}(a,1,u)$ are
constructed, the other $T$-functions can be expressed in terms of
$Q_{k,m}(u)$ by either iterating eq.~(\ref{T-KM-a}) with respect to
$a$ or eq.~(\ref{T-KM-s}) with respect to $s$. Therefore, setting
successively $a=1,2, \ldots $ ($s=1,2, \ldots$) in
eq.~(\ref{T-KM-a}) (eq.~(\ref{T-KM-s}))) starting with $a=1$
($s=1$), one can step by step express $T_{K,M}(a,s,u)$ with
different $a$ and $s$ in terms of $Q_{k,m}(u)$. Equations
(\ref{T-KM-a}) and (\ref{T-KM-s}) solve the problem of the
integration of the Hirota equation for the case of rectangular
paths. Using zero curvature conditions (\ref{z-c1}), one can
easily generalize these equations to the case of an arbitrary zigzag path,
along the lines of section 4.2.

\section{Higher representations in the quantum space}

In the previous sections, it is implied that zeros of the polynomial
function $\phi (u)$ are in general position. This corresponds to an
inhomogeneous spin chain in the vector representation of $gl(K|M)$
at each site. In principle, this includes all other cases such as
spins in higher representations in the quantum space. Indeed, the
higher representations can be constructed by fusing elementary ones
according to the fusion procedure outlined in Section 2. There, we
have considered fusion in the auxiliary space but for the quantum
space the construction is basically the same. To get a higher
representation at a site of the spin chain, one should combine
several sites of the chain carrying the vector representations, with
the corresponding $\theta$'s being chosen in a specific
``string-like" way, and then project onto the higher representation.
Before the projection, the spin chain looks exactly like the ones
dealt with in the previous sections. However, zeros of the function
$\phi (u)$ are no longer in general position.

The string-type boundary values impose certain requirements on the
location of zeros of the polynomials $T(a,s,u)$. Indeed, the Hirota
equation implies that some of these polynomials must contain similar
``string-like" factors. From the Hirota equation point of view,
the projection
onto a higher representation means selecting a class of polynomial
solutions divisible by factors of this type. In fact, given the
boundary values, different schemes of extracting such factors are
possible. They correspond to different types of fusion in the
quantum space.

\subsection{Symmetric representations in the quantum space}

To be more specific, consider the simplest case of symmetric tensor
representations (one-row Young diagrams). Fix an integer $\ell \geq
1$ and consider a combined site $i$ consisting of $\ell$ sites
(labeled by the double index as $(i,1), \ldots , (i, \ell )$)
carrying the vector representation. According to the fusion
procedure, the corresponding parameters $\theta_{i,r}$ form a
``string": $\theta_{i,r}=\theta_i - 2(r-1),\quad r=1,\dots,\ell$.
Therefore, the boundary values of the $T$-functions are given by
$T(0,s,u)=\phi_{\ell}^{+}(u-s)$, $T(a,0,u)=\phi_{\ell}^{+}(u+a)$,
where
\begin{equation}\label{high1}
\begin{array}{lll}
\phi_{\ell}^{+} (u)& = & \phi (u)\phi (u+2)\ldots \phi (u+2(\ell
-1))
\\ &&\\ &=& \displaystyle{2^{\ell N}
\prod_{i=1}^{N}\frac{\Gamma
 \left (\frac{u-\theta_i}{2}+\ell \right )}{\Gamma
\left (\frac{u-\theta_i}{2}\right )} } \,, \quad \ell =0,1,2, \ldots
\,.
\end{array}
\end{equation}
Here $\phi (u)=\prod_{i=1}^{N} (u-\theta_i )$, as before. The
representation through the $\Gamma$-function is useful for the
analytic continuation in $\ell$.

A thorough inspection shows that the Hirota equation is consistent
with extracting the following polynomial factor\footnote{We note
that this factor here and in (\ref{high2a}) below can be obtained
directly from the fusion procedure in the quantum space as the
product of ``trivial zeros" of the fused $R$-matrices.} from
$T(a,s,u)$ for $s=0,1, \ldots , \ell$ and $a\geq 1$:
\begin{equation}\label{high2}
\begin{array}{lll}
T(a,s,u)&=& \phi_{\ell -s}^{+}(u+s+a)\tilde T(a,s,u)
\\ &&\\
&=&\displaystyle{ 2^{(\ell -s)N}\left [ \prod_{i=1}^{N} \frac{\Gamma
 \left (\frac{u-s+a-\theta_i}{2}+\ell \right )}{\Gamma
\left (\frac{u+s+a-\theta_i}{2}\right )}\right ] \tilde T(a,s,u)\,.}
\end{array}
\end{equation}
Here $\tilde T(a,s,u)$ is a polynomial of degree $sN$ (if $s=0,1,
\ldots , \ell$). We extend the definition of $\tilde T(a,s,u)$ to
higher values of $s$ by setting $\tilde T (a,s,u)= T(a,s,u)$ ($s\geq
\ell$, $a\geq 1$). Note that the factor in the right
hand side of (\ref{high2})
is a product of functions depending separately on $u+s+a$ and
$u-s+a$. Therefore, if this relation between $T$ and $\tilde T$ was
valid for all values of $s$, then the $\tilde T$'s would obey the
same Hirota equation in the whole $(a,s,u)$ space (see
(\ref{gaugef})). However, since the definition of $\tilde T$ is
changed when $s\geq \ell$, the Hirota equation breaks down in the
plane $s=\ell$. It is easy to see that it gets modified as follows:
\begin{equation}\label{high3}
\begin{array}{c}
\tilde T(a,s, u\! + \! 1) \tilde T(a,s, u\! - \! 1) \, - \, \left [
\phi (u\! + \! s \! + \! a \! - \! 1)\right ]^{\delta_{s,\ell}}
\tilde T(a,s\! + \! 1,u) \tilde T(a,s\! - \! 1,u)
\\  \\
=\, \left [ \phi_{s\! +\! (\ell \! -\! s)
\theta (s\geq \ell )}^{+}(u\! -\! s)
\right ]^{\delta_{a,1}}
\tilde T(a\! + \! 1,s,u) \tilde T(a\! - \! 1,s,u)\,.
\end{array}
\end{equation}
The function $\theta (s\geq \ell )$ is defined as
$\theta (s\geq \ell )=1$ if $s\geq \ell$ and $0$ otherwise, so
the pre-factor in the right hand side equals $\phi_s^+ (u-s)$
if $0\leq s \leq \ell$ and $\phi_{\ell}^+ (u-s)$ if $s\geq \ell$
(at $a=1$).
The boundary conditions are $\tilde T(a,0,u)=\tilde T(0,s,u)=1$
($a,s \geq 0$). If the point $(a,s)$ belongs to the interior
boundary, then the function $\tilde T$ defined by eq.~(\ref{high2})
(and by eq.~(\ref{high2a}) below)
may contain some additional zeros of a similar string-like type.

 \subsection{Antisymmetric representations in the quantum space}

In the case of the antisymmetric fusion (one-column Young diagrams)
the parameters $\theta_{i,r}$ also form a ``string":
$\theta_{i,r}=\theta_i + 2(r-1),\quad  r=1,\dots,\ell$. Comparing to
the symmetric fusion, this string ``looks" to the opposite
direction, i.e., the sequence of $\theta$'s increases rather than
decreases. The two types of strings are actually equivalent since
they are obtained one from the other by an overall shift of
$\theta_i$'s. Our convention here is chosen to be consistent with
the general case outlined below.

The boundary values of the $T$-functions are given by
$T(0,s,u)=\phi_{\ell}^{-}(u-s)$, $T(a,0,u)=\phi_{\ell}^{-}(u+a)$,
where
\begin{equation}\label{high1a}
\begin{array}{lll}
\phi_{\ell}^{-} (u)& = & \phi (u)\phi (u-2)\ldots \phi (u-2(\ell
-1))
\\ &&\\ &=& \displaystyle{2^{\ell N}
\prod_{i=1}^{N}\frac{\Gamma
 \left (\frac{u-\theta_i}{2}+1 \right )}{\Gamma
\left (\frac{u-\theta_i}{2}-\ell +1\right )} } \,, \quad \ell
=0,1,2, \ldots \,.
\end{array}
\end{equation}
A thorough inspection shows that the Hirota equation is consistent
with extracting the following polynomial factor from $T(a,s,u)$ for
$a=0,1, \ldots , \ell$ and $s\geq 1$:
\begin{equation}\label{high2a}
\begin{array}{lll}
T(a,s,u)&=& \phi_{\ell -a}^-(u-s-a)\tilde T(a,s,u)
\\ &&\\
&=&\displaystyle{ 2^{(\ell -a)N}\left [ \prod_{i=1}^{N} \frac{\Gamma
 \left (\frac{u-s-a-\theta_i}{2}+1 \right )}{\Gamma
\left (\frac{u-s+a-\theta_i}{2}-\ell +1\right )}\right ] \tilde
T(a,s,u)\,. }
\end{array}
\end{equation}
Here $\tilde T(a,s,u)$ is a polynomial of degree $aN$ (if $a=0,1,
\ldots , \ell$). We extend the definition of $\tilde T(a,s,u)$ to
higher values of $a$ by setting $\tilde T (a,s,u)= T(a,s,u)$ ($a\geq
\ell$, $s\geq 1$).
The modified Hirota equation for $\tilde T$ reads as follows:
\begin{equation}\label{high3a}
\begin{array}{c}
\tilde T(a,s, u\! + \! 1) \tilde T(a,s, u\! - \! 1) \, - \,
\left [ \phi_{a\! +\! (\ell \! -\! a)
\theta (a\geq \ell )}^{-}(u\! +\! a)
\right ]^{\delta_{s,1}}
\tilde T(a,s\! + \! 1,u)
 \tilde T(a,s\! - \! 1,u)
 \\  \\
 =\,
 \left [\phi (u\! - \! s \! - \! a \! + \! 1)\right ]^{\delta_{a,\ell}}
 \tilde T(a\! + \! 1,s,u) \tilde T(a\! - \! 1,s,u)\,.
\end{array}
\end{equation}
The pre-factor in the second term in the
left hand side equals $\phi_a^- (u+a)$
if $0\leq a \leq \ell$ and $\phi_{\ell}^- (u+a)$ if $a\geq \ell$
(at $s=1$).
The boundary conditions are $\tilde T(a,0,u)=\tilde T(0,s,u)=1$
($a,s \geq 0$).

\subsection{The general case: remarks and conjectures}

Let us consider the case of a general covariant representation $\mu$
at each site of the chain. We construct such a site $i$ by fusing
$|\mu |$ ``elementary" sites with $\theta$-parameters $\theta_i +
2(p-q)$, where $(p,q)\in \mu$. We remind the reader that the integer
coordinates $(p,q)\in {\Bbb Z}^2$ on a Young diagram are such that
the row index $p$ increases as one goes from top to bottom and the
column index $q$ increases as one goes from left to right, and the
top left box of $\mu$ has the coordinates $(1,1)$.

Given a diagram $\mu$, we define the polynomial function
\begin{equation}\label{gen1}
\phi_{\mu}(u)=\prod_{i=1}^{N}\prod_{(p,q)\in \mu} (u-2(p-q)-\theta_i
) =\prod_{(p,q)\in \mu}\phi (u-2(p-q))\,,
\end{equation}
then the boundary values of the $T$-functions are:
\begin{equation}\label{gen3}
T(0,s,u)=\phi_{\mu}(u-s)\,, \quad \quad T(a,0,u)=\phi_{\mu}(u+a)\,.
\end{equation}

\begin{figure}[t]
    \centering
        \includegraphics[angle=-90,scale=0.5]{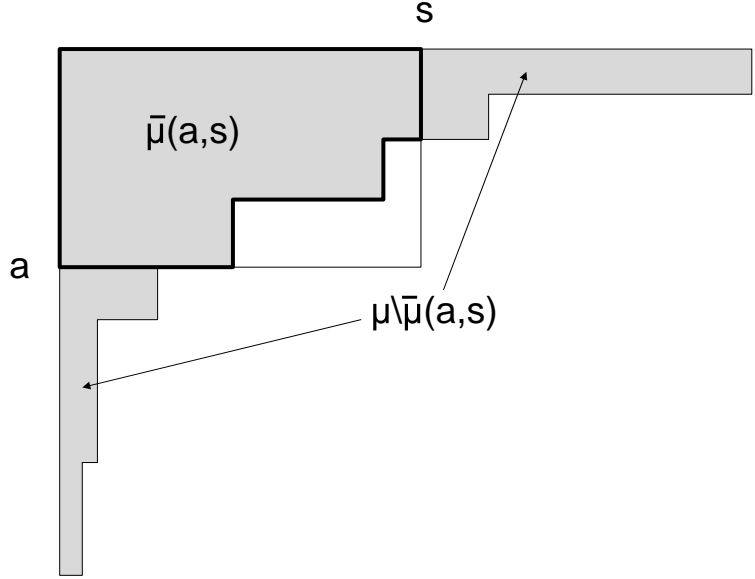}
    \caption{\it The intersection of Young diagrams $\mu$
and $(s^a)$.  }
    \label{fig:young}
\end{figure}

Let $\bar \mu (a,s)$ be the Young diagram obtained as the
intersection of $\mu$ and the rectangular diagram $(s^a)$
(Fig.~\ref{fig:young}):
$$
\bar \mu (a,s)\equiv \mu \cap (s^a)\,.
$$
For brevity, we will sometimes denote the cut diagram $\bar \mu
(a,s)$ simply $\bar \mu$ dropping the dependence on $a,s$. Let us
introduce the polynomial function
\begin{equation}\label{gen2}
\phi_{\mu \setminus \bar \mu (a,s)}(u)=
\frac{\phi_{\mu}(u)}{\phi_{\bar \mu (a,s)}(u)} =\prod_{(p,q)\in \mu
\setminus \bar \mu (a,s)}\phi (u-2(p-q))\,,
\end{equation}
where the product goes over boxes of the skew diagram $\mu \setminus
\bar \mu (a,s)$. If $\mu$ is contained in the rectangle $(s^a)$,
then we set $\phi_{\mu \setminus \bar \mu (a,s)}(u)=1$.

Now we are ready to present our first conjecture. We expect that the
projection onto the representation $\mu$ in the quantum space means,
for the Hirota equation, that we consider the solutions such that
the polynomial $T(a,s,u)$ is divisible by the polynomial $\phi_{\mu
\setminus \bar \mu (a,s)}$:
\begin{equation}\label{gen4}
T(a,s,u)=\phi_{\mu \setminus \bar \mu (a,s)}(u\! -\! s\! +\! a) \,
\tilde T(a,s,u)\,.
\end{equation}
For one-row or one-column diagrams this formula yields equations
(\ref{high2}) and (\ref{high2a}). Presumably, this conjecture can be
proved by means of the technique developed in \cite{NazTar}.
If the point $(a,s)$ belongs to the interior boundary, then the
function $\tilde T$ defined by eq.~(\ref{gen4})
may contain some additional zeros of a
similar string-like type.

We note that the functions $\tilde T(a,s,u)$ defined by (\ref{gen4})
obey the modified Hirota equation:
\begin{equation}\label{gen5}
\begin{array}{ll}
&\tilde T(a,s, u+1)\, \tilde T(a,s, u-1)
\\&\\
=& \displaystyle{\left [\prod_{j=\bar \mu'_{s+1}}^{\bar \mu'_{s}-1}
\phi (u\! +\! s\! +\! a \! - \! 2j \! - \! 1)\right ]\, \tilde
T(a,s\! + \! 1, u)\tilde T(a,s\! - \! 1, u)}
\\&\\
+& \displaystyle{\left [\prod_{j=\bar \mu_{a+1}}^{\bar \mu_{a}-1}
\phi (u\! -\! s\! -\! a \! + \! 2j \! + \! 1)\, \right ]\tilde
T(a\! + \! 1,s, u)\tilde T(a\! - \! 1,s, u)}\,.
\end{array}
\end{equation}
Here the products like $\prod_{j=n}^{n-1}$ are understood to be $1$
and
$$
\bar \mu'_{s} =\mbox{min}\, ( \mu'_{s}, a)\,, \quad \quad \bar
\mu_{a} =\mbox{min}\, ( \mu_{a}, s)
$$
are respectively the lengths of the $s$-th column and the $a$-th row
of the diagram $\bar \mu =\bar \mu (a,s)$.
The boundary conditions are $\tilde T(a,0,u)=\tilde T(0,s,u)=1$
($a,s \geq 0$).
Note that the pre-factors
in the modified Hirota equation are equal to $1$ if the rectangle
$(s^a)$ is contained in the diagram $\mu$. In the opposite case,
when the diagram $\mu$ is contained in the rectangle $(s^a)$, the
functions $\tilde T (a,s,u)$ coincide with $ T (a,s,u)$ and the
pre-factors are again equal to $1$. Given equation (\ref{gen4}), the
derivation of the modified Hirota equation for $\tilde T$'s is
straightforward. We present here two simple identities which appear
to be useful:
\begin{eqnarray}\label{PHIRA}
\frac{\phi_{\bar \mu (a,s+1)}(u)}{\phi_{\bar \mu (a,s)}(u)}&=& \phi
(u\! +\! 2s)\, \phi (u\! +\! 2s \! -\! 2)\, \ldots \, \phi \left
(u\! +\! 2s \! -\! 2 (\bar
\mu'_{s+1} \! -\! 1)\right )\,,\nn\\
\frac{\phi_{\bar \mu (a+1,s)}(u)}{\phi_{\bar \mu (a,s)}(u)}&=& \phi
(u\! -\! 2a)\, \phi (u\! -\! 2a \! +\! 2)\, \ldots \, \phi \left
(u\! -\! 2a \! +\! 2 (\bar \mu_{a+1} \! -\! 1)\right )\,.
\end{eqnarray}

The next challenge is to find out how the polynomial factor
extracted from the $T$-functions behaves under the B\"acklund
transformations. Here is our second conjecture. At each step $(k,m)$
of the chain of the transformations BT1 and BT2 the same relation
(\ref{gen4}) holds,
\begin{equation}\label{gen6}
T_{k,m}(a,s,u)=\phi_{\mu_{k,m} \setminus \bar \mu_{k,m} (a,s)}(u\!
-\! s\! +\! a) \, \tilde T_{k,m}(a,s,u)\,,
\end{equation}
where the diagram $\mu_{k,m}$ is obtained from $\mu =\mu_{K,M}$ by
cutting off $K-k$ upper rows and $M-m$ left columns. If $K-k$
exceeds the number of rows in the diagram $\mu$, or $M-m$ exceeds
the number of columns in $\mu$, then we set $\mu_{k,m}=\emptyset$.
In other words, the transformation BT1 cuts off the upper row while
BT2 cuts off the left column. The coordinates $(p,q)$ on the
diagrams $\mu_{k,m}$ are such that the top left box of any diagram
has coordinates $(1,1)$.

At last, we would like to remark that instead of the function
$\phi_{\mu}(u)$ one could use the function
\begin{equation}\label{gen1a}
\bar \phi_{\mu}(u)= \prod_{(p,q)\in \mu}\phi (u+2(p-q))
\end{equation}
and arrive to similar formulas. This can probably be explained by
invoking the representation theory of (super)Yangians. It
suggests\footnote{We are grateful to V.Tarasov for a discussion on
this point.} that the function $\bar \phi_{\mu}(u)$ corresponds to
the representation associated with the usual Young diagram $\mu$
while the function $\phi_{\mu}(u)$ comes from the representation
associated with the ``reversed" diagram $\mu^{\dag}$ regarded as a
skew diagram, see the first reference in \cite{NazTar}. (The
reversed diagram is obtained from $\mu$ by inversion with respect to
the left top corner.) This point needs further clarification.

\subsection{Bethe equations with non-trivial vacuum parts}

It is known that when ``spins" in the quantum space belong to a
higher representation of the symmetry algebra, Bethe equations
(\ref{Bethe1})-(\ref{Bethe6}) acquire non-trivial right hand sides
(sometimes called vacuum parts). We are going to show that they
actually follow from the equations with trivial vacuum parts if one
partially fixes the roots of the polynomials $Q_{k,m}(u)$ in a
special way. Specifically, we set
\begin{equation}\label{gen7}
Q_{k,m}(u)=\phi_{\mu_{k,m}} (u)\, \tilde Q_{k,m}(u)\,,
\end{equation}
which can be regarded as an ansatz suggested by the fusion
procedure. Note that it is the particular case of eq. (\ref{gen6})
at $a=0$ or $s=0$. Accepting this, we are going to substitute it
into the $QQ$-relation to get an equation for $\tilde Q$'s. This
allows us to derive the system of Bethe equations for the roots of
$\tilde Q$'s. The derivation itself does not depend on the validity
of the conjectures given above.

\begin{figure}[t]
    \centering
        \includegraphics[angle=-90,scale=0.5]{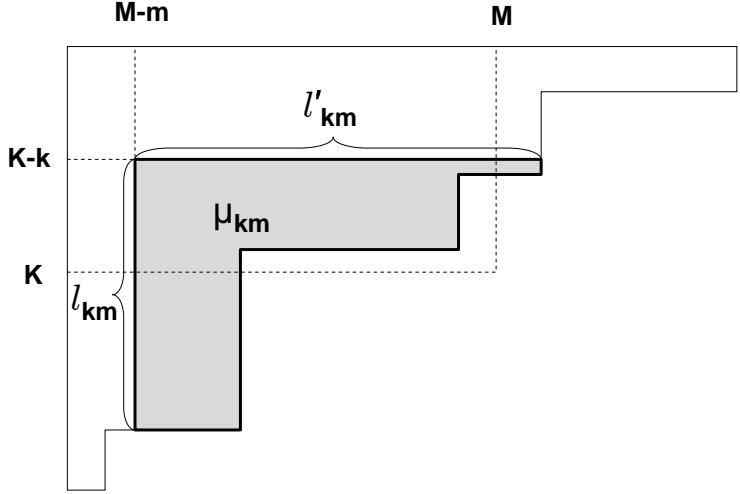}
    \caption{\it Reduction of the Young diagram $\mu$ to
 $\mu_{k,m}$ under $K-k$ applications of BT1 and $M-m$ applications of BT2.
  The full diagram $\mu$ is depicted by thin lines (including the axis),
  the reduced one,
  $\mu_{k,m}$, by the thick lines. }
    \label{fig:MUREDUC}
\end{figure}

To proceed, we need some more notation. Given a Young diagram $\mu$,
let $l(\mu )=\mu'_{1}$ be the number of its rows or, equivalently,
the length of the first row of the transposed diagram $\mu'$. The
short hand notation
\begin{equation}\label{gen8}
l_{k,m}=l(\mu_{k,m})\,, \quad \quad l_{k,m}'=l(\mu_{k,m}')
\end{equation}
is convenient. In other words, $l_{k,m}\times l'_{k,m}$ is the
minimal rectangle (of height $l_{k,m}$ and length $l_{k,m}'$)
containing the diagram $\mu_{k,m}$. It is obvious that if
$\mu_{k,m}\neq \emptyset$, then (see Fig.\ref{fig:MUREDUC})
\begin{equation}\label{gen9}
\begin{array}{l}
l_{k,m}=(\mu'_{k,m})_1 =\mu_{M-m+1}' -K +k
\\ \\
l_{k,m}'=(\mu_{k,m})_1 =\mu_{K-k+1}' -M +m\,.
\end{array}
\end{equation}

\begin{figure}[t]
    \centering
        \includegraphics[angle=-90,scale=0.5]{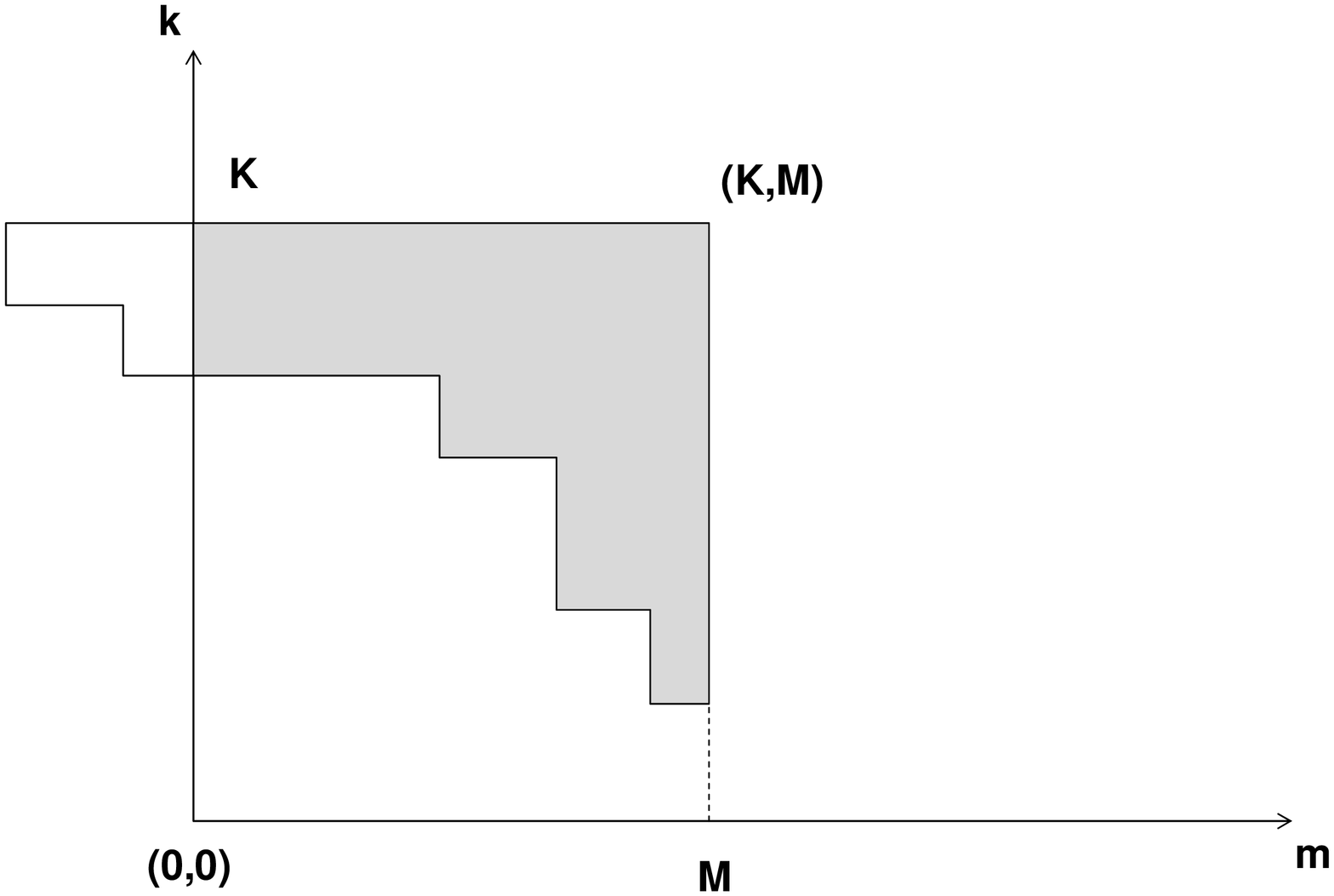}
    \caption{\it The domain (shadowed region), where the Hirota
equation for $\tilde Q$'s is modified in the rectangle $0\leq k\leq
K$, $0\leq m\leq M$. The diagram $\mu$ is reflected with respect to
its vertical boundary. }
    \label{fig:domainmu}
\end{figure}

Now we are ready to substitute (\ref{gen7}) into the Hirota equation
for $Q$'s (\ref{QHIROTA}). We have:
$$
\begin{array}{lll}
\phi_{\mu_{k,m}}(u)&=& \phi_{\mu_{k-1, m-1}}(u)\,
\phi_{l'_{k,m}}^{+}(u)\, \phi_{l_{k,m}}^{-}(u)\, \phi ^{-1} (u)
\\&&\\
&=&\phi_{\mu_{k-1, m}}(u-2) \, \phi_{l'_{k,m}}^{+}(u)
\\&&\\
&=&\phi_{\mu_{k, m-1}}(u+2) \, \phi_{l_{k,m}}^{-}(u)\,,
\end{array}
$$
provided $\mu_{k,m}$ is not empty (otherwise the right hand sides
are equal to 1), where the ``string polynomials" $\phi^{\pm}_{l}(u)$
are defined in (\ref{high1}) and (\ref{high1a}). Using these obvious
identities, it is straightforward to obtain the $\tilde Q\tilde
Q$-relation:
\begin{equation}\label{gen10}
\begin{array}{c}
\phi \bigl (u+ 2l'_{k,m}\bigr )\, \tilde Q_{k\! -\! 1, m\! -\!
1}(u)\, \tilde Q_{k,m}(u+2)
\\ \\
- \,\,\,\, \phi \bigl (u \! -\!  2l_{k,m}\! +\! 2 \bigr )\, \tilde
Q_{k\! -\! 1, m\! -\! 1}(u+2) \, \tilde Q_{k,m}(u)
\\\\
= \,\,\, \tilde Q_{k-1,m}(u) \, \tilde Q_{k,m-1}(u+2)\,.
\end{array}
\end{equation}
In this form it is valid if $\mu_{k,m}\neq \emptyset$, otherwise the
functions $\tilde Q$ obey the standard Hirota equation
(\ref{QHIROTA}). In other words, the Hirota equation gets modified
in the region shown in Fig.\ref{fig:domainmu}. The boundary of this
region in the $(k,m)$ plane is exactly the boundary of the diagram
$\mu$. The functions $\tilde Q_{0,0}$ and $\tilde Q_{K,M}$ are fixed
to be 1: $\tilde Q_{0,0}(u)= \tilde Q_{K,M}(u)=1$.

The Bethe equations are derived from (\ref{gen10}) in the same way
as in section 5. They acquire non-trivial right hand
sides which can be compactly written in terms of the quantities
$l_{k,m}$ and $l'_{k,m}$:

\begin{equation}\label{Bethe1a}
\frac{\tilde Q_{k-1,m}\left (u_{j}^{(k,m)}\right ) \tilde
Q_{k,m}\left (u_{j}^{(k,m)}-2\right ) \tilde Q_{k+1,m}\left
(u_{j}^{(k,m)}+2\right )} {\tilde Q_{k-1,m}\left
(u_{j}^{(k,m)}-2\right ) \tilde Q_{k,m}\left (u_{j}^{(k,m)}+2\right
) \tilde Q_{k+1,m}\left (u_{j}^{(k,m)}\right )}=\, -\, \frac{\phi
\left (u_{j}^{(k,m)} + 2l'_{k,m} \right )}{\phi \left (u_{j}^{(k,m)}
+ 2l'_{k+1,m} \right )} \,,
\end{equation}

\vspace{3mm}

\begin{equation}\label{Bethe2a}
\frac{\tilde Q_{k,m+1}\left (u_{j}^{(k,m)}\right ) \tilde
Q_{k,m}\left (u_{j}^{(k,m)}-2\right ) \tilde Q_{k,m-1}\left
(u_{j}^{(k,m)}+ 2\right )}{\tilde Q_{k,m+1}\left
(u_{j}^{(k,m)}-2\right ) \tilde Q_{k,m}\left (u_{j}^{(k,m)}+2\right
) \tilde Q_{k,m-1} \left (u_{j}^{(k,m)}\right )}=\, -\, \frac{\phi
\left (u_{j}^{(k,m)} - 2l_{k,m+1} \right )}{\phi \left
(u_{j}^{(k,m)} - 2l_{k,m} \right )} \,,
\end{equation}

\vspace{3mm}

\begin{equation}\label{Bethe5a}
\frac{\tilde Q_{k+1,m}\left (u_{j}^{(k,m)}\right ) \tilde
Q_{k,m-1}\left (u_{j}^{(k,m)}+2\right )}{ \tilde Q_{k+1,m}\left
(u_{j}^{(k,m)}+2\right ) \tilde Q_{k,m-1}\left (u_{j}^{(k,m)}\right
)}=\, \frac{\phi \left (u_{j}^{(k,m)} + 2l_{k+1,m}' \right )}{\phi
\left (u_{j}^{(k,m)} - 2l_{k,m} \right )} \,,
\end{equation}

\vspace{3mm}

\begin{equation}\label{Bethe6a}
\frac{\tilde Q_{k,m+1}\left (u_{j}^{(k,m)}\right ) \tilde
Q_{k-1,m}\left (u_{j}^{(k,m)}-2\right )}{\tilde Q_{k,m+1}\left
(u_{j}^{(k,m)}-2\right ) \tilde Q_{k-1,m}\left (u_{j}^{(k,m)}\right
)}=\, \frac{\phi \left (u_{j}^{(k,m)} - 2l_{k,m+1} \right )}{\phi
\left (u_{j}^{(k,m)} + 2l_{k,m}' \right )} \,,
\end{equation}

\vspace{3mm}

\noindent These equations are listed here in the same order as
equations (\ref{Bethe1})-(\ref{Bethe6}). For empty diagrams,
$l_{k,m}$ and $l'_{k,m}$ are put equal to $0$. Using formulas
(\ref{gen9}), we can represent the right hand sides in a more
explicit but less compact form:

\begin{equation}\label{Bethe1b}
\frac{\tilde Q_{k-1,m}\left (u_{j}^{(k,m)}\right ) \tilde
Q_{k,m}\left (u_{j}^{(k,m)}-2\right ) \tilde Q_{k+1,m}\left
(u_{j}^{(k,m)}+2\right )} {\tilde Q_{k-1,m}\left
(u_{j}^{(k,m)}-2\right ) \tilde Q_{k,m}\left (u_{j}^{(k,m)}+2\right
) \tilde Q_{k+1,m}\left (u_{j}^{(k,m)}\right )}=\, -\, \frac{\phi
\left (u_{j}^{(k,m)}\! +\! 2m \! - \! 2M \! +\! 2\mu_{K-k+1}\right
)}{\phi \left (u_{j}^{(k,m)}\! +\! 2m \! - \! 2M \! +\!
2\mu_{K-k}\right )} \,,
\end{equation}

\vspace{3mm}

\begin{equation}\label{Bethe2b}
\frac{\tilde Q_{k,m+1}\left (u_{j}^{(k,m)}\right ) \tilde
Q_{k,m}\left (u_{j}^{(k,m)}-2\right ) \tilde Q_{k,m-1}\left
(u_{j}^{(k,m)}+2\right )}{ \tilde Q_{k,m+1}\left
(u_{j}^{(k,m)}-2\right ) \tilde Q_{k,m}\left (u_{j}^{(k,m)}+2\right
) \tilde Q_{k,m-1}\left (u_{j}^{(k,m)}\right )}=\, -\, \frac{\phi
\left (u_{j}^{(k,m)}\! -\! 2k \! + \! 2K \! -\! 2\mu_{M-m}'\right
)}{\phi \left (u_{j}^{(k,m)}\! -\! 2k \! + \! 2K \! -\!
2\mu_{M-m+1}'\right )} \,,
\end{equation}

\vspace{3mm}

\begin{equation}\label{Bethe5b}
\frac{\tilde Q_{k+1,m}\left (u_{j}^{(k,m)}\right ) \tilde
Q_{k,m-1}\left (u_{j}^{(k,m)}+2\right )}{\tilde Q_{k+1,m}\left
(u_{j}^{(k,m)}+2\right ) \tilde Q_{k,m-1}\left (u_{j}^{(k,m)}\right
)}=\, \frac{\phi \left (u_{j}^{(k,m)}\! +\! 2m \! - \! 2M \! +\!
2\mu_{K-k}\right )} {\phi \left (u_{j}^{(k,m)}\! -\! 2k \! + \! 2K
\! -\! 2\mu_{M-m+1}'\right )} \,,
\end{equation}

\vspace{3mm}

\begin{equation}\label{Bethe6b}
\frac{\tilde Q_{k,m+1}\left (u_{j}^{(k,m)}\right ) \tilde
Q_{k-1,m}\left (u_{j}^{(k,m)}-2\right )}{\tilde Q_{k,m+1}\left
(u_{j}^{(k,m)}-2\right ) \tilde Q_{k-1,m}\left (u_{j}^{(k,m)}\right
)}=\, \frac{\phi \left (u_{j}^{(k,m)}\! -\! 2k \! + \! 2K \! -\!
2\mu_{M-m}'\right )} {\phi \left (u_{j}^{(k,m)}\! +\! 2m \! - \! 2M
\! +\! 2\mu_{K-k+1}\right )}\,.
\end{equation}

\vspace{3mm}

\noindent These equations are the building blocks to make up the
chain of Bethe equations for any undressing path.

For example, the chain of the Bethe equations for the simplest path
$(K,M)\longrightarrow (0,M) \longrightarrow (0,0)$ is as follows.
Moving down from $(K,M)$ to $(0,M)$, we have the equations
\begin{equation}\label{Bethe1c}
\frac{\tilde Q_{k-1,M}\left (u_{j}^{(k,M)}\right ) \tilde
Q_{k,M}\left (u_{j}^{(k,M)}-2\right ) \tilde Q_{k+1,M}\left
(u_{j}^{(k,M)}+2\right )} {\tilde Q_{k-1,M}\left
(u_{j}^{(k,M)}-2\right ) \tilde Q_{k,M}\left (u_{j}^{(k,M)}+2\right
) \tilde Q_{k+1,M}\left (u_{j}^{(k,M)}\right )}=\, -\, \frac{\phi
\left (u_{j}^{(k,M)}\! +\! 2\mu_{K-k+1}\right )}{\phi \left
(u_{j}^{(k,M)} \! +\! 2\mu_{K-k}\right )} \,,
\end{equation}
where $k=1,2, \ldots , K-1$. They agree with the chain of Bethe
equations presented in \cite{KulResh2,MTV} for the bosonic case. At
the corner, the equation is
\begin{equation}\label{Bethe5c}
\frac{\tilde Q_{1,M}\left (u_{j}^{(0,M)}\right ) \tilde
Q_{0,M-1}\left (u_{j}^{(0,M)}+2\right )}{\tilde Q_{1,M}\left
(u_{j}^{(0,M)}+2\right ) \tilde Q_{0,M-1}\left (u_{j}^{(0,M)}\right
)}=\, \frac{\phi \left (u_{j}^{(0,M)}\! +\!  2\mu_{K}\right )} {\phi
\left (u_{j}^{(0,M)}\! + \! 2K \! -\! 2\mu_{1}'\right )} \,,
\end{equation}
Finally, moving to the left from $(0,M)$ to $(0,0)$, we have the
equations
\begin{equation}\label{Bethe2c}
\frac{\tilde Q_{0,m+1}\left (u_{j}^{(0,m)}\right ) \tilde
Q_{0,m}\left (u_{j}^{(0,m)}-2\right ) \tilde Q_{0,m-1}\left
(u_{j}^{(0,m)}+2\right )}{ \tilde Q_{0,m+1}\left
(u_{j}^{(0,m)}-2\right ) \tilde Q_{0,m}\left (u_{j}^{(0,m)}+2\right
) \tilde Q_{0,m-1}\left (u_{j}^{(0,m)}\right )}=\, -\, \frac{\phi
\left (u_{j}^{(0,m)}\! + \! 2K \! -\! 2\mu_{M-m}'\right )}{\phi
\left (u_{j}^{(0,m)}\! + \! 2K \! -\! 2\mu_{M-m+1}'\right )} \,,
\end{equation}
where $m=1,2, \ldots , M-1$. One can see that the differences of
arguments of the $\phi$-functions in the numerator and the
denominator are equal to the doubled Kac-Dynkin labels
(\ref{KacDynkin}) corresponding to the diagram $\mu$.

One can also write down the chain of equations for an arbitrary
path, similarly to eq. (\ref{Bethe7}). However, the general form of
the right hand sides is not very illuminating.

The last remark concerns the form of the ansatz (\ref{gen7}).
Instead of (\ref{gen7}) one could use a similar ansatz
\begin{equation}\label{gen11}
Q_{k,m}(u)=\bar \phi_{\mu_{k,m}}\left
(u \! +\! 2(K\! -\! k)\! -\! 2(M\! -\! m)\right )
\, \bar Q_{k,m}(u)
\end{equation}
with the function $\bar \phi_{\mu}(u)$ defined in (\ref{gen1a}).
This leads to a similar system of Bethe equations with non-trivial
vacuum parts. There is a one-to-one correspondence between
solutions of the both systems.

\section{Examples}

\subsection{  Baxter and Bethe equations for $gl(1|1)$ }

\begin{figure}[t]
   \centering
        \includegraphics[angle=-90,scale=0.4]{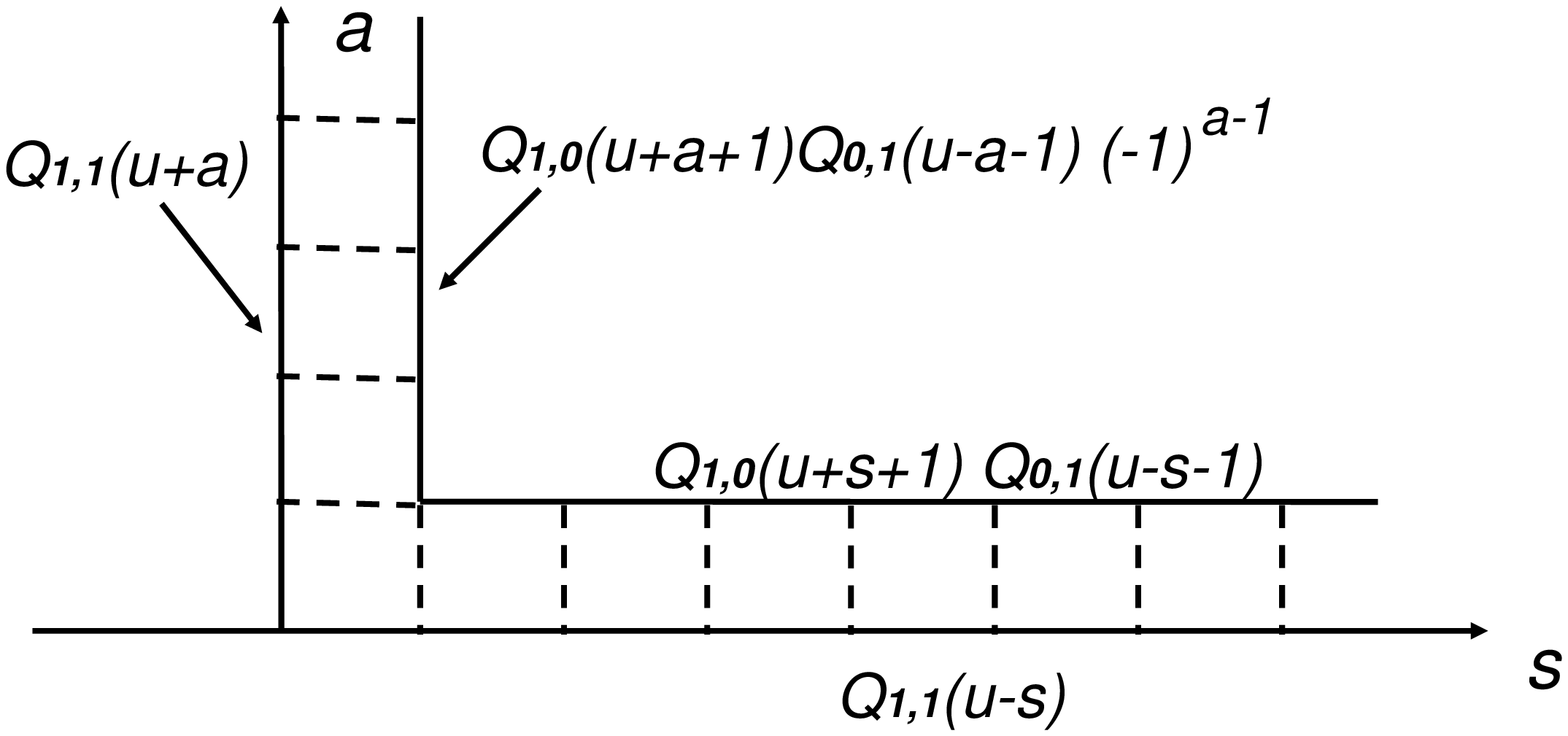}
    \caption{Boundary conditions for the transfer matrix
    $T(a,s,u)\equiv T_{1,1}(a,s,u)$ for the $gl(1|1)$
    superalgebra. }
    \label{fig:U11}
\end{figure}

Let us consider in detail the $gl(1|1)$ case. In this case all
elements of transfer matrix $T_{1,1}(a,s,u)\equiv T(a,s,u)$ lay on
the boundaries of the fat hook domain, as we see in
Fig.~\ref{fig:U11}. We read off from this figure:
\begin{eqnarray}
\label{QRELS}
T(0,0,u)&=&Q_{1,1}(u)\equiv \phi(u)=\prod_{j=1}^N\(u-\th_j\)\nn\\
T(a,0,u)&=&Q_{1,1}(u+a),\qquad a\ge 0\nn\\
T(0,s,u)&=&Q_{1,1}(u-s),\qquad -\infty<s<\infty \nn\\
T(a,1,u)&=&(-1)^{a-1}Q_{1,0}(u+a+1)Q_{0,1}(u-a-1),\qquad a\ge 1\nn\\
T(1,s,u)&=&Q_{1,0}(u+s+1)Q_{0,1}(u-s-1),\qquad s\ge 1\,.
\end{eqnarray}
Using the Hirota equation (\ref{QHIROTA}) we get the only nontrivial
$QQ$-relation:
\begin{eqnarray*}
Q_{0,0}(u)Q_{1,1}(u+2)-Q_{1,1}(u)Q_{0,0}(u+2)&=&Q_{0,1}(u)Q_{1,0}(u+2),
\end{eqnarray*}
which gives, together with $Q_{0,0}(u)=1$,
\begin{equation}\label{HIR11}
Q_{1,0}(u)Q_{0,1}(u-2)=\phi(u)-\phi(u-2).
\end{equation}
This relation allows us to exclude, for example, $Q_{0,1}$  or
$Q_{1,0}$ from \eq{QRELS} and to obtain, in particular, the
Baxter $TQ$-relations:
\begin{eqnarray}\label{TQREL}
T(1,s,u)&=&\frac{Q_{0,1}(u-s-1)}{Q_{0,1}(u+s-1)}\(\phi(u+s+1)-
\phi(u+s-1)\)\,,\nn\\
T(1,s,u)&=&\frac{Q_{1,0}(u+s+1)}{Q_{1,0}(u-s+1)}\(\phi(u-s+1)-\phi(u-s-1)\).
\end{eqnarray} From the second relation, using
regularity of the left hand side, we obtain the following Bethe
equation:
\begin{eqnarray}\label{BAE01}
1=\frac{\phi(u_j^{(1,0)})}{\phi(u_j^{(1,0)}-2)}.
\end{eqnarray}
It is a free fermion equation. We can also obtain it from the
undressing procedure $Q_{1,1}\to Q_{1,0}\to Q_{0,0}=1$, putting
$u=u^{(1,0)}+s-1$ in \eq{HIR11}. For another  undressing procedure,
$Q_{1,1}\to Q_{0,1}\to Q_{0,0}=1$, or by canceling poles at
$u=u^{(0,1)}-s+1$ in the first equation in (\ref{TQREL}), we get the
Bethe equation
\begin{eqnarray}\label{BAE10}
1=\frac{\phi(u_j^{(0,1)})}{\phi(u_j^{(0,1)}+2)}.
\end{eqnarray}

Note that the Bethe equations (\ref{BAE01}), (\ref{BAE10}) do
not depend on the representation parameter $s$ whereas the
eigenvalues $T(1,s,u)$ do depend on it. In principle,  nothing
prevents this parameter to be continued analytically to the complex
plane. This is precisely the continuous label of typical (long)
irreducible representations  \cite{BMR} of the superalgebra  in
auxiliary space.

Let us consider higher rectangular irreps in the physical space. To
cover representations with an  arbitrary spin $\ell$ (one-row
diagrams with $\ell$ boxes), we take a special form of the function
$\phi (u)$, as in eq.~(\ref{high1}):
\begin{eqnarray}\label{PHIF}
  \phi(u)=\varphi(u)\varphi(u+2)\dots \varphi(u+2\ell)\varphi\(u+2(\ell-1)\),
\end{eqnarray}
describing a ``string" of length $\ell$.
(Comparing to the previous section, we have changed the notation slightly
denoting the polynomial function with roots
$\theta_i$ by $\varphi (u)=\prod_j (u-\theta_j)$.)
According to (\ref{gen7}), we have:
$$
\begin{array}{lll}
Q_{0,1}(u)& =& \tilde Q_{0,1}(u)\\ &&\\
Q_{1,0}(u)& =& \varphi (u)\varphi
(u+2)\ldots \varphi (u+2(\ell -2))\, \tilde Q_{1,0}(u)\,.
\end{array}
$$From the first
equation in (\ref{TQREL})
we obtain the transfer matrix $T_{1,1}(1,s,u)\equiv T(1,s,u)$:
\begin{eqnarray}\label{TS1}
T(1,s,u+1)=
\frac{\tilde Q_{0,1}(u-s)}{\tilde Q_{0,1}(u+s)}
\[\varphi(u+s+2\ell )-\varphi(u+s)\]\,\,\,
\prod_{j=1}^{\ell -1}\varphi(u+s+2j).
\end{eqnarray}
The last factor represents trivial zeros of $T(a,s,u)$. It has the
same origin as the one in eq.~(\ref{high2}).
However, one should note that because $T(1,s,u)$ lives on the
interior boundary, this factor contains more zeros than
the one in eq.~(\ref{high2}) (see the remark in the end of
section 7.1).
The Bethe equation (\ref{BAE10}) becomes:
\begin{eqnarray}\label{BAER}
1=\frac{\varphi(u_j^{(0,1)})}{\varphi(u_j^{(0,1)}+2\ell )}.
\end{eqnarray}

It is easy to generalize these results to the case of an arbitrary
physical spin $\ell_l$ at each site $l=1,2, \ldots , N $ of the chain.  In
this case we take
\begin{eqnarray*}
  \phi(u)=\prod_{l=1}^N  (u-\theta_l)(u-\theta_l+2)
\ldots (u-\theta_l+2\ell_l-2).
\end{eqnarray*}
This yields the transfer matrix
\begin{eqnarray}\label{TS2}
T(1,s,u+1) =\frac{\tilde Q_{0,1}(u-s)}{\tilde Q_{0,1}(u+s)}
\(\prod_{l=1}^N\(u-\theta_l+s+2\ell_l\)-
\prod_{l=1}^N\(u-\theta_l+s\)\)
\prod_{l=1}^N\prod_{j=0}^{\ell_l-2}(u-\theta_l+s+2j).
\end{eqnarray}
The last factor represents the trivial zeros which can be absorbed
by normalization.
Introducing a new renormalized  function $T'$ with the trivial
zeroes removed,
\begin{equation}
T(1,s,u+1)=T'(1,s,u+1)\,\,
 \prod_{l=1}^N\prod_{j=1}^{\ell_l-1}(u-\theta_l+s+2j)
\end{equation}
we obtain, redefining the parameters $\theta_l= \hat \theta_l+\ell_l$,
\begin{equation}
T'(1,s,u+1)=\frac{\tilde Q_{0,1}(u-s)}{\tilde
Q_{0,1}(u+s)}\(\prod_{l=1}^N\(u-\hat \theta_l+s+\ell_l\)-
\prod_{l=1}^N\(u-\hat \theta_l+s-\ell_l\)\)
\end{equation}
which is essentially the same transfer matrix eigenvalue
as for  the rational limit of the equations (A.9), (A.10) from
\cite{Fendley:1992dm}, with the definition of the $gl(1|1)$
$S$-matrix (3.1), (3.2) from \cite{Fendley:1991ve}.
The Bethe equations are
\begin{eqnarray}\label{BAERA}
1=\prod_{l(\ne j)=1}^N \frac{ u_j^{(0,1)}-\hat \theta_l+s+
\ell_l}{u_j^{(0,1)}-\hat \theta_l+s-\ell_l}.
\end{eqnarray}
For the complete comparison  one should exchange in these papers the
parameters $\ell\rightarrow s,\,\, s\to r$ of the auxiliary and
quantum spaces and take the rational limit. Some simple shifts in
the arguments and definitions of Bethe roots are also necessary.
Note that unlike \cite{Fendley:1992dm},  where they are the soliton
charges, both representation labels $\ell$ and $s$ can be now
considered as continuous parameters of the typical representation of
$gl(1|1)$. This corresponds to the limit of large charges and large
period of these soliton charges.

\subsection{  Baxter and Bethe equations for $gl(2|1)$ }

\begin{figure}[t]
    \centering
        \includegraphics[angle=-90,scale=0.4]{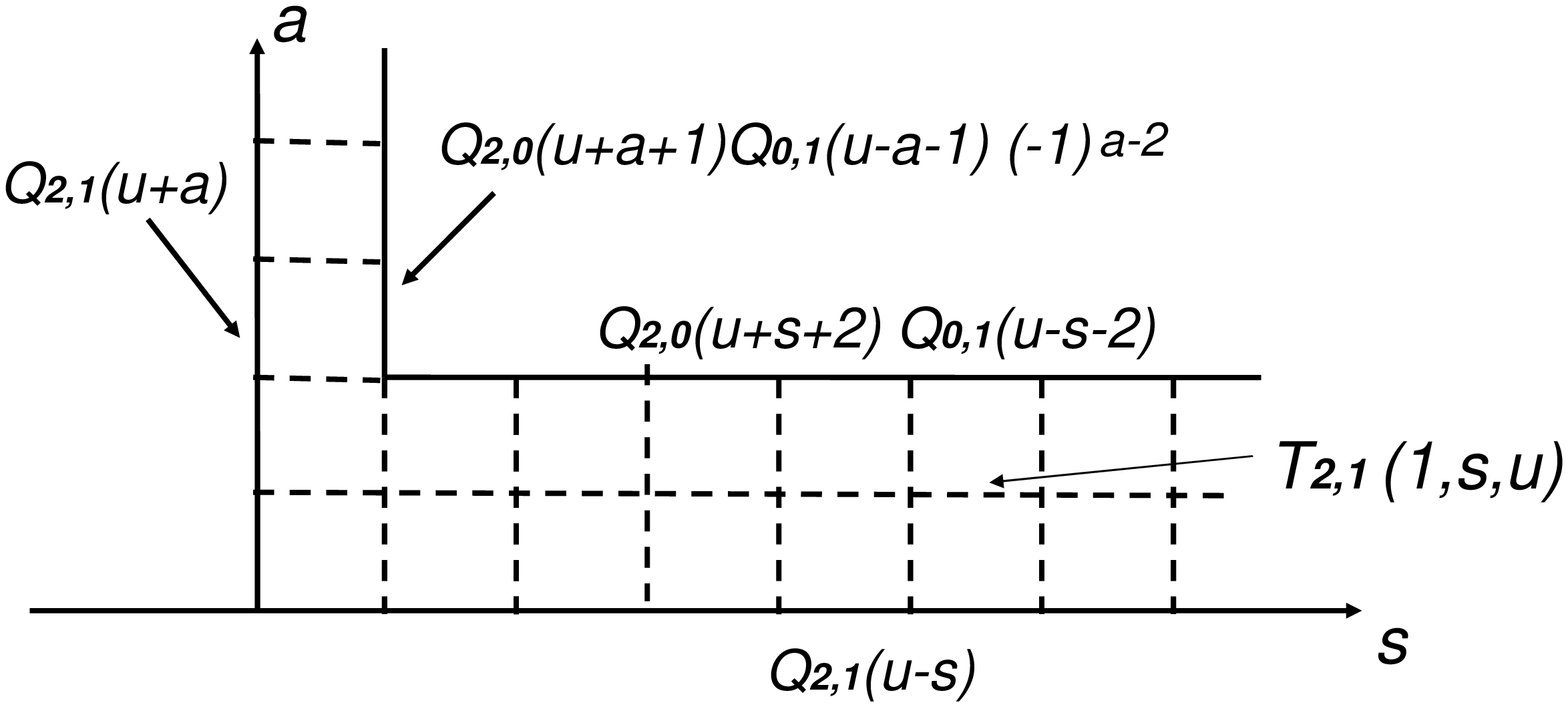}
    \caption{\it  Boundary conditions for $T(a,s,u)\equiv T_{2,1}(a,s,u)$
    for the $gl(2|1)$    superalgebra. }
    \label{fig:SU21}
\end{figure}

In the $gl(2|1)$ case\footnote{The construction of the Baxter $Q$-operators
and $TQ$-relations for the models based on $gl(2|1)$
(and $U_q (\widehat{gl(2|1)})$) were recently discussed in
\cite{Belitsky}, \cite{Bazh-Tsub}.
We thank Z.~Tsu\-boi for bringing these works to
our attention.}, the  transfer matrices
$T_{2,1}(a,s,u)\equiv T(a,s,u)$ are on the boundaries of the fat hook
except $T(1,s,u)$ in the middle row,
as we see in Fig.~\ref{fig:SU21}. There are six $Q$-functions $Q_{k,m}(u)$
($k=0,1,2;\,m=0,1$), two of them being fixed by the boundary conditions. The
$T$-functions are expressed through them in the following way:
\begin{eqnarray}\label{TTTQ}
T(0,0,u)&=&Q_{2,1}(u)\equiv \phi(u)=\prod_{j=1}^N\(u-\th_j\)\nn\\
T(a,0,u)&=&Q_{2,1}(u+a),\qquad a\ge 0\nn\\
T(0,s,u)&=&Q_{2,1}(u-s),\qquad -\infty<s<\infty\nn\\
T(a,1,u)&=&Q_{2,0}(u+a+1)Q_{0,1}(u-a-1)(-1)^{a-2},\qquad a\ge 2\nn\\
T(2,s,u)&=&Q_{2,0}(u+s+2)Q_{0,1}(u-s-2),\qquad s\ge 1\,.
\end{eqnarray}

\subsubsection{Baxter equations for the Kac-Dynkin diagram
$_1\bigotimes\!\!-\!\!\!-\!\!\!-\!\!\bigodot$}

(Here and below the subscript $1$ means the fundamental
representation in the quantum space.)

The operator generating series (\ref{BAXTERARB}) reads
\begin{eqnarray}\label{BAXT21}
&&\sum_{a=0}^\infty\frac{T(a,1,u+a+1)}{Q_{2,1}(u+2a+2)}\,
e^{2a\p_u}= \hat U_{1,0}  \hat U_{2,0}\hat V^{-1}_{2,0}=
\nn\\
&=& \(\frac{Q_{1,0}(u)}{Q_{1,0}(u+2)}-e^{2\p_u}\)
\(\frac{Q_{2,0}(u)}{Q_{2,0}(u+2)}\frac{Q_{1,0}(u+2)}{Q_{1,0}(u)}-e^{2\p_u}\)
\(\frac{Q_{2,0}(u)}{Q_{2,0}(u+2)}\frac{Q_{2,1}(u+2)}{Q_{2,1}(u)}-e^{2\p_u}\)^{-1}.
\end{eqnarray}
In particular,
\begin{eqnarray}\label{TT21}
\frac{T(2,1,u+3)}{\phi(u+4)}=&&\frac{Q_{2,0}(u+6)}{Q_{2,0}(u+4)}
\frac{\phi(u+4)}{\phi(u+6)}
-\frac{Q_{1,0}(u+4)}{Q_{1,0}(u+2)}\frac{Q_{2,0}(u+6)}{Q_{2,0}(u+4)}
\frac{\phi(u+2)}{\phi(u+6)}\nn\\
&-&\frac{Q_{1,0}(u)}{Q_{1,0}(u+2)}\frac{Q_{2,0}(u+6)}{Q_{2,0}(u+2)}
\frac{\phi(u+2)}{\phi(u+6)}
+\frac{Q_{2,0}(u+6)}{Q_{2,0}(u+2)}\frac{\phi(u)}{\phi(u+6)}\,.
\end{eqnarray}From this equation and
equation
$$
T(2,s,u)=\frac{Q_{2,0} (u+s+2)}{Q_{2,0}(u-s+4)}\, T(2,1,u-s+1)
$$
which follows from \eq{TTTQ}, we obtain
\begin{eqnarray}
\label{TSUQ}
\frac{T(2,s,u)}{\phi(u-s+2)}&=&\frac{Q_{2,0}(u+s+2)}{Q_{2,0}(u-s+2)}
\frac{\phi(u-s+2)}{\phi(u-s+4)}
-\frac{Q_{1,0}(u-s+2)}{Q_{1,0}(u-s)}\frac{Q_{2,0}(u+s+2)}{Q_{2,0}(u-s+2)}
\frac{\phi(u-s)}{\phi(u-s+4)}\nn\\
&-&\frac{Q_{1,0}(u-s-2)}{Q_{1,0}(u-s)}\frac{Q_{2,0}(u+s+2)}{Q_{2,0}(u-s)}
\frac{\phi(u-s)}{\phi(u-s+4)}
+\frac{Q_{2,0}(u+s+2)}{Q_{2,0}(u-s)}\frac{\phi(u-s-2)}{\phi(u-s+4)}
\,.
\end{eqnarray}
Note that the representation index $s$ can be treated as a
continuous parameter. It corresponds to the continuous label of
typical representations of the $gl(2|1)$ superalgebra.

\subsubsection{Spins in higher irreps in the quantum space  }

As an illustrative example to section~5, as well as for the
purpose of comparison with the result of \cite{pfannmuller-1996-479}
(\eq{FRPF} obtained from the specific $S$-matrix by
the algebraic Bethe Ansatz), we consider here spins in the quantum space
in the irrep $(r^2)$ (Young diagrams with
two rows of length $r$). At the end, the number
$r$ will be treated as a continuous label.

Let us introduce new definitions, following
eq.~(\ref{gen1}) of the previous section:
\begin{equation}\label{PHID2}
\phi(u)\equiv\phi_{2,1} (u)=
\varphi(u-2)\[\varphi^2(u)\varphi^2(u+2)\dots
\varphi^2\(u+2r-4\)\]\varphi\(u+2r-2\),
\end{equation}
\begin{equation}\label{PHID3}
\phi_{2,0} (u)= \varphi(u-2)\[\varphi^2(u)\varphi^2(u+2)\dots
\varphi^2\(u+2r-6\)\]\varphi\(u+2r-4\)\,,
\end{equation}
\begin{equation}\label{PHID1}
\phi_{1,0} (u)=\varphi(u)\varphi(u+2)\dots\varphi\(u+2r-4\).
\end{equation}
Here $\varphi (u)=\prod_j (u-\theta_j)$.
Then, according to eq.~(\ref{gen7}),
\begin{equation}\label{Q2TIL}
Q_{2,0}(u)=\phi_{2,0} (u) \, \tilde Q_{2,0}(u)\,,
\end{equation}
\begin{equation}\label{Q1TIL}
Q_{1,0}(u)=\phi_{1,0} (u) \, \tilde Q_{1,0}(u)\,.
\end{equation}
In this notation, \eq{TSUQ} takes the form
\begin{eqnarray}\label{TSUQREW}
T'(2,s,u)&=&\frac{\tilde
Q_{2,0}(u+s+2)}{\tilde Q_{2,0}(u-s+2)}\nn\\
&-&\frac{\tilde Q_{1,0}(u-s+2)}{\tilde Q_{1,0}(u-s)}\frac{\tilde
Q_{2,0}(u+s+2)}{\tilde Q_{2,0}(u-s+2)}\frac{ \phi_{1,0}(u-s+2)}{
\phi_{1,0}(u-s)}
\frac{\phi(u-s)}{\phi(u-s+2)}\nn\\
&-&\frac{\tilde Q_{1,0}(u-s-2)}{\tilde Q_{1,0}(u-s)}\frac{\tilde
Q_{2,0}(u+s+2)}{\tilde Q_{2,0}(u-s)}
\frac{\phi_{1,0}(u-s-2)}{\phi_{1,0}(u-s)}\frac{\phi_{2,0}
(u-s+2)}{\phi_{2,0}(u-s)}\frac{\phi(u-s)}{\phi(u-s+2)}\nn\\
&+&\frac{\tilde Q_{2,0}(u+s+2)}{\tilde Q_{2,0}(u-s)}
\frac{\phi_{2,0}(u-s+2)}{\phi_{2,0}(u-s)}\frac{\phi(u-s-2)}{\phi(u-s+2)}\nn
\,,
\end{eqnarray}
where we redefined the transfer matrix extracting trivial zeros:
\begin{eqnarray}\label{PHIT}
T(2,s,u)&=& T'(2,s,u)\, \frac{\phi^2(u-s+2)}{\phi(u-s+4)}
\frac{\phi_{2,0}(u+s+2)}{\phi_{2,0}(u-s+2)}=\nn\\
&=&  T'(2,s,u)\,\,
\[\varphi(u-s)\varphi(u-s+2)\phi_{2,0}(u+s+2)\].
\end{eqnarray}
For the reason already discussed
in the case of $gl(1|1)$, $T'$ is not equal to $\tilde T$ of \eq{gen4}.

Calculating the ratios  of $\phi$-functions in \eq{TSUQREW}, we find
for the following result for the
transfer matrix in physical irrep $(r^2)$ and the auxiliary
irrep $(s^2)$:
\begin{eqnarray}\label{TSR}
T'_r(2,s,u)&=&\frac{\tilde
Q_{2,0}(u+s+2)}{\tilde Q_{2,0}(u-s+2)}\nn\\
&-&\frac{\tilde Q_{1,0}(u-s+2)}{\tilde Q_{1,0}(u-s)}\frac{\tilde
Q_{2,0}(u+s+2)}{\tilde Q_{2,0}(u-s+2)}
\frac{\varphi(u-s-2)}{\varphi(u-s+2r)}\nn\\
&-&\frac{\tilde Q_{1,0}(u-s-2)}{\tilde Q_{1,0}(u-s)}\frac{\tilde
Q_{2,0}(u+s+2)}{\tilde
Q_{2,0}(u-s)}\frac{\varphi(u-s-2)}{\varphi(u-s+2r)}
\nn\\
&+&\frac{\tilde Q_{2,0}(u+s+2)}{\tilde Q_{2,0}(u-s)}
\frac{\varphi(u-s-4)}{\varphi(u-s+2r-2)}\frac{\varphi(u-s-2)}{\varphi(u-s+2r)}
\,.
\end{eqnarray}
Using the notation of \eq{Bethe88}
$\tilde Q_{k+m}(u)=\tilde Q_{k,m}(u+k-m)$ and setting $\tilde
\varphi(u)=\varphi(u+r-1)$ we bring \eq{TSR} to the form
\begin{eqnarray}\label{TSRTIL}
T'_r(2,s,u)&=&\frac{\tilde
Q_{2}(u+s)}{\tilde Q_{2}(u-s)}\nn\\
&-&\frac{\tilde Q_{1}(u-s+1)}{\tilde Q_{1}(u-s-1)}\frac{\tilde
Q_{2}(u+s)}{\tilde Q_{2}(u-s)}
\frac{\tilde \varphi(u-s-r-1)}{\tilde \varphi(u-s+r+1)}\nn\\
&-&\frac{\tilde Q_{1}(u-s-3)}{\tilde Q_{1}(u-s-1)}\frac{\tilde
Q_{2}(u+s)}{\tilde Q_{2}(u-s-2)}\frac{\tilde
\varphi(u-s-r-1)}{\tilde \varphi(u-s+r+1)}
\nn\\
&+&\frac{\tilde Q_{2}(u+s)}{\tilde Q_{2}(u-s-2)} \frac{\tilde
\varphi(u-s-r-3)}{\tilde \varphi(u-s+r-1)}\frac{\tilde
\varphi(u-s-r-1)}{\tilde \varphi(u-s+r+1)} \,.
\end{eqnarray}
This transfer matrix eigenvalue coincides, after shifting $r\to r-1$
and making an easy generalization to
the inhomogeneous chain (i.e., to spins in arbitrary irreps at each site
of the chain), with \eq{BAX} taken from
\cite{pfannmuller-1996-479} (see Appendix~E, where their result is
rewritten in our notation).
Note also that the spin label $r$, as
well as $s$, can be treated as continuous parameters here. Hence
our method is general enough to describe the transfer matrices in
all possible  typical and atypical irreps in quantum and auxiliary
spaces.

Canceling  the poles at $u=u^{(2,0)}_j+s-2$ and
$u=u^{(1,0)}_j+s$ in \eq{TSR}, we write the following Bethe equations:
\begin{eqnarray}\label{BAEX0}
\frac{\varphi\(u^{(2,0)}_j+2r-2\)}{\varphi\(u^{(2,0)}_j-4\)}&=&
\frac{\tilde Q_{1,0}\(u^{(2,0)}_j\)}{\tilde
Q_{1,0}\(u^{(2,0)}_j-2\)}
\nn\\
-1&=&\frac{\tilde Q_{1,0}\(u^{(1,0)}_j+2\)}{\tilde
Q_{1,0}\(u^{(1,0)}_j-2\)} \frac{\tilde
Q_{2,0}\(u^{(1,0)}_j\)}{\tilde Q_{2,0}\(u^{(1,0)}_j+2\)},
\end{eqnarray}
or, in the notation of eq.~(\ref{TSRTIL}),
\begin{eqnarray}\label{BAEST}
\frac{\tilde \varphi\(u^{(2)}_j+r+1\)}{\tilde \varphi\(u^{(2)}_j-r-1\)}&=&
\frac{\tilde Q_{1}\(u^{(2)}_j+1\)}{\tilde Q_{1}\(u^{(2)}_j-1\)}
\nn\\
-1&=&\frac{\tilde Q_{1}\(u^{(1)}_j+2\)}{\tilde Q_{1}\(u^{(1)}_j-2\)}
\frac{\tilde Q_{2}\(u^{(1)}_j-1\)}{\tilde Q_{2}\(u^{(1)}_j+1\)}.
\end{eqnarray}

As an example, let us also consider two other equations for
different choices of the Kac-Dynkin diagram.

\subsubsection{Bethe equations for the Kac-Dynkin diagram
$\bigotimes\!\!-\!\!\!-\!\!\!-\!\!\bigodot_1$}

Let us express $T(1,s,u)$ through $Q$'s using \eq{BAXTER}:
\begin{eqnarray}\label{BAXTER21}
&&\sum_{a=0}^\infty\frac{T(a,1,u+a+1)}{Q_{2,1}(u+2a+2)}\,
e^{2a\p_u}=\hat V^{-1}_{0,0} \hat U_{1,1}  \hat U_{2,1}=
\nn\\ \\
&=&\(\frac{Q_{0,1}(u+2)}{Q_{0,1}(u)}-e^{2\p_u}\)^{-1}
\(\frac{Q_{1,1}(u)}{Q_{1,1}(u+2)}\frac{Q_{0,1}(u+2)}{Q_{0,1}(u)}-e^{2\p_u}\)
\(\frac{Q_{2,1}(u)}{Q_{2,1}(u+2)}\frac{Q_{1,1}(u+2)}{Q_{1,1}(u)}-e^{2\p_u}\).\nn
\end{eqnarray}
In particular,
\begin{equation}\label{T21one}
\frac{T(1,1,u+2)}{\phi(u+4)}=-\frac{Q_{1,1}(u)}{Q_{1,1}(u+2)}
-\frac{Q_{0,1}(u)}{Q_{0,1}(u+2)}
\frac{\phi(u+2)}{\phi(u+4)}\frac{Q_{1,1}(u+4)}{Q_{1,1}(u+2)}+
\frac{Q_{0,1}(u)}{Q_{0,1}(u+2)}
\frac{\phi(u+2)}{\phi(u+4)}.
\end{equation}
Canceling the poles in \eq{T21one}, it is straightforward to write
the following Bethe equations for $gl(2|1)$ superalgebra:
\begin{eqnarray}\label{BAEX021}
\frac{\phi\(u^{(1,1)}_j+4\)}{\phi\(u^{(1,1)}_j+2\)}&=&
\frac{Q_{0,1}\(u^{(1,1)}_j\)}{Q_{0,1}\(u^{(1,1)}_j+2\)}
\frac{Q_{1,1}\(u^{(1,1)}_j+4\)}{Q_{1,1}\(u^{(1,1)}_j\)}
\nn\\
1&=&\frac{Q_{1,1}\(u^{(0,1)}_j+4\)}{Q_{1,1}\(u^{(0,1)}_j+2\)}.
\end{eqnarray}

\subsubsection{Bethe equations for the Kac-Dynkin diagram
$\bigotimes\!\!-\!\!\!-\!\!\!-\!\!\bigotimes_{1}$}

Using \eq{VUREL} or \eq{BAXTERARB}, we write the operator
generating series:
\begin{eqnarray}\label{FBAX21}
&&\sum_{a=0}^\infty\frac{T(a,1,u+a+1)}{Q_{2,1}(u+2a+2)}\,
e^{2a\p_u}=\hat U_{1,0} \hat V^{-1}_{1,0}  \hat U_{2,1}=
\nn\\ \\
&=&\(\frac{Q_{1,0}(u)}{Q_{1,0}(u+2)}-e^{2\p_u}\)
\(\frac{Q_{1,0}(u)}{Q_{1,0}(u+2)}\frac{Q_{1,1}(u+2)}{Q_{1,1}(u)}-e^{2\p_u}\)^{-1}
\(\frac{Q_{2,1}(u)}{Q_{2,1}(u+2)}\frac{Q_{1,1}(u+2)}{Q_{1,1}(u)}-e^{2\p_u}\).\nn
\end{eqnarray}
In particular,
\begin{equation}\label{T21two}
\frac{T(1,1,u+2)}{\phi(u+3)}=-\frac{Q_{1,0}(u+4)}{Q_{1,0}(u+2)}
\frac{\phi(u+2)}{\phi(u+4)}
+\frac{Q_{1,0}(u+4)}{Q_{1,0}(u+2)}\frac{Q_{1,1}(u)}{Q_{1,1}(u+2)}
\frac{\phi(u+2)}{\phi(u+4)} -\frac{Q_{1,1}(u)}{Q_{1,1}(u+2)}.
\end{equation}
Canceling the poles in \eq{T21two}, we write the following Bethe
equations:
\begin{eqnarray}\label{BAEOX}
1&=&\frac{Q_{1,1}\(u^{(1,0)}_j\)}{Q_{1,1}\(u^{(1,0)}_j-2\)}\nn\\
\frac{\phi\(u^{(1,1)}_j\)}{\phi\(u^{(1,1)}_j+2\)}&=&
\frac{Q_{1,0}\(u^{(1,1)}_j\)}{Q_{1,0}\(u^{(1,1)}_j+2\)}.
\end{eqnarray}

\subsection{Examples of the integration algorithm}

To illustrate the  general algorithm
of section 6, we apply it to
$gl(2)$, $gl(2|1)$ and $gl(2|2)$ (super)algebras.

\subsubsection{$gl(2)$ algebra}

In this case equation (\ref{T1s}) reads
\begin{eqnarray}
\label{T21-sl2} T_{2,0}(1,s,u) =  \widehat
H^{-1}_{2^-,0}(0,s,u)~T_{1,0}(1,s,u), \quad 0 \leq s < \infty.
\end{eqnarray}
Substituting explicit expressions for $\widehat
H^{-1}_{2^-,0}(0,s,u)$ (\ref{H1-n-inv}) and $T_{1,0}(1,s,u)$
(\ref{Ta0}), we immediately obtain the explicit formula for
non-vanishing functions which do not belong to the boundaries:
\begin{eqnarray}
\label{T21-sl2-1} T_{2,0}(1,s,u+s-1) = Q_{1,0}(u+2s) Q_{1,0}(u-2)
\sum_{j=0}^{s}\frac{Q_{2,0}(u+2j)}{Q_{1,0}(u+2j-2)Q_{1,0}(u+2j)}.
\end{eqnarray}

\subsubsection{$gl(2|1)$ superalgebra}

In this case equation (\ref{T1s}) reads
\begin{eqnarray}
\label{T21-s} T_{2,1}(1,s,u) =  \widehat H_{2,0^+}(0,s,u)~\widehat
H^{-1}_{2^-,0}(0,s,u)~T_{1,0}(1,s,u) \equiv  \widehat
H_{2,0^+}(0,s,u)~T_{2,0}(1,s,u), \quad 0 \leq s < \infty.
\end{eqnarray}
Substituting explicit expressions for $\widehat H_{2,0^+}(0,s,u)$
(\ref{H;a,s=0}) and $T_{2,0}(1,s,u)$ (\ref{T21-sl2-1}),
we obtain, after straightforward calculations, the following result
for non-vanishing functions which do not belong to the boundaries:
\begin{eqnarray}
\label{T21-s-1} T_{2,1}(1,s,u\! +\! s\! -\! 1) = Q_{1,0}(u\! +\! 2s)
\left (\frac{Q_{2,1}(u)}{Q_{1,0}(u)} +
Q_{1,1}(u\! -\! 2)\sum_{j=0}^{s-1}\frac{Q_{2,0}(u+2j+2)}
{Q_{1,0}(u+2j)Q_{1,0}(u+2j+2)}\, \theta(s\! -\! 1)
\right ).
\end{eqnarray}
The step function $\theta (s)=1$ at $s\geq 0$ and
$0$ otherwise.
In the calculation, we have used
eq.~(\ref{bilEqQ}) to substitute
$$\frac{Q_{1,0}(u-2)Q_{2,1}(u)-Q_{1,0}(u)Q_{2,1}
(u-2)}{Q_{2,0}(u)}=Q_{1,1}(u-2)\,.$$

\subsubsection{$gl(2|2)$ superalgebra}

In this case equations (\ref{T1s}) and (\ref{Ta1}) read
\begin{eqnarray}
\label{T22-1} T_{2,2}(1,s,u) &=& \widehat H_{2,1^+}(0,s,u)~\widehat
H_{2,0^+}(0,s,u)~
\widehat H^{-1}_{2^-,0}(0,s,u)~T_{1,0}(1,s,u)\nn\\
&\equiv&  \widehat H_{2,1^+}(0,s,u)~\widehat
H_{2,0^+}(0,s,u)~T_{2,0}(1,s,u) \equiv  \widehat
H_{2,1^+}(0,s,u)~T_{2,1}(1,s,u), \quad 0 \leq s < \infty~~~~
\end{eqnarray}
and
\begin{eqnarray}
\label{T22-2} T_{2,2}(a,1,u)=\widehat H_{1^+,2}(a,0,u)~\widehat
H_{0^+,2}(a,0,u) ~\widehat H^{-1}_{0,2^-}(a,0,u)~T_{0,1}(a,1,u),
\quad 0 \leq a < \infty\,.
\end{eqnarray}
Substituting explicit expressions for the shift operators, we
finally obtain:
\begin{eqnarray}
\label{T22-3} &&T_{2,2}(1,s,u+s-1) = Q_{1,0}(u+2s)
\Big(\frac{Q_{2,2}(u)}{Q_{1,0}(u)}
+ \frac{Q_{2,2}(u)Q_{1,1}(u-2)Q_{2,0}(u+2)}{Q_{2,1}(u)Q_{1,0}(u)Q_{1,0}(u+2)}
\, \theta(s-1)\nn\\
&-&
\frac{Q_{2,2}(u-2)Q_{2,1}(u+2)}{Q_{2,1}(u)Q_{1,0}(u+2)}\, \theta(s-1) +
Q_{1,2}(u-2)\sum_{j=0}^{s-2}
\frac{Q_{2,0}(u+2j+4)}{Q_{1,0}(u+2j+2)Q_{1,0}(u+2j+4)}\, \theta(s-2)\Big)~~~
\end{eqnarray}
and
\begin{eqnarray}
\label{T22-4} &&T_{2,2}(a,1,u-a+1) = Q_{0,1}(u-2a)
\Big(\frac{Q_{2,2}(u)}{Q_{0,1}(u)}
+ \frac{Q_{2,2}(u)Q_{1,1}(u+2)Q_{0,2}(u-2)}{Q_{1,2}(u)Q_{0,1}(u)Q_{0,1}(u-2)}\,
\theta(a-1)\nn\\
&+&
\frac{Q_{2,2}(u+2)Q_{1,2}(u-2)}{Q_{1,2}(u)Q_{0,1}(u-2)}\theta(a-1) +
Q_{2,1}(u+2)\sum_{j=0}^{a-2}
\frac{(-1)^jQ_{0,2}(u-2j-4)}{Q_{0,1}(u-2j-2)Q_{0,1}(u-2j-4)}\, \theta(a-2)\Big).~~~~
\end{eqnarray}
In the calculation, we have used eq.~(\ref{bilEqQ}) to substitute
$$Q_{1,1}(u-2)Q_{2,2}(u)-Q_{1,1}(u)Q_{2,2}(u-2)=Q_{1,2}(u-2)Q_{2,1}(u)$$
and
$$\frac{Q_{0,1}(u)Q_{1,2}(u+2)-Q_{0,1}(u+2)Q_{1,2}(u)}{Q_{0,2}(u)}=Q_{1,1}(u+2).$$
The functions $T_{2,2}(1,s,u)$ and
$T_{2,2}(a,1,u)$ form a complete set of non-vanishing $T$-functions which
do not lie on the boundaries of the fat hook domain for the case of
the $gl(2|2)$ superalgebra.

\section{ Discussion  }

In this paper we have dealt with finite dimensional representations
of $gl(K|M)$ and the periodic spin chains based on the rational
$R$-matrices. We believe that the method is powerful enough to
incorporate various generalizations, such as extension to twisted
and open spin chains, to infinite-dimensional representations of
non-compact version of the symmetry group as well as to models with
more general $R$-matrices, including various exotic ones, as the
Hubbard $R$-matrix or the recently constructed AdS/CFT S-matrix. We
expect that the Hirota relation should be the same for all these
problems, with only difference being in the boundary and analyticity
conditions.

For example, the extension to spin chains with twisted boundary conditions
can be accomplished by replacing the polynomial $Q$-functions
by ``Bloch polynomials", i.e., functions of
the form $Q(u)=Ae^{\kappa u}\prod_j (u-u_j)$, where $\kappa$
is related to the twisted boundary condition.
This leads to appropriate simple modifications in the Bethe equations.

The extension to supersymmetric spin chains with trigonometric
$R$-matrices is also straightforward. The Hirota equation and the
boundary conditions remain the same. The only change is again in the
analytical properties of the $T$-functions: in the trigonometric
case they are ``trigonometric polynomials", i.e., finite products of
$\sin (\eta (u-u_j))$ or $\mbox{sinh}\, (\eta (u-u_j))$. A
hypothetical generalization to the elliptic case is much more
interesting. As far as we know, supersymmetric quantum spin chains
with elliptic $R$-matrices were never discussed in the literature.
On the other hand, the bosonic case suggests \cite{KLWZ} that the
functional relations between commuting integrals of motion are given
by the same Hirota equation. The solutions are sought in the form of
``elliptic polynomials", i.e., finite products of Jacobi
theta-functions $\theta (\eta (u-u_j)| \tau)$. From this point of
view, it would be interesting to analyze elliptic polynomial
solutions to the Hirota equation with boundary conditions of the fat
hook type. They might solve (as yet hypothetical) supersymmetric
quantum integrable models with elliptic $R$-matrices.

It is also important to elaborate,
within the framework of Hirota equations, a more
direct approach to spin chains in typical
representations of superalgebras (the ones having a continuous label).
One would like to understand them
not only in the sense of analytic continuation with respect to
the representation label (as we demonstrated in this paper by
simple examples) but also to find their place on other levels
of the construction. A characterization of solutions to the
Hirota equation that are responsible for spin chains
in typical representations would be of particular importance.

\section*{Acknowledgements}

We would like to thank I.~Cherednik, S.~Kho\-rosh\-kin, I.~Kos\-tov,
I.~Kri\-che\-ver, P.~Ku\-lish, M.~Na\-za\-rov, K.~Sa\-kai, D.~Ser\-ban,
V.~Ta\-ra\-sov, V.~Tol\-stoy, Z.~Tsu\-boi and P.~Vie\-ira  for
discussions at different stages of
this work. The work of V.K. has been partially supported by
European Union under the RTN contracts MRTN-CT-2004-512194 and by
the ANR program INT-AdS/CFT -ANR36ADSCSTZ. A.S. would like to thank
the LPTENS for the hospitality during his visit. The work of A.S.
was partially supported by the RFBR Grant No. 06-01-00627-a,
RFBR-DFG Grant No. 06-02-04012-a, DFG Grant 436 RUS 113/669-3, the
Program for Supporting Leading Scientific Schools (Grant No.
NSh-5332.2006.2), and the Heisenberg-Landau Program. A.~Z. thanks
LPTENS, where most of the work was done, for the hospitality during
his visit. The work of A.~Z. was partially supported by RFBR grant
06-02-17383, by grant RFBR-06-01-92054-$\mbox{CE}_{a}$, by the grant
for support of scientific schools Nsh-8004.2006.2, by the ANR
project GIMP No. ANR-05-BLAN-0029-01, by INTAS 03-51-6346 and by
grant NWO 047.017.015.

\section*{Appendix A. Superalgebras and their representations}

\def\theequation{A\arabic{equation}}
\setcounter{equation}{0}

For completeness, we list here some basic objects and notation
related to Lie superalgebras $gl(K|M)$ and their representations.
The main references are \cite{Kac}-\cite{GouldZhang}.

We use the notation $p(\alpha )$ for the ${\Bbb Z}_2$-grading of the
index $\alpha$: $p(\alpha ) =0$ if $\alpha$ is bosonic   and
$p(\alpha )=1$ if $\alpha$ is fermionic. The same notation is used
to denote the grading of any objects (vectors, operators,...) with
definite parity. Objects with definite parity are called
homogeneous.

A matrix $A$ with matrix elements $A^{i}_{j}$ (we imply that they
are usual numbers, i.e., $p(A^{i}_{j})=0$) is said to be even (odd)
if $p(i)+p(j)$ is even (odd) for all non-vanishing elements of $A$.
The $R$-matrix (\ref{F1}) is an even matrix.

Let $X$, $Y$ be two graded spaces, and let $e_i$, $f_{j}$ be the
corresponding (homogeneous) basis vectors. For any two vectors
$x=\sum_i x_i e_i$, $y=\sum_i y_i f_i$ we have
$$
x\otimes y = \sum_{i,j}(x_i e_i)\otimes (y_j f_j)=\sum_{i,j}
(-1)^{p(i)p(j)} x_i y_j \, (e_i \otimes f_j) \,,
$$
so the components of the vector $x\otimes y$ in the basis $e_i
\otimes f_j$ are $(-1)^{p(i)p(j)} x_i y_j$. The action of the
operator $A\otimes B$ in $X\otimes Y$ is defined on homogeneous
objects as $A\otimes B \, (x\otimes y)=(-1)^{p(x)p(y)} A(x)\otimes
B(y)$.  Matrix elements of the tensor product of even matrices
$A^{i}_{j}$, $B^{\alpha}_{\beta}$ are:
$$
(A\otimes B)^{i\alpha}_{j\beta}=(-1)^{p(\alpha ) (p(i)+p(j))}
A^{i}_{j} B^{\alpha}_{\beta}\,.
$$
This rule explains the origin of the sign factors in the graded
Yang-Baxter equation.

The Lie superalgebra $gl(K|M)$ can be most transparently defined
through its matrix realization: it is the set of block matrices
\begin{equation}\label{ABCD}
g= \left (
\begin{array}{cc}
A&B\\ C&D \end{array}\right )
\end{equation}
such that $A,B,C,D$ are respectively $K\times K$, $K\times M$,
$M\times K$ and $M\times M$ matrices. The even subalgebra
$gl(K|M)_0$ has $B=C=0$, the odd subalgebra $gl(K|M)_1$ has $A=D=0$.
For homogeneous elements, the bracket is given by
$$ [g,g']=gg' -(-1)^{p(g)p(g')}g' g\,. $$
Note that $gl(K|M)_0 = gl(K)\oplus gl (M)$. For elements $g$
realized as above the supertrace is defined by $\mbox{str}\, g =
\mbox{tr}\, A - \mbox{tr}\, D$.

A basis for $gl(K|M)$ consists of matrices $E_{ij}$ with entry $1$
at position $(i,j)$ and $0$ otherwise. A Cartan subalgebra of
$gl(K|M)$ is spanned by the elements $E_{ii}$, $i=1, \ldots , K+M$.
The set of generators of $gl(K|M)$ consists of the $E_{ii}$ and the
elements $E_{i, i+1}$ and $E_{i+1, i}$, $i=1, \ldots , K+M-1$. The
space dual to the Cartan subalgebra is spanned by the linear forms
$\epsilon_i :  g\mapsto A_{ii}$ ($i=1, \ldots , K$) and $\delta_i :
g\mapsto D_{ii}$ ($i=1, \ldots , M$), where $g$ is given by
(\ref{ABCD}). Let us choose the basis in the space dual to the
Cartan subalgebra to be $\epsilon_1 , \ldots , \epsilon_K, \delta_1
, \ldots , \delta_M$. On this space there is a bilinear form induced
by the supertrace in the superalgebra:
$$
(\epsilon_i | \epsilon_j )= \delta_{ij}\,, \quad (\epsilon_i |
\delta_j ) = (\delta_i | \epsilon_j ) =0\,, \quad (\delta_i |
\delta_j )=-\delta_{ij}\,.
$$
Even (bosonic) roots are $\epsilon_i - \epsilon_j$ and $\delta_i -
\delta_j$ ($i\neq j$), odd (fermionic) roots are $\pm (\epsilon_i
-\delta_j)$. There are several choices of simple root systems. The
distinguished simple root system has the form $\alpha_i =\epsilon_i
- \epsilon_{i+1}$ for $i=1, \ldots , K-1$,
$\alpha_{K}=\epsilon_{K}-\delta_1$, $\alpha_{K+j}= \delta_j
-\delta_{j+1}$ for $j=1, \ldots , M-1$. All other simple root
systems are obtained from this one by reflections with respect to
odd roots $\alpha$ with $(\alpha |\alpha )=0$.

Elements of the space dual to the Cartan subalgebra are called the
weights. The weight is an expression of the form
$$
\Lambda = \sum_{i=1}^{K}\Lambda_i \epsilon_i  + \sum_{j=1}^{M}\bar
\Lambda_j \delta_j\,.
$$
Let $\lambda$ be a Young diagram formed by a sequence of
non-negative non-increasing integers: $\lambda = (\lambda_1,
\lambda_2, \ldots )$, $\lambda_1 \geq \lambda_2 \geq \ldots \geq 0$.
To such a diagram one assigns the weight $\Lambda_i =\lambda_i$
($i=1, \ldots , K)$, $\bar \Lambda_j = \bar \lambda_j$ $(j= 1,
\ldots , M)$, where $\bar \lambda_j =\mbox{max}\, (\lambda_j' -K, \,
0)$ and $\lambda'_j$ is the height of the $j$-th column of the
diagram $\lambda$. It is implied that $\lambda_{K+1} \leq M$.

A Kac-Dynkin label is a sequence $b_1, b_2, \ldots , b_K, \ldots ,
b_{K+M-1}$, where all $b_j$ except $b_K$ are non-negative integers
while $b_K$ may be any real number. There is a one-to-one
correspondence between finite-dimensional irreducible
representations (irreps) of the superalgebra $gl(K|M)$ and the
Kac-Dynkin labels \cite{Kac}. We consider covariant tensor irreps of
the superalgebra $gl(K|M)$. One can assign the following Kac-Dynkin
label to any highest weight $\Lambda$ associated with a Young
diagram $\lambda$ as above:
\begin{equation}\label{KacDynkin}
\begin{array}{ll}
b_i =\lambda_i - \lambda_{i+1}\,, & i=1, \ldots , K-1\,,
\\&\\
b_K = \lambda_K +\bar \lambda_1\,, &
\\&\\
b_{j+K} =\bar \lambda_j - \bar \lambda_{j+1}\,, & j=1, \ldots ,
M-1\,.
\end{array}
\end{equation}
Therefore, one associates a tensor irrep of $gl(K|M)$ to any Young
diagram. (The diagrams containing a rectangular subdiagram with
$K+1$ rows and $M+1$ columns correspond to vanishing
representations, so such diagrams are illegal, similarly to diagrams
containing $K+1$ rows for $gl(K)$.) However, for superalgebras this
correspondence is not one-to-one. Different Young diagrams may
correspond to equivalent irreps (i.e., to the same Kac-Dynkin
label). In particular, this is the case for rectangular
diagrams\footnote{For brevity, we call irreps corresponding to the
Young diagrams of rectangular shape {\it rectangular irreps}.} when
$M+n$ columns of $K$ boxes are replaced by $M$ columns of $K+n$
boxes \cite{BMR}.

There is a large class of finite-dimensional irreps of $gl(K|M)$
which cannot be associated with a Young diagram. Given an
irreducible tensor representation with the highest weight $\Lambda$,
there is a one-parametric family of finite-dimensional irreps  with
the highest weight $\Lambda (c)=\Lambda +
c\sum_{i=1}^{K}\epsilon_i$, where $c$ is a real parameter. This
yields the Kac-Dynkin label with a non-integer $b_K= \lambda_K +\bar
\lambda_1 +c$.

One distinguishes typical and atypical irreps \cite{Kac} (in
physical terminology, long and short irreps, respectively).
An irrep with the highest weight $\Lambda$ is atypical iff there
exists at least one pair $(i,j)$, $i=1, \ldots , K$,
$j=1, \ldots , M$, such that
\begin{equation}\label{ATYP}
(\Lambda +\rho , \epsilon_i - \delta_j )=0\,,
\end{equation}
where
$$
\rho = \frac{1}{2}\sum_{i=1}^{K}(K-M-2i+1)\epsilon_i
+\frac{1}{2}\sum_{j=1}^{M}(K+M-2j+1)\delta_j\,.
$$
For typical irreps, there is no such pair $(i,j)$ that
(\ref{ATYP}) holds.
All irreps with $b_K \geq M$ are typical. All rectangular irreps
corresponding to diagrams having $a$ rows and $s$ columns with $a<K$
or $s<M$ are atypical.

\section*{Appendix B. Auxiliary linear problems for the Hirota
equation}

\def\theequation{B\arabic{equation}}
\setcounter{equation}{0}

Let $\tau = \tau (p_1 , p_2 , p_3)$ be a function of three
variables. For brevity, we denote
$$
\tau (p_1 +1, p_2 , p_3):= \tau_1,\quad  \tau (p_1, p_2 +1 , p_3):=
\tau_2, \quad \tau (p_1 +1, p_2 +1 , p_3):= \tau_{12}, \quad
\mbox{etc.}
$$
Let $\alpha \beta \gamma $ be any cyclic permutation of $123$.
Consider the following system of three linear equations for a
function $\psi =\psi (p_1 , p_2 , p_3)$:
\begin{equation}\label{A1}
\left ( e^{\p_{\alpha}}+\lambda_{\gamma}\, \frac{\tau \,
\tau_{\alpha \beta}}{\tau_{\alpha}\tau_{\beta}} \right ) \psi =
e^{\p_{\beta}}\psi\,, \quad \alpha \beta \gamma = 123, \; 231, \;
312,
\end{equation}
where $\lambda_{\alpha}$ are parameters and $\p_{\alpha}\equiv \p/
\p_{p_{\alpha}}$. Their compatibility implies the Hirota equation
for $\tau$:
\begin{equation}\label{AH}
\lambda_1 \tau_1 \tau_{23}+\lambda_2 \tau_2 \tau_{13}+ \lambda_3
\tau_3 \tau_{12}=0\,.
\end{equation}

To see this, consider  the first and the second linear equations.
Their compatibility means that the difference operators
$$
e^{-\p_{1}} \left ( e^{\p_{2}} - \lambda_{3}\, \frac{\tau \,
\tau_{12}}{\tau_{1}\tau_{2}} \right ) \quad \mbox{and} \quad
e^{-\p_{3}} \left ( e^{\p_{2}} + \lambda_{1}\, \frac{\tau \,
\tau_{23}}{\tau_{2}\tau_{3}} \right )
$$
commute. A simple calculation shows that their commutativity is
equivalent to the condition that the function
$$
\frac{\lambda_1 \tau_1 \tau_{23}+ \lambda_3 \tau_3 \tau_{12}}{\tau_2
\tau_{13}}:= h(p_1, p_3)
$$
does not depend on $p_2$. Compatibility with the third linear
problem implies that the function $h$ must be a constant equal to
$-\lambda_3$, whence the Hirota equation follows. In the case when
the function $h$ can be fixed in some other way (for example, from
the boundary conditions), just two linear problems are enough to
represent the Hirota equation as a discrete zero curvature
condition. Moreover, the specific form of the coefficient functions
in the difference operators (\ref{A1}) implies that the
compatibility follows from the existence of {\it just one}
non-trivial common solution (cf. \cite{Krichever06}).

In terms of the function $\varphi = \psi \tau$ the linear problems
acquire the form
\begin{equation}\label{A2}
\tau_{\gamma}\varphi_{\beta}- \tau_{\beta}\varphi_{\gamma}
+\lambda_{\alpha}\tau_{\beta \gamma}\varphi =0 \,, \quad \quad
\alpha \beta \gamma = 123, \; 231, \; 312\,.
\end{equation} From the first and the second ones we have
$$
\tau_2 = \frac{\tau_3 \varphi_2 + \lambda_1
\tau_{23}\varphi}{\varphi_3}\,, \quad \tau_1 = \frac{\tau_3
\varphi_1 - \lambda_2 \tau_{13}\varphi}{\varphi_3}\,.
$$
Plugging this into the Hirota equation, we obtain another linear
problem compatible with the previous ones:
\begin{equation}\label{A3}
\lambda_1 \tau_{23}\varphi_1 + \lambda_2 \tau_{13}\varphi_2 +
\lambda_3 \tau_{12}\varphi_3 =0\,.
\end{equation}
The four linear problems can be combined into a single matrix
equation as follows \cite{Shinzawa-2}:
\begin{equation}\label{A4}
\left (
\begin{array}{cccc}
0 & \tau_3 & -\tau_2 & \lambda_1 \tau_{23}\\ &&&\\
-\tau_3 &0& \tau_1 &  \lambda_2 \tau_{13}\\ &&&\\
\tau_2 & -\tau_1 & 0 & \lambda_3 \tau_{12}\\ &&&\\
-\lambda_1 \tau_{23} & -\lambda_2 \tau_{13} & -\lambda_3 \tau_{12} &
0
\end{array} \right )
\left (
\begin{array}{c}
\varphi_1 \\ \\    \varphi_2 \\ \\ \varphi_3 \\ \\
\varphi
\end{array} \right ) =0\,.
\end{equation}
The determinant of the antisymmetric matrix in the left hand side is
equal to $(\lambda_1 \tau_1 \tau_{23} +  \lambda_2 \tau_2 \tau_{13}
+\lambda_3 \tau_3 \tau_{12} )^2$. It vanishes if $\tau$ obeys the
Hirota equation, and the rank of the matrix is 2 in this case, so
only two of the four equations are linearly independent.

One may regard system (\ref{A2}) as linear equations for $\tau$ with
coefficients $\varphi$. Shifting the variables $p_{\beta} \to
p_{\beta}-1$, $p_{\gamma} \to p_{\gamma}-1$, and then passing to the
new variables $p_{1,2,3}\to -p_{1,2,3}$, one sees that the form of
this system remains the same. Since the Hirota equation is invariant
under the simultaneous change of the signs of all variables, the
compatibility condition is the same Hirota equation for $\varphi$:
\begin{equation}\label{A5}
\lambda_1 \varphi_{23}\varphi_1 + \lambda_2 \varphi_{13}\varphi_2 +
\lambda_3 \varphi_{12}\varphi_3 =0\,.
\end{equation}

Let us pass to the ``laboratory" variables $x_1, x_2 , x_3$
according to the formulas
$$
\begin{array}{l}
p_1 =\frac{1}{2}(- \varepsilon_1 x_1 \! +\! \varepsilon_2
x_2 \! +\! \varepsilon_3 x_3) \\ \\
p_2 =\frac{1}{2}(\, \varepsilon_1 x_1 -\varepsilon_2 x_2 +
\varepsilon_3 x_3) \\ \\
p_3 =\frac{1}{2}(\, \varepsilon_1 x_1 +\varepsilon_2 x_2 -
\varepsilon_3 x_3)\,,
\end{array}
$$
where $\varepsilon_{\alpha}=\pm 1$ is a fixed set of signs (clearly,
there are $2^3 = 8$ possible choices). The inverse transformation
reads
$$
x_1 =\varepsilon_1 (p_2 + p_3)\,, \quad x_2 =\varepsilon_2 (p_1 +
p_3)\,, \quad x_3 =\varepsilon_3 (p_1 + p_2) \,.
$$
We introduce the functions $T(x_1, x_2 , x_3)= \tau (p_1 , p_2 , p_3
)$ and $F(x_1, x_2 , x_3)= \varphi (p_1 , p_2 , p_3 )$, where the
variables $x_{\alpha}$ and $p_{\alpha}$ are connected by the
formulas given above. In the ``laboratory" variables, the system of
four linear problems takes the form
\begin{equation}\label{A6}
\left (
\begin{array}{cccc}
0 & T_{12} & -T_{13} & \lambda_1 T_{1123}\\ &&&\\
-T_{12} &0& T_{23} &  \lambda_2 T_{1223}\\ &&&\\
T_{13} & -T_{23} & 0 & \lambda_3 T_{1233}\\ &&&\\
-\lambda_1 T_{1123} & -\lambda_2 T_{1223} & -\lambda_3 T_{1233} & 0
\end{array} \right )
\left (
\begin{array}{c}
F_{23} \\ \\  F_{13} \\ \\ F_{12} \\ \\ F
\end{array} \right ) =0\,,
\end{equation}
where we denote $T_1 \equiv T(x_1 \! +\! \varepsilon_1 , x_2 ,
x_3)$, $T_{12} \equiv T(x_1 \! +\! \varepsilon_1 , x_2 \! +\!
\varepsilon_2, x_3)$, $T_{1123} \equiv T(x_1 \! +\! 2\varepsilon_1 ,
x_2 \! +\! \varepsilon_2, x_3\! +\! \varepsilon_3)$, etc (and
similarly for $F$).

The compatibility of these linear problems implies the Hirota
equation
$$
\lambda_1 T_{1123}T_{23} + \lambda_2 T_{1223}T_{13} + \lambda_3
T_{1233}T_{12} =0\,.
$$
Shifting the variables ($x_{\alpha}\to
x_{\alpha}-\varepsilon_{\alpha}$), we get the equation
$$
\lambda_1 T(x_1 \! +\! \varepsilon_1, x_2 , x_3) T(x_1 \! -\!
\varepsilon_1, x_2 , x_3) + \lambda_2 T(x_1, x_2 \! +\!
\varepsilon_2 , x_3) T(x_1, x_2 \! -\!\varepsilon_2, x_3) +
\lambda_3 T(x_1, x_2, x_3 \! +\! \varepsilon_3) T(x_1, x_2, x_3 \!
-\! \varepsilon_3) =0
$$
Note that it is the same for any choice of the $\varepsilon$'s.
Linear equations (\ref{A6}) provide inequivalent B\"acklund
transformations for it. However, only four of them (corresponding,
say, to the choices $\varepsilon_1 = \varepsilon_2 =\varepsilon_3
=1$, $-\varepsilon_1 = \varepsilon_2 =\varepsilon_3 =1$,
$\varepsilon_1 = -\varepsilon_2 =\varepsilon_3 =1$, $\varepsilon_1 =
\varepsilon_2 =-\varepsilon_3 =1$) are actually different because
the simultaneous change of signs of all $\varepsilon$'s means
passing to the ``conjugate" system of linear problems, where the
roles of $T$ and $F$ are interchanged. Equation (\ref{LP53}) in the
main text corresponds to the choice $x_1 =a$, $x_2 =s$, $x_3 =u$,
$\lambda_1 =\lambda_2 =-\lambda_3 =1$, $\varepsilon_1 =
-\varepsilon_2 =\varepsilon_3 =1$.

\section*{Appendix C. On bilinear
equations (\ref{BT1,2-1})-(\ref{BT1,2-3n})
and their compatibility with
(\ref{LINPRT1}), (\ref{LINPRT2})}

\def\theequation{C\arabic{equation}}
\setcounter{equation}{0}

Here we complete the proof of equation (\ref{BT1,2-1})
and present some more details on
compatibility of bilinear equations (\ref{LINPRT1}), (\ref{LINPRT2})
and (\ref{BT1,2-1})-(\ref{BT1,2-3n}).

As we have seen, the pair of equations (\ref{alt1}), (\ref{alt2})
implies the relation
\begin{equation}\label{bileqTap}
\begin{array}{c}
T_{k,m}(a, s\! +\! 1, u)T_{k+1, m+1}(a,s, u\! +\! 1) - T_{k,m}(a,s,
u\! +\! 1) T_{k+1, m+1}(a, s\! +\! 1, u)
\\ \\
=\,\,\, f_{k,m}(a, u\! +\! s)\, T_{k+1, m}(a,s, u\! +\! 1) T_{k,
m+1}(a, s\! +\! 1, u)\,,
\end{array}
\end{equation}
where $f_{k,m}(a, u\! +\! s)$ is an arbitrary function of $k,m$ and
$a, \, u+s$. In the same way, the pair of equations
(\ref{alt3}), (\ref{alt4}) implies the relation
\begin{equation}\label{bileqTap1}
\begin{array}{c}
T_{k,m}(a\! -\! 1,s,u)T_{k+1,m+1}(a,s,u\! +\! 1)-
T_{k,m}(a,s,u\! +\! 1)T_{k+1,m+1}(a\! -\! 1,s,u)
\\ \\
=\,\,\, g_{k,m}(s, u\! -\! a) \, T_{k,m+1}(a\! -\! 1,s,u)
T_{k+1,m}(a,s,u\! +\! 1)\,,
\end{array}
\end{equation}
where $g_{k,m}(s, u\! -\! a)$ is an arbitrary function of $k,m$ and
$s, \, u-a$. On the other hand, we have the identity
\begin{eqnarray}
&&\frac{T_{k,m}(a,s+1,u)
T_{k+1,m+1}(a,s,u+1)-T_{k,m}(a,s,u+1)T_{k+1,m+1}(a,s+1,u)}
{T_{k,m+1}(a,s+1,u)T_{k+1,m}(a,s,u+1)}\\
&=&\frac{T_{k,m}(a-1,s,u)T_{k+1,m+1}(a,s,u+1)-
T_{k,m}(a,s,u+1)T_{k+1,m+1}(a-1,s,u)}
{T_{k,m+1}(a-1,s,u)T_{k+1,m}(a,s,u+1)}\,,
\end{eqnarray}
which is straightforwardly proved by passing to the common
denominator, grouping together similar terms and using equations
(\ref{LINPRT1}), (\ref{LINPRT2}). Its left hand side is
$f_{k,m}(a, u\! +\! s)$, and the right hand side is
$g_{k,m}(s, u\! -\! a)$. Therefore, we conclude that
$$
f_{k,m}(a, u\! +\! s)= g_{k,m}(s, u\! -\! a)\,.
$$
This equality is possible only if $f_{k,m}(a, u\! +\! s)$
actually depends on the difference of its arguments,
i.e., on $u+s-a$. This fact has been used in section
3.4 to prove that $f_{k,m}(a, u\! +\! s)=1$.


The system of bilinear equations (\ref{BT1,2-1})-(\ref{BT1,2-3n})
can be represented in the form
\begin{eqnarray}
\label{BT1,2-1e}
(e^{\partial_u} -
e^{\partial_s})~\frac{T_{k+1,m+1}(a,s,u)}{T_{k,m}(a,s,u)}
&=&\frac{T_{k,m+1}(a,s+1,u)~T_{k+1,m}
(a,s,u+1)}{T_{k,m}(a,s+1,u)~T_{k,m}(a,s,u+1)},\\
\label{BT1,2-4e} (e^{\partial_a} -
e^{-\partial_u})~\frac{T_{k+1,m+1}(a,s,u)}{T_{k,m}(a,s,u)}
&=&\frac{T_{k,m+1}(a,s,u-1)~T_{k+1,m}
(a+1,s,u)}{T_{k,m}(a,s,u-1)~T_{k,m}(a+1,s,u)},\\
\label{BT1,2-3e} (e^{\partial_a} -
e^{\partial_s})~\frac{T_{k+1,m+1}(a,s,u)}{T_{k,m}(a,s,u)}
&=&\frac{T_{k,m+1}(a,s+1,u)~T_{k+1,m}
(a+1,s,u)}{T_{k,m}(a,s+1,u)~T_{k,m}(a+1,s,u)},\\
\label{BT1,2-5e} (e^{\partial_s} -
e^{-\partial_u})~\frac{T_{k+1,m+1}(a,s,u)}{T_{k,m}(a,s,u)} &=&
\frac{T_{k,m+1}(a-1,s+1,u-1)~T_{k+1,m}
(a+1,s,u)}{T_{k,m}(a,s,u-1)~T_{k,m}(a,s+1,u)}\,,\\
\label{BT1,2-2e}
(e^{\partial_a} -
e^{\partial_u})~\frac{T_{k+1,m+1}(a,s,u)}{T_{k,m}(a,s,u)}
&=&\frac{T_{k,m+1}(a,s+1,u)~T_{k+1,m}
(a+1,s-1,u+1)}{T_{k,m}(a,s,u+1)~T_{k,m}(a+1,s,u)},\\
\label{BT1,2-3ne} (e^{\partial_s} -
e^{-\partial_a})~\frac{T_{k+1,m+1}(a,s,u)}{T_{k,m}(a,s,u)}
&=&\frac{T_{k,m+1}(a-1,s+1,u-1)~T_{k+1,m}
(a,s,u+1)}{T_{k,m}(a-1,s,u)~T_{k,m}(a,s+1,u)}\,.
\end{eqnarray}
These equations have a similar structure:
different commuting difference operators in the left hand sides
act on the same fraction of the $T$-functions.
It is a simple exercise to verify that these
equations are self-consistent if equations
(\ref{LINPRT1}), (\ref{LINPRT2}) are satisfied.
The self-consistency conditions obviously follow from the
commutativity of the difference operators in the left hand sides.
Now, let us demonstrate that all the equations
from (\ref{LINPRT1}), (\ref{LINPRT2}) are encoded in
the system (\ref{BT1,2-1e})-(\ref{BT1,2-3ne}). To this end, we
subtract the sum of equations (\ref{BT1,2-1e}) and (\ref{BT1,2-2e})
(equations (\ref{BT1,2-3e}) and (\ref{BT1,2-5e}))
from eq.~(\ref{BT1,2-3e})
(eq.~(\ref{BT1,2-4e})) and find that the resulting equation
reproduces the first equation from (\ref{LINPRT2})
(respectively, the second
equation from (\ref{LINPRT1})).
The remaining
second (first) equation from BT2 (\ref{LINPRT2}) (BT1 (\ref{LINPRT1}))
results from the consistency condition of equations (\ref{BT1,2-4e}) and
(\ref{BT1,2-5e}) (respectively,
equations (\ref{BT1,2-1e}) and (\ref{BT1,2-2e})).

The connections between the different bilinear relations
for the transfer
matrices $T_{k,m}(a,s,u)$
become more transparent if one represents them in a matrix
form. Namely, two pairs of equations
(\ref{BT1,2-1}), (\ref{BT1,2-2}) and (\ref{BT1,2-4}), (\ref{BT1,2-5})
can be rewritten as
\begin{eqnarray}
&&\label{matrix1} \left ( \begin{array}{cc}
T_{k,m}(a, s\! +\! 1, u)& -T_{k\!+\!1,m\!+\!1}(a, s\! +\! 1, u)\\ &\\
-T_{k,m}(a\!+\!1, s, u) & T_{k\!+\!1,m\!+\!1}(a\! +\! 1, s, u)
\end{array} \right )
\left ( \begin{array}{c} T_{k\!+\!1,m\!+\!1}(a,\, s,\, u\! +\! 1)\\
\\ T_{k,m}(a,s,u\! +\! 1)
\end{array} \right ) \nn\\
&&= \,\, T_{k,m\!+\!1}(a,s\!+\!1,u) \left ( \begin{array}{c}
T_{k\!+\!1,m}(a, s, u\!+\!1)\\ \\
T_{k\!+\!1,m}(a \! +\! 1, s\!-\!1,u\!+\!1)
\end{array} \right )
\end{eqnarray}
and
\begin{eqnarray}
&&\label{matrix2} \left ( \begin{array}{cc}
-T_{k,m}(a, s\! +\! 1, u)& T_{k\!+\!1,m\!+\!1}(a, s\! +\! 1, u)\\ &\\
-T_{k,m}(a\!+\!1, s, u) & T_{k\!+\!1,m\!+\!1}(a\! +\! 1, s, u)
\end{array} \right )
\left ( \begin{array}{c} T_{k\!+\!1,m\!+\!1}(a,\, s,\, u\! -\! 1)\\
\\ T_{k,m}(a,s,u\! -\! 1)
\end{array} \right ) \nn\\
&&= \,\, T_{k\!+\!1,m}(a\!+\!1,s,u) \left ( \begin{array}{c}
T_{k,m\!+\!1}(a \! -\! 1, s\!+\!1, u\!-\!1)\\ \\
T_{k,m\!+\!1}(a, s, u\!-\!1)
\end{array} \right ),
\end{eqnarray}
respectively. Then, multiplying both sides of equations (\ref{matrix1})
and (\ref{matrix2}) by the matrices inverse to the
ones in their left hand sides
(for the calculation of the determinant eq.~(\ref{BT1,2-3})
is used), the resulting
equations reproduce respectively the first and the second equations
from the linear systems BT1 and BT2
(\ref{LINPRT1}), (\ref{LINPRT2}). Similarly,
representing the two equations of
the last line of the matrix equation (\ref{LP531}) in the matrix
form
\begin{eqnarray}
&&\label{matrix3} \left ( \begin{array}{cc}
-T_{k\! +\! 1,m\! +\! 1}(a\! -\! 1, s, u)&
T_{k\!+\!1,m\!+\!1}(a,s, u\! -\! 1)\\ &\\
-T_{k,m}(a\!-\!1, s, u) & T_{k,m}(a,s,u\! -\! 1)
\end{array} \right )
\left ( \begin{array}{c} T_{k\!+\!1,m}(a\! +\! 1,\, s,u)\\
\\ T_{k\! +\! 1,m}(a,s,u\! +\! 1)
\end{array} \right ) \nn\\
&&= \,\, T_{k\!+\!1,m}(a,s\!-\!1,u) \left ( \begin{array}{c}
T_{k\! +\! 1,m\!+\!1}(a, s\!+\!1, u)\\ \\
T_{k,m}(a, s\!+\!1,u)
\end{array} \right ),
\end{eqnarray}
multiplying the both sides by the matrix inverse to the one in the
l.h.s., and using eq.~(\ref{BT1,2-2}) to calculate the determinant
of this matrix, one arrives at equations
(\ref{BT1,2-5}), (\ref{BT1,2-3n}).

\section*{Appendix D. An alternative derivation of
the Bethe equations}

\def\theequation{D\arabic{equation}}
\setcounter{equation}{0}

Here we give a direct derivation of the Bethe equations from the
pair of linear problems (B\"acklund transformation) (\ref{LINPRT1}).
It does not use the Hirota equation for the $Q$-functions.

We know that we can use the $k$ (respectively, $m$) flow of the BT1
(resp., BT2) to decrease rank $K$ (resp., $M$) of the original
problem to $K-1$ (resp., $M-1$). For the step by step passing from
$(K,M)$ to $(0,0)$ we can choose any zigzag path of the type drown
in Fig.~\ref{fig:UNDRESS}. The interior boundaries of the domain of
non-vanishing $T$'s move towards the exterior ones until the domain
collapses to the horizontal and vertical lines, so that the original
problem gets ``undressed". The final solution can be formulated in
terms of Bethe equations. Let us derive them for the simplest path
with just one turn.

First, using the transformation BT1 at each step, we move the
horizontal interior boundary  $a=K,\, s\geq M$ to the half-line
$a=0,\,  s\geq M$, as shown in Fig.~\ref{fig:SHIFT1}. Consider the
second equation from \eq{LINPRT1} at $s=0$ (Fig.~\ref{fig:LINPAP1},
position 4):
\begin{equation}
\label{LINPRTQ}
  T_{k,M}(a,1,u\! +\! 1)Q_{k-1,M}(u\! +\! a)
-Q_{k,M}(u\! +\! a)T_{k-1,M}(a,1,u\! +\! 1) = Q_{k,M}(u\! +\! a +\!
2) T_{k-1,M}(a\! -\! 1,1,u),
\end{equation}
where $k=1, \dots , K$, or, denoting $P^a_k(u)\equiv
T_{k,M}(a,1,u-a+1)$, $Q_{k,M}(u)\equiv Q_k(u)$:
\begin{equation}
\label{LINPRTP}
  P^a_k(u)Q_{k-1}(u)-P^a_{k-1}(u)Q_k(u) =
P^{a-1}_{k-1}(u-2)Q_{k}(u+2),\quad
   k=1,\dots, K.
\end{equation}
We know that $P^a_k(u)$ and $Q_k(u)$ should be polynomials in $u$.
We set:
\begin{equation}
\label{QPOL}
  Q_k(u) =  \prod_{j=1}^{J_k}\(u-u^{(k)}_j\)\,.
\end{equation}
Taking equation (\ref{LINPRTP}) at zeroes of $Q_{k-1}(u)$,
$Q_{k}(u)$ and $Q_{k}(u+2)$, we obtain the equations
\begin{eqnarray}
\label{LINPRTPP}
 -P^a_{k-1}\(u^{(k-1)}_j\)Q_k\(u^{(k-1)}_j\)
&=& P^{a-1}_{k-1}\(u^{(k-1)}_j-2\)Q_{k}\(u^{(k-1)}_j+2\),\nn\\
 P^a_{k}\(u^{(k)}_j\)Q_{k-1}\(u^{(k)}_j\)
&=& P^{a-1}_{k-1}\(u^{(k)}_j-2\)Q_{k}\(u^{(k)}_j+2\),\nn\\
 P^a_{k}\(u^{(k)}_j-2\)Q_{k-1}\(u^{(k)}_j-2\)
&=& P_{k-1}^a\(u^{(k)}_j -2\)Q_{k}\(u^{(k)}_j-2\).
\end{eqnarray}
Dividing the second equation (with the shift $a\to a+1$) by the
third one and excluding the ratio of $P$'s with the help of the
first equation, we arrive at the following standard system of Bethe
equations:
\begin{equation}
\label{BETHER}
\frac{Q_{t+1}\(u^{(t)}_j+2\)Q_{t}\(u^{(t)}_j-2\)Q_{t-1}\(u^{(t)}_j\)}
{Q_{t+1}\(u^{(t)}_j\)Q_{t}\(u^{(t)}_j+2\)Q_{t-1}\(u^{(t)}_j-2\)}=-1
\quad (j=1, \ldots J_t \,, \; t=1, \ldots, K-1).
\end{equation}

Then, using the transformation BT2 \eq{LINPRT2} at each step, we
move (as shown in Fig.~\ref{fig:SHIFT2}) the vertical interior
boundary $s=M,\, a\geq 0$ to the half-line $s=0,\, a\geq 0$.
Consider the second equation of (\ref{LINPRT2}) at $s=0$
(Fig.~\ref{fig:LINPAP1}, position 4). In a similar way, we get the
Bethe equations for the roots $u_{j}^{(m-M)}$ of $Q_{0,m}(u):=
Q_{m-M}(u)$:
\begin{equation}
\label{BETHEL}
\frac{Q_{t+1}\(u^{(t)}_j\)Q_{t}\(u^{(t)}_j-2\)Q_{t-1}\(u^{(t)}_j+2\)}
{Q_{t+1}\(u^{(t)}_j -2\)Q_{t}\(u^{(t)}_j+2\)Q_{t-1}\(u^{(t)}_j\)}=-1
\quad (j=1, \ldots J_t \,, \; t=-1, \ldots, -M+1).
\end{equation}
To get the missing equation for zeros $u^{(0)}_j$ of the function
$Q_0(u)= Q_{0,M}(u)$, we use the second equation of BT1 at $k=1,
m=M$, and the second equation of BT2 at $k=0, m=M$:
\begin{eqnarray}
\label{LINPRFERMI}
  P_1^{a}(u)Q_{0}(u)-P_0^{a}(u)Q_{1}(u) &=& P_0^{a-1}(u-2)Q_{1}(u+2),\nn\\
 R_{1}^{a}(u)Q_{0}(u)- P_0^{a}(u)Q_{-1}(u) &=&
 P_0^{a-1}(u-2)Q_{-1}(u+2),
\end{eqnarray}
where $P_{1}^{a}(u)=T_{1,M}(a,1,u-a+1)$,
$R_{1}^{a}(u)=T_{0,M-1}(a,1,u-a+1)$. At the ``fermionic" roots of
$Q_{0}(u)= \prod_{j=1}^{J_0}\(u-u^{(0)}_j\)$ the ratio of these two
 equations gives the missing system of
 Bethe equation:
\begin{equation}\label{BETHEF}
\frac{Q_{1}\(u^{(0)}_j\)Q_{-1}\(u^{(0)}_j+2\)}
{Q_{1}\(u^{(0)}_j+2\)Q_{-1}\(u^{(0)}_j\)}=1\,, \quad j=1, \ldots
J_0\,.
\end{equation}
The ``boundary conditions" for the system of nested Bethe equations
are: $Q_{K}(u)=Q_{K,M}(u) =\phi (u)$ (a given polynomial function),
$Q_{-M}(u)=Q_{0,0}(u)=1$.

Note that our numbering of the $Q$'s here differs slightly from the
one in section 4.5: the $Q$'s here are numbered from $-M$ to $K$
while the same $Q$'s there have the numbers from $0$ to $K+M$.
Taking this into account, it is easy to see that the chain of Bethe
equations obtained here coincides with (\ref{Bethe7}), where the
signs are chosen as $p_1=p_2=\ldots =p_M =-1$,
$p_{M+1}=p_{M+2}=\ldots =p_{M+K}=1$.

This is only one  of possible sets of Bethe equations for the same
model. We could also apply the elementary moves BT1 and BT2 in any
other order. At each change of the direction we obtain an equation
of the fermionic type (\ref{BETHEF}). The different sets of
equations lead to the same solution for the $T$-functions
$T(a,s,u)\equiv T_{K,M}(a,s,u)$ at the highest level of the
hierarchy. The different systems of Bethe equations are known to be
related by certain ``duality transformations"
\cite{Tsuboi-3,Beisert:2005di}. As we have seen in the main text,
they admit a natural explanation in terms of a discrete ``zero
curvature condition" on the $(k,m)$ lattice.

\section*{Appendix E.  Comparison of the results for $gl(2|1)$
algebra with  the results of \cite{pfannmuller-1996-479}}

\def\theequation{E\arabic{equation}}
\setcounter{equation}{0}

Let us compare eq.(6.1) of \cite{pfannmuller-1996-479}, with
the notation $\mu=- \half u,\, \lambda_k=\half u^{(2)}_k,\, \nu_k=\half
u^{(1)}_k,\, b=s+\half$,
\begin{eqnarray*}
 Q_1(u)=\prod_{j=1}^{J_u}\(u-u^{(1)}_j\),\qquad
 Q_2(u)=\prod_{j=1}^{J_v}\(u-u^{(2)}_j\),\qquad
 \phi_\pm(u)=\prod_{j=1}^N\(u\pm r_j\),
\end{eqnarray*}
with our eq.~(\ref{TSR}).
The Baxter $TQ$-relation from \cite{pfannmuller-1996-479} for the
$(b,1/2)$ irrep in the auxiliary space and $(b_i,1/2)$ irreps
at the sites of the chain now reads as follows:
\begin{eqnarray}\label{FRPF}
 T_{ s |\{\r_l \}}(u)&=&\qquad\qquad\qquad\frac{Q_2(u+s)}{Q_2(u-s)}
 \nn\\
 &-&
 \frac{Q_1(u-s+1)}{Q_1(u-s-1)}\frac{Q_2(u+s)}{Q_2(u-s)}\qquad\quad
\frac{\phi_+(u-s)}{\phi_-(u-s)}
 \nn\\
&-& \frac{Q_1(u-s-3)}{Q_1(u-s-1)} \frac{Q_2(u+s)}{Q_2(u-s-2)}\quad
\frac{\phi_+(u-s)}{\phi_-(u-s)}
 \nn\\
&+&\qquad\qquad\qquad
 \frac{Q_2(u+s)}{Q_2(u-s-2)}\quad
 \frac{\phi_+(u-s)}{\phi_-(u-s)} \frac{\phi_+(u-s-2)}{\phi_-(u-s-2)}\,.
 \label{BAX}\end{eqnarray}
We see that it coincides
with our \eq{TSR}  up to the  redefinition of
$\varphi$-functions.

Let us give also the Bethe equations ensuring the polynomiality of
$T_{s|\{r_l\}}(u)$:
\begin{itemize}
\item[1.]  Canceling poles at $u=u^{(2)}_j+s$,
\begin{eqnarray}\label{FRAHM1}
 \frac{\phi_-(u^{(2)}_j)}{\phi_+(u^{(2)}_j)}=
\frac{Q_1(u^{(2)}_j+1)}{Q_1(u^{(2)}_j-1)}
\end{eqnarray}
we arrive at the Bethe equations for the fermionic node.
\item[2.] Canceling the poles at $u=u^{(1)}_j+s+1$,
\begin{eqnarray}\label{FRAHM2}
-1=\frac{Q_1(u^{(1)}_j+2)}{Q_1(u^{(1)}_j-2)}
\frac{Q_2(u^{(1)}_j-1)}{Q_2(u^{(1)}_j+1)}
\end{eqnarray}
 we get the Bethe equations for the bosonic node. (We note
 a mistake in the
denominator of the r.h.s. of eq.~(6.3) of \cite{pfannmuller-1996-479}.)
\item[3.] The condition of canceling
poles at $u=u^{(2)}_j+b+3$ is the same as the first one.
\end{itemize}



\end{document}